\documentclass[aps,showpacs,floatfix,superscriptaddress,a4paper,12pt]{book}
\usepackage{graphicx,amsmath,bbm,mathrsfs,amssymb,psfrag}

\newtheorem{theorem}{Theorem}
\newtheorem{corollary}{Corollary}
\newtheorem{lemma}{Lemma}
\newtheorem{definition}{Definition}
\newtheorem{proposition}{Proposition}
\newtheorem{remark}{Remark}
\newtheorem{example}{Example}

\newcommand{\qed}{\hfill $\Box$\medskip}
\newcommand{\proof}{\noindent{\bf Proof:\quad }}

\newcommand{\cI}{{\mathcal{I}_{spec}}}
\newcommand{\cQ}{{\mathcal{Q}}}
\newcommand{\cV}{{\mathcal{V}}}

\newcommand{\cP}{{\mathcal{P}}}
\newcommand{\SI}{{\mathbb{S}}}
\newcommand{\cCP}{{\mathcal{CP}}}

\newcommand{\cC}{{\mathcal{C}}}
\newcommand{\Tr}{{{\rm Tr}}}
\newcommand{\QI}{{{\rm Q}}}
\newcommand{\CI}{{\mathbb{C}}}
\newcommand{\NI}{{\mathbb{N}}}

\newcommand{\RI}{{\mathbb{R}}}
\newcommand{\id}{{{\rm id}}}
\newcommand{\rT}{{\rm T}}
\newcommand{\1}{{\mathbbm{1}}}
\newcommand{\HI}{{\mathbbm{H}}}

\newcommand{\hD}{\widehat{\Delta}}
\newcommand{\fI}{{\mathfrak{I}}}
\newcommand{\fJ}{{\mathfrak{J}}}

\begin{document}

\begin{titlepage}
  \begin{center}
  \hspace{-35pt}
    \vbox to0pt{%
    \vbox to\textheight{\vfil
      \vspace{-6.5cm}
    
    \includegraphics[width=15cm]{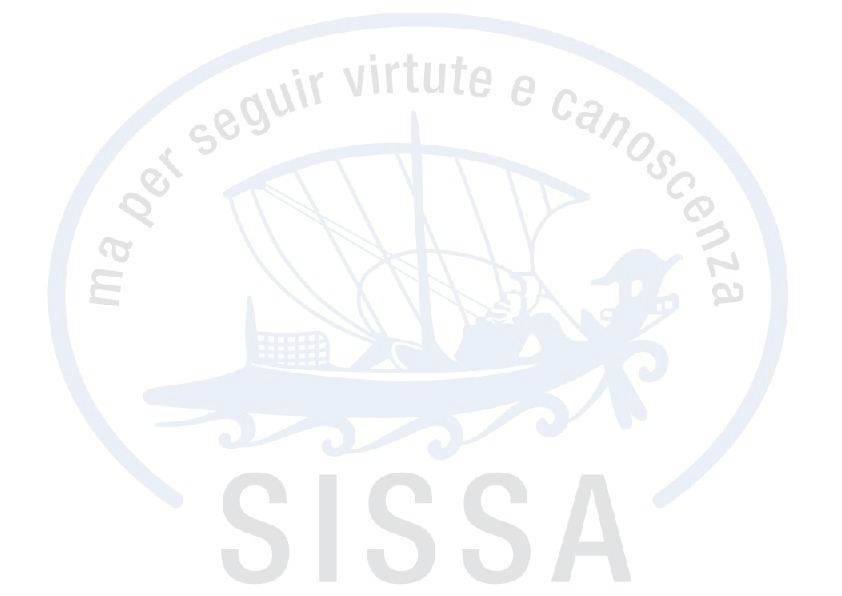}%
    \vfil}\vss}
  \end{center}
\vspace{-4cm}
\begin{center}
\hspace{-2.25cm} 
\begin{minipage}{.23\textwidth}%
  \includegraphics[height=2.25cm]{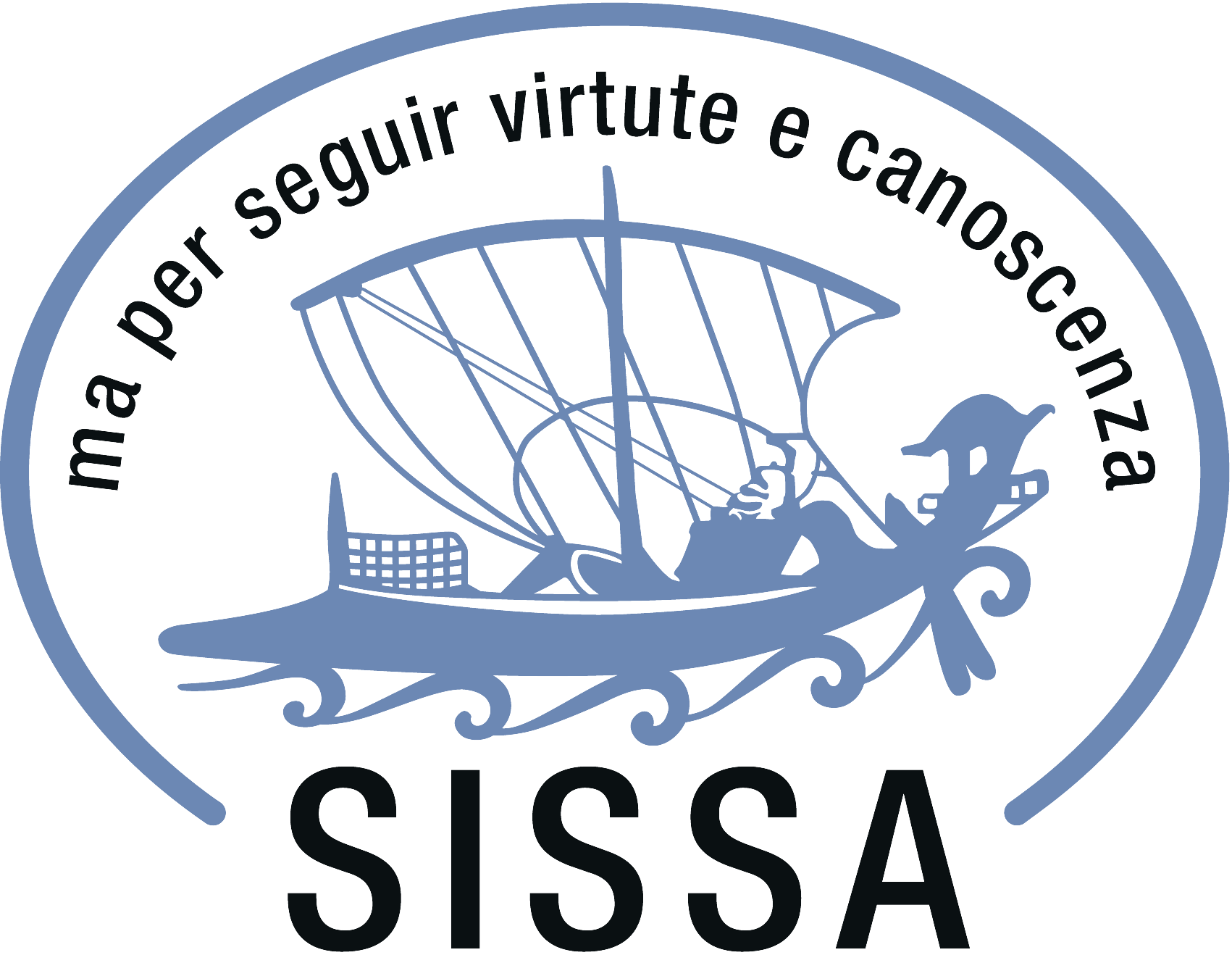}
\end{minipage}%
\begin{minipage}{.77\textwidth}%
	\begin{tabular}{l}
\scshape{{INTERNATIONAL SCHOOL FOR ADVANCED STUDIES}}\\
\hline
\scshape{
\large{Mathematical Physics Sector}}\\ 
\end{tabular}%
\end{minipage}
\end{center}

\begin{center}
\vspace{1cm}
\upshape{\Large{Academic Year 2011/2012}}\\

\vspace{3.3cm}
\linespread{1}
{\Large{\underline{\textbf{Entanglement Witnessing based on Positive Maps}}}} \\ \vspace{0.2cm}
{\Large{\textbf{ Characterization of a Class of Bipartite \\  \vspace{0.2cm} n $\times$ n qubit Systems}}}\\ 
\linespread{1}
\vspace{4.5cm}
\Large{thesis submitted for the degree of}\\
\Large{\textit{Doctor of Philosophy}}
\end{center}

\vspace{1cm}

\begin{tabbing}
\=\large{Advisor:} \hspace{7.7cm} \=\large{Candidate:}\\
\vspace{0.5cm}

\>\large{\bfseries{Prof. Fabio Benatti}} \>\large{\bfseries{Rayhane Karbalaii}}\\
\end{tabbing}

\begin{center}
\large{\today}
\end{center}

\end{titlepage}
\clearpage
\thispagestyle{empty}
\phantom{a}

\thispagestyle{empty}

\vskip 4cm

\hfill
To my mother,

\hfill
For her unbounded love and patience.

\newpage

\thispagestyle{empty}
\begin{center}

\includegraphics[scale=0.60]{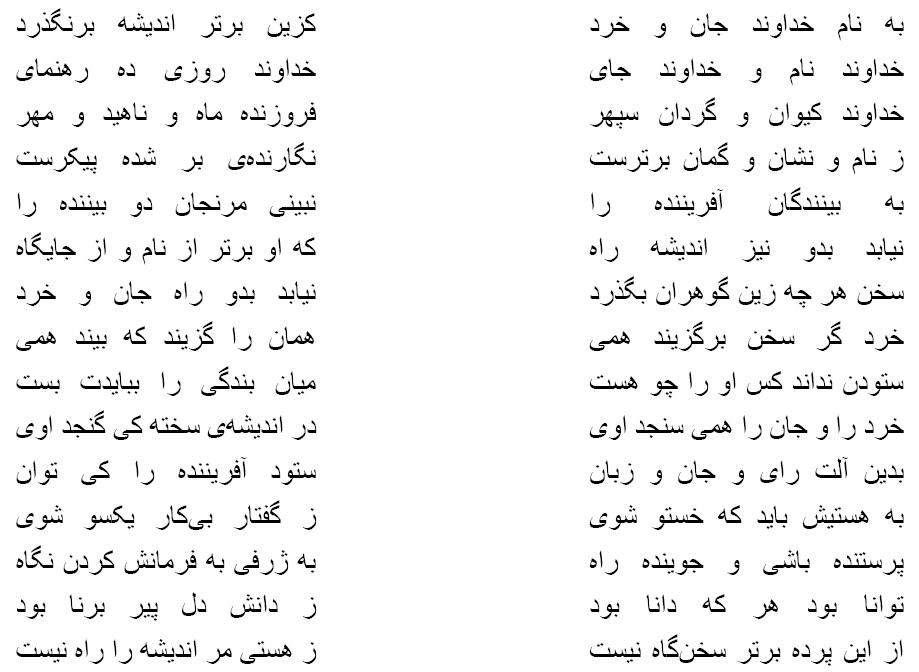}

\end{center}

\newpage

\thispagestyle{empty}

In the name of the Lord of both wisdom and mind,

To nothing sublimer can thought be applied,

The Lord of whatever is named or assigned

A place, the Sustainer of all and the Guide,

The Lord of Saturn and the turning sky,

Who causeth Venus, Sun, and Moon to shine,

Who is above conception, name, or sign,

The Artist of the heaven's jewellery!

Him thou canst see not though thy sight thou strain,

For thought itself will struggle to attain

To One above all name and place in vain,

Since mind and wisdom fail to penetrate

Beyond our elements, but operate

On matters that the senses render plain.

None then can praise God as He is. Observe

Thy duty: 'tis to gird thyself to serve.

He weigheth mind and wisdom; should He be

Encompassed by a thought that He hath weighed?

Can He be praised by such machinery

As this, with mind or soul or reason's aid?

Confess His being but affirm no more,

Adore Him and all other ways ignore,

Observing His commands. Thy source of might

Is knowledge: thus old hearts grow young again,

But things above the Veil surpass in height

All words: God's essence is beyond our ken.

\vskip 2cm

\hfill From Shahname
\vskip 1cm

\hfill
By Hakim Abu'l-Qasim Ferdowsi Tusi (940-1020 CE)

\newpage
\thispagestyle{empty}

\chapter*{Thanks to}
\thispagestyle{empty}

Fabio Benatti, who has been my advisor, in its very meaning, for the last three years. In these years, Fabio helped me to grow up personally and mathematically, without me being able to judge which one was more difficult! I regard myself extremely lucky for getting him to know; he has such a great personality that makes him resisting against all my attempts to drive him angry!
\newline His patience in keeping our discussions in Italian and also his wellfounded recommendations on Italian literature were some of my greatest sources to become linguistically fluent.
\newline He taught me to consider the difficulties in life as cycling steep uphills.
\newline I shall give him a thousand thanks!

\paragraph*{}
I would like to thank Mathematical Physics Sector of SISSA for giving me the opportunity of pursuing my PhD studies by them.
In particular, I would like to thank Gianfausto Dell'Antonio, for his scientific care of my research work.

I would like to thank Mohammadreza Abolhasani, Vahid Karimipour, Giancarlo Ghirardi and Andreas Buchleitner for teaching me Math and Physics.

My very especial thanks goes to Reinhold and Renata Bertlmann from Vienna, who were the first persons who welcomed me in Trieste and introduced me to this lovely city, which by now has become my second hometown, and where, without exaggeration, at least $2/3$ of the people are my friends!

I would like to thank Laura Safred, for giving me the opportunity to live in one of the loveliest buildings in Trieste, where my neighbours Zia Donatella (particularly during the hard time of writing this thesis), Zia Caterina, Zia Lili, Zia Anna Maria and Zio Carlo, Nonna Silvana, Claudio and Piero (the funniest person in the building), made me feel like being home. I thank them all.

\newpage
\thispagestyle{empty}

Since the very first days of my staying in Trieste, I've been introduced to the people of Studio Tommaseo: Giuliana Carbi and Franco Jesurun; my especial gratitude goes to Giuliana: during all these years she was taking care of me like a real mom! I could never be happier in Trieste without her love!

I would like to thank all the lovely people I got to know through Giuliana and Franco: Zia Anita, Zio Enrico, Zio Nicolo', Zia Andreina, Nonna Marassi and Nonno Lucio, Zia Lola and Aldo (for his fantastic Parmigiana!) Ermanna (for being a lovely companion during writing this thesis), Lorenzo, Nonna Pinuccia and Nonno Pierachille, Massimo, Manu and Fabrizio, Costanza, Fiora and Zia Grazia, Chiara and Flavio.

I would like to thank my colleagues and friends from the university of Trieste and SISSA, for their friendship and fun we had together; this gang is: Angus, Nicola, Sara, Romina, Simona, Angelo, Urs, Sandro, Luca, Giancarlo, Kelvin, Giuseppe, Raffaello, Ugo (specially his parents: Mamma Laura and papa' Guglielmo), Giacomo.

I wish to thank all the people working at SISSA, among them, particularly, Riccardo Iancer and Federica Tuniz because of making all those horrible paper works so easy, I shall miss them forever! And Zia Lucia from the bar, for her fantastic sandwiches!

I would like to thank all my friends working at ICTP, particularly: Mamma Kotou, Zio Pierre, Rosana, Zia Gordana, Valerio, Lucio, Rosita, Sabina, Mary Ann, Stefano Pistocco, Uncle Robert and Tiziana, Freddy, Anna, Aga and Tiziano, Massimo Reja (for playing ping pong with me), Gabriel, Marino and my lovely French teacher Licia.

A very special thank to Erich Jost (Zio Erich); probably, without his encouragement and support I would never have taken cycling seriously. Thanks for all the sportive fun we had together.

I would like to thank my cycling team Cottur for giving me enthusiasm and encouragement; specially Francesco Seriani, Zio Daniele, Capitano Gianni, Fabrizio, Giacomo, Michela, Manuel (a real hero I got to know!) Luca, Dario, Marcello.

I would like to thank my friends from the team Generali for taking me in to their trainings, particularly: Zio Ezio, Andrea, Massimo.

My thanks goes to my cycling trainer Sasko for all his helps and also my friends from the team Flamme Rouge: Carolina and Luca, Roberto, Igor, whom I was racing with, even outside Trieste. Especially, for their friendship and above all for their encouragement to ride more uphills!

\newpage
\thispagestyle{empty}

I particularly thank Giovanni Iacono and Andrea Canton, not only for their friendship, but also for having taught me playing ping pong. I also thank Sonja Milic and the table tennis team in Sgonico for training me and improving my skills.

I would like to thank Paola D'Andrea for her patience and fantastic ideas about how to make a friend happy!

I appreciated my Iranian friends here, for giving me the opportunity to keep in practice with my mother tongue! These people are: Maryam and Mahdi, Athena, Arash, Zhaleh, Fahimeh, Shima, Mahmoud, Mohammad, Ladan and Houman.

I would like to thank my lovely long-term friends for not forgetting me because of being far from each other: Laleh (for being my older sister), Nazanin, Mahboubeh, Ida, Alireza, Mehdi, Mohammad, Nima and Nafiseh.

A very special thank to Barry Sanders, who is one of my greatest supporters. I thank him for encouraging me to work, by keeping in touch through Skype and inspiring me with his positive and energetic personality!

My sincere gratitudes to Hamidreza Sepangi and Mojgan Hosseini, who have been always supporting me and making me feel like part of their family.

My deepest gratitude goes to my family: Manzar and Saeed, Mina and Pirouz, Amir Ashraf Hosseini and my beloved sister Samira. I also thank my little nephews because of sending me letters full of love: Mandana, Negin, Nikoo and Mohamamd Hossein.

My most sincere and deepest gratitude goes to my uncle: Amoo Wolf. He is not only the best uncle on earth, but also my best friend. I shall never be able to thank him for all he has been doing for me!

And the best comes in the end: My heart is with my mother.

There will never be enough words to thank my mother, as there will never be anyone who loves me more than her!

\newpage
\thispagestyle{empty}

{\huge\textbf{Declaration}}

\vskip 2cm

I hereby declare that this thesis is my own work and effort and that it has not been submitted anywhere for any award. Where other sources of information have been used, they have been acknowledged.

\tableofcontents

\chapter{Introduction}

\paragraph*{}
It is not exactly clear when the very first idea of a computer was born: It seems that the first idea of an automatic computing machine was proposed by Charls Babbage in 1822. In the middle of 1930, Alan Turing by introducing the universal Turing machine, made it clear what characteristics computing machines needs to have, and how important the programming is. Shortly after his paper, the first computer was built.

\paragraph*{}
From 1947, onward transistors made the computer hardware develop very rapidly. So steadily the miniaturization pace that, in the middle of the sixties, Gordon Moore claimed that every two years the number of transistors used in a specific volume would doubl. This obviously had increased the speed of processing and the memory of the computers. This technology development went on until now, that we are not using the 30 tons computers anymore, but much smaller and lighter. However making computers and their hardware smaller, requires new considerations: when facing atomic scales, quantum effects play the major role instead of classical ones. Therefore if we want to have smaller computers, we might need to replace the classical computer technology by a quantum one.

\paragraph*{}

In 1982, Richard Feynman, mentioned that simulating a quantum system by a classical computer is very difficult~\cite{feyn}. Instead he suggested to use a computer which is ruled by Quantum Mechanics. Three years later, in 1985, David Deutsch proposed a model of quantum Turing machine~\cite{deut}. This was one of the first attempts to reveal the extreme power of quantum computing. The interesting point is that quantum technology not only permits the use of very small size microchips, but it also gives the possibility to tackle calculations by a new generation of algorithms based on Quantum Mechanics. It was Peter Shor, in 1994, who made one of the biggest steps in this field and showed how powerful such a quantum algorithm can be in factorizing a given number into its prime factors~\cite{shor}.

\paragraph*{}

In order to understand the main idea behind such an algorithm, let us mention some simple facts. From a physical point of view, a bit is a system which can be in two different states; these states can be yes or no, right or wrong, 0 or 1. A classical bit can be a mechanical on/off switch, or an electronic device capable of distinguishing a voltage difference. But if we choose an atom as a bit, in this case called qubit, then quantum mechanics tells us that the state of the atom can be a superposition the only two states available to a classical bit. Mathematically we write such a state as:
$$
\vert\psi\rangle=\alpha\vert0\rangle+\beta\vert1\rangle, \qquad |\alpha|^2+|\beta|^2=1.
$$
Now, consider a classical memory of three bits: at a given time, it can be in only one of the eight possible states $000$, $001$, $010$,...,$111$. On the other hand, a quantum memory of three qubits, can instead be in all the eight states at the same time:$\vert\psi\rangle=a_1\vert000\rangle+a_2\vert001\rangle+\cdots+a_8\vert111\rangle$. Therefore if we increase the number of qubits, the information stored in them increases exponentially: a memory of $n$ qubits stores $2^n$ classical binary strings simultaneously, while its classical partner does store only one string. This gives the possibility of performing the computation on the superposition of all the possible states in one step. This is called parallelism: a quantum computer can do a certain computation on $2^n$  numbers using $n$ qubits in one operation, while a classical computer has to do the same computation $2^n$ times with different entries, or it has to have $2^n$ processors operating at the same time. This means that quantum computers can in line of principle perform exponentially faster than the classical ones.
Of course, after a quantum computation of sort, the solution of a given problem will be contained in just one of the classical strings appearing in the quantum superposition;  indeed, the very core of quantum algorithmics is not only to write a sequence of quantum gates in order to solve the problem, but also to devise protocols able to read the superoposition so that it gets projected onto the solution string with very high probability.

\paragraph*{}

However this is not the only spectacular feature of the quantum computers. The other aspect which makes them very much different from their classical counterparts is the quantum effect called \textit{entanglement}. In 1935, Einstein, Podolski and Rosen, tried to explain the odd characters of the \textit{entangled quantum states}, using the hidden variables~\cite{eins}. It was however, John Bell in the sixties, the one who deeply investigated, the borders between Classical and Quantum Mechanics. His celebrated inequalities~\cite{bell}, that are to be satisfied by any local classical probabilistic theory, are violated by \textit{entangled states}. That means that entanglement has no classically local counterpart.

\paragraph*{}

Consider the following state:
\begin{equation}
 \label{eq01}
\vert\psi\rangle=\frac{1}{\sqrt{2}}(\vert01\rangle-\vert10\rangle),
\end{equation}
 where Alice has the first qubit and Bob owns the second one. Here the states $\vert0\rangle$ and $\vert1\rangle$ are the eigenstates of the spin along the $Z$ axis, with eigenvalues $+1$ and $-1$ respectively. If Alice does no measurements on her qubit, she can only guess that Bob's qubit is in the state $\vert0\rangle$ with probability $\frac{1}{2}$ or, it is in the state $\vert1\rangle$ with the same probability. However if she does a measurement along the $Z$ axis on her qubit, according to what she gets, she can say exactly in which state Bob's qubit is. For instance, if after measurement, she gets $+1$, then she can say that Bob's qubit is in state $\vert1\rangle$. While if she would have got $-1$, then Bob's state would have been $\vert0\rangle$.

\paragraph*{}

Now consider the state
\begin{equation}
\label{eq02}
\vert\psi\rangle=\frac{1}{\sqrt{2}}(\vert0\rangle+\vert1\rangle)\otimes\vert0\rangle.
\end{equation}
 In this case Alice can say that Bob's qubit is in the state $\vert0\rangle$ without doing any measurement on her qubit. Indeed Bob's qubit state, before and after the measurement on Alice's qubit does not change. While in the first case the state of Bob's qubit was modified by Alice's measurement, in the second case it did not.

\paragraph*{}

This gives an idea that entangled states such as~\eqref{eq01} are correlated in a way that the separable states as in~\eqref{eq02} are not and that these correlations have no classical counterparts.

\paragraph*{}

As we mentioned before entanglement plays a very important rule in quantum computation and quantum information. Quantum dense coding and quantum cryptography are examples of that. However the most astonishing application of entanglement is quantum teleportation, which concerns sending a qubit from one party to another one.

Suppose Alice wants to send the following state to Bob:
$$
\vert\phi\rangle_C=\alpha\vert0\rangle+\beta\vert1\rangle.
$$

\medskip

All Alice knows about the coefficients $\alpha$ and $\beta$ is that $|\alpha|^2+|\beta|^2=1$; that is the state $\vert\phi\rangle_C$ is unknown to her. In order to send this state, they share one of the following maximally entangled states (later we will see that these states are called Bell states, and form and orthonormal basis in $\CI^2\otimes\CI^2$):
\begin{eqnarray*}
\vert\Phi^{\pm}\rangle&=&\frac{1}{\sqrt{2}}(\vert 00\rangle\pm\vert 11\rangle),\\
\vert\Psi^{\pm}\rangle&=&\frac{1}{\sqrt{2}}(\vert 01\rangle\pm\vert 10\rangle).
\end{eqnarray*}
\medskip

Let Alice and Bob share the state $\vert\Phi^+\rangle$. Then the overall state would be:
$$
\vert\phi\rangle_C\otimes\vert\Phi\rangle_{AB}=\frac{1}{\sqrt{2}}(\alpha\vert000\rangle+\alpha\vert011\rangle+
\beta\vert100\rangle+\beta\vert111\rangle)
$$

\medskip
This can be re-written as:
\begin{eqnarray*}
\vert\phi\rangle_C\otimes\vert\Phi\rangle_{AB}&=&\frac{1}{2}(\vert\Phi^+\rangle_{AC}\otimes(\alpha\vert0\rangle+\beta\vert1\rangle)_B+
\vert\Phi^-\rangle_{AC}\otimes(\alpha\vert0\rangle-\beta\vert1\rangle)_B\\
&+&\vert\Psi^+\rangle_{AC}\otimes(\alpha\vert1\rangle+\beta\vert0\rangle)_B+
\vert\Psi^-\rangle_{AC}\otimes(\alpha\vert1\rangle-\beta\vert0\rangle)_B).
\end{eqnarray*}

\medskip

Now if Alice does a measurement of the Bell basis on her first two qubits, according to whether she finds them in $\vert\Phi^{\pm}\rangle$, $\vert\Psi^{\pm}\rangle$ the state would collapse into one of the following states:
\begin{eqnarray*}
\vert\Phi^+\rangle_{AC}\otimes(\alpha\vert0\rangle+\beta\vert1\rangle)_B\\
\vert\Phi^-\rangle_{AC}\otimes(\alpha\vert0\rangle-\beta\vert1\rangle)_B\\
\vert\Psi^+\rangle_{AC}\otimes(\alpha\vert1\rangle+\beta\vert0\rangle)_B\\
\vert\Psi^-\rangle_{AC}\otimes(\alpha\vert1\rangle-\beta\vert0\rangle)_B.
\end{eqnarray*}

\medskip

Therefore if she calls Bob through a classical channel (phone, e-mail, skype, etc.) and tell him what was the outcome of her measurement on Bell states basis, then Bob can do one of the following operations on his qubit in order to get it into the initial $\vert\phi\rangle_C$ state:
\begin{itemize}
\item
If Alice gets $\vert\Phi^+\rangle$, Bob does nothing.
\item
If Alice gets $\vert\Phi^-\rangle$, Bob acts with $\sigma_3$ on his qubit.
\item
If Alice gets $\vert\Psi^+\rangle$, Bob acts with $\sigma_1$ on his qubit.
\item
If Alice gets $\vert\Psi^-\rangle$, Bob acts with $\sigma_3\sigma_1$ on his qubit.
\end{itemize}
Where $\sigma_1$ is the first Pauli matrix, and $\sigma_3$ is the third one. This way Alice has sent her unknown qubit to Bob. This process is called quantum teleportation: it amounts to sending one qubit of information using two classical bits and an entangled state as transmission channel.

\paragraph*{}
Ever since, quantum entanglement has been one of the the main ingredients of quantum information theory, and the focus of many studies~\cite{horo8}. However very fundamental questions have still partial answers: when is it that a generic quantum state is entangled? How much entangled is it? Or, is its entanglement useful for quantum information processing or not? Indeed, just detecting entanglement in physical systems is one of the most challenging issues.

\medskip

Interestingly, beside rejuvenating the whole of Quantum Mechanics, quantum
information has also stimulated a more abstract interest in an old mathematical
problem: the characterization of positive linear maps on (the states on) $C^*$
algebras of operators~\cite{horo7}.

\medskip
From a physical point of view, all actually occurring processes must be described by completely positive (CP) maps whose structure was fully characterized by Stinespring in the 50ties and whose central role in Quantum Mechanics
was clarified by Kraus in the 70ies [4]. Their importance is due to the fact that
they are much more than positive: let the system on which they act (A) be
coupled to an ancillary system (B) completely inert and extend the action of
the map from A to A+B by composing it with the identity action on B. Then,
CP maps preserve positivity when they act on the states of the bipartite system
A+B. Instead, only positive maps do not because there surely exist entangled
states of the compound system A+B whose spectrum does not remain positive
under a positive, but not completely positive map composed with the identity.
\medskip

Since one cannot physically exclude such statistical couplings, complete positivity is the only guarantee against the appearance of unphysical negative probabilities in the spectrum of a transformed entangled state in the case of couplings
to ancillas. The most renown amongst positive, but not completely positive
maps is matrix-transposition: as transposition, it preserves positivity, but
fails to do so as partial-transposition, that is when acting on one party only
of a bipartite compound system~\cite{horo7}.
\medskip

Thus, positive maps are not fully consistent as descriptions of actual physical processes; however, exactly because they do not preserve the positivity of all entangled states, they can then be used as entanglement witnesses. Transposition is such a witness, which is exhaustive only in lower dimension; that is for two qubits (2-dim. system), or one qubit and one qutrit (three-dim. system). Otherwise, for instance already for two pairs A and B of two
qubits each, there are entangled states of A+B which remain positive under
partial-transposition: they are called PPT entangled states, namely positive
under partial transposition, but entangled nevertheless.

\medskip
The physical interest of PPT entangled states is that their entanglement
content cannot be augmented by any distillation protocol operating on great
numbers of them: they are also called bound entangled states as their entanglement is somewhat locked in and cannot be extracted. The mathematical interest of these states is instead related to the notion of decomposable
positive maps. It was introduced by St{\o}rmer and Choi in order to isolate a
particular sub-class of positive maps and used by Woronowicz to characterize all
positive maps in low dimension (this makes transposition an exhaustive witness
in such cases)~\cite{horo7}.
\medskip

Only a sub-class of higher dimensional positive maps is decomposable, that
is can be split into the sum of a CP map and a CP map composed with transposition and these cannot detect the entanglement of PPT states. Therefore, one
is interested in constructing new families of non-decomposable positive maps
able to witness certain bound-entangled states. This activity has been pursued in recent years, among others, by the Horodeccy~\cite{horo7}, by Kossakowski and
Chru{\'s}ci{\'n}ski~\cite{{koss3},{koss4}} in Poland.

\medskip

In this thesis work we have studied of $16\times  16$ density
matrices of two pairs of qubits, equidistributed over particularly symmetric
orthogonal one-dimensional projections $P_{\alpha\beta}$,$\alpha,\beta = 0, 1, 2, 4$, for which Benatti et al. have
been able to classify all PPT states, but only a few entangled ones among them.
As PPT states are related to specific discrete geometric structures associated
to a 16 point lattice, the issue is to understand whether bound
entanglement might be related to a further specification of such structures and,
if so, to characterize all PPT entangled lattice states.
\medskip

The organization of the thesis will be as follows:

\paragraph*{}
\textbf{Chapter 2:} We present an essential introduction of Functional Analysis as the basic mathematical tools dealing with Quantum Mechanics: Hilbert spaces and operators acting on them with a couple of most important norms on them will be defined.
In this work, we shall consider only finite dimensional Hilbert spaces, $\HI=\CI^d$; therefore the algebra of bounded linear operators defined on them will be isomorphic to the full matrix algebra $M_d(\CI)$.

\paragraph*{}
\textbf{Chapter 3:} We explain how quantum systems can be described using the mathematical tools of Chapter 2. Density matrices will be introduced as quantum states, which are, in general, mixtures of pure states with corresponding weights.
 Pure states, that is projectors onto Hilbert space vectors, are  a particular case when the physical state of a quantum system is completely known.

The von Neumann entropy as a counterpart of Shannon entropy will also be introduced; in compound systems, this quantity is used to compare the mixedness of density matrices of subsystems, called reduced density matrices, to the one of whole system. Throughout this work, we shall be interested in bipartite systems only. The definition of entangled and separable states in such systems will follow. We shall see, that as far as one is dealing with pure states, thanks to the Schmidt decomposition, distinguishing entangled and separable states is an easy task, which is not the case for mixed states. We will present Bell states and Werner states as examples of entangled states.

\paragraph*{}
\textbf{Chapter 4:} We explain what are the requirements for a linear map which is to describe a physical transformation. We start by giving some mathematical definitions of positive maps and completely positive maps. Using the Kraus representation we will study the structure of CP maps. Then Choi matrix will be introduced, as it provides a technique based on its block positivity or positivity to decide whether the corresponding map is positive or completely positive. We shall see that, though positive maps are not suitable candidates for describing physical transformations, nevertheless they can be used as mathematical tools to detect the entanglement.

Many examples will be given throughout the chapter; particularly the Transposition Map, which will be used to define decomposable maps. We shall also see, how a generic positive map can be written in terms of the Trace Map and a suitable completely positive map.

\paragraph*{}
\textbf{Chapter 5:} This chapter focuses on variety of entanglement detecting methods. Partial Transposition, which is exhaustive in low dimensions, and PPTness as a necessary condition for separability will be presented. Using the Hahn-Banach separation theorem, we introduce entanglement witnesses and how they can be connected to positive maps through the Jamio{\l}kowski isomorphism. The Reduction Criterion, which is based on the Reduction Map, already introduced in chapter 4 as a decomposable map, will be presented.
Further, we shall rapidly overview the Range Criterion, which was used to detect the first PPT entangled state in $\CI^2\otimes\CI^4$, it will be followed by the notion of  Unextendible Product Bases (UPB). We end this chapter by explaining the Realignment Method, which, unlike the other methods which manipulate only one subsystem, involves both subsystems to detect the entanglement. Many examples will be given throughout this chapter.

\paragraph*{}
\textbf{Chapter 6:} The last chapter contains new results. They concern a class of states over two parties consisting of $n$ qubits each. The states studied are diagonal in the basis generated by the action of tensor products of the form
$\1_{2^n}\otimes\sigma_{\vec{\mu}}$, $\sigma_{\vec{\mu}}=\otimes_{i=1}^n\sigma_{\mu_i}$, on the totally symmetric state $\vert\Psi^{2^n}_+\rangle\in  \CI^{2^n}\otimes\CI^{2^n}$.
We first characterize the structure of positive maps  detecting the entangled ones among them; we will show that their possible entanglement will be witnessed using a subclass of positive maps, namely diagonal positive maps. Then, the result will be illustrated by examining some entanglement witnesses for the case  $n=2$. Further, we will show how, for pairs of two qubits, being separable, entangled and bound entangled are state properties related to the geometric patterns of subsets of a $16$ point square lattice.

\newpage

\chapter{Preliminaries}

\section{Banach and Hilbert spaces}

In this chapter we focus on the essential background of functional analysis~\cite{re-si,take,krey}, and review the basic mathematical tools which are used in Quantum Mechanics as we are interested to describe it using the vector and operators over vector spaces, in particular Hilbert spaces.

\begin{definition}\textbf{Banach Space}

A Banach space is a vector space with a norm defined on it, with respect to which, it is complete.
\end{definition}

\begin{definition}\textbf{Hilbert Space}

A Hilbert space $\HI$ is a Banach space with respect to a norm given by the scalar product $\langle .\vert .\rangle$. For every vector $\psi\in\HI$, $\parallel \psi\parallel=\sqrt{\langle \psi\vert \psi\rangle}$ denotes its norm.
\end{definition}

We shall consider Hilbert spaces with countable orthonormal bases $\{\psi_i\}_{i\in \NI}\subset\HI$, the corresponding projectors $P_i=\vert\psi_i\rangle\langle\psi_i\vert$ satisfying the completeness relation: $\sum_iP_i=\1$.

\begin{definition}\textbf {Operator Norm}
\label{norm-operator}

The norm of an operator $O$ on a Hilbert space $\HI$ is defined as:
\begin{equation}
\parallel O\parallel:=\sup_{\parallel\psi\parallel=1} \parallel O\vert\psi\rangle\parallel.
\end{equation}
It satisfies:
\begin{equation}
\parallel O^{\dagger}\parallel=\parallel O\parallel,\qquad\parallel O^{\dagger} O\parallel=\parallel O\parallel^2.
\end{equation}
$O$ is said to be bounded if $\parallel O\parallel<\infty$. We denote the set of all bounded linear operators acting on $\HI$ by $B(\HI)$.
\end{definition}

The linear combination and the product of two bounded operator is also bounded. Moreover all Cauchy type sequences with respect to the norm\eqref{norm-operator}, converge to an element in $B(\HI)$ ~\cite{benatti1} which makes this space complete. That means that $B(\HI)$ is a $C^*-algebra$.

\begin{definition}\textbf{Spectrum}
\label{specrtum-infi}

The spectrum of a bounded operator $O$ denoted by $\sigma(O)$, is the set of all complex numbers $\lambda$ such that $\lambda\id-O$ has no inverse in $B(\HI)$, where $\id$ is the identity operator.
\end{definition}

The projectors onto vectors of Hilbert spaces are the simplest examples of bounded operators on $\HI$ with the spectrum $\sigma(P_i)=\{0,1\}$.

In Quantum Mechanics positive linear operators are highly important, they are bounded operators with non-negative spectrum, indeed they can be considered as observables whose measured outcomes are positive.

\begin{definition}\textbf{Positive Operator}
\label{positive-operator}

An operator $O\in B(\HI)$ acting on $\HI$, is positive if $\langle\psi\vert O \psi\rangle\geq0$ for all $\vert \psi\rangle\in\HI$.
\end{definition}

$B^+(\HI)$ denotes the set of all bounded linear positive operators acting on a Hilbert space $\HI$: $B^+(\HI)\subset B(\HI)$.

A set is said to be a cone if it is invariant to non-negative scaling. For any positive operator $O\in B^+(\HI)$ it holds that $\alpha O\in B^+(\HI)$, where $\alpha\geq 0$, therefore the set $B^+(\HI)$ is a cone.

\section{Operators on a Finite Dimensional Hilbert Space}
\paragraph*{}

Every finite Hilbert space $\HI$ of dimension $d$ is isomorphic to vector space of d-dimensional complex vectors $\CI^d$. Fixing the orthonormal basis set $\{\vert i\rangle\}$,
 $$
 \vert i\rangle=\begin{pmatrix} &0&  \\ &\vdots& \\ &1& \\ &\vdots&\\ &0& \end{pmatrix} \hbox{i-th},
$$
any vector $\vert\psi\rangle\in\HI$ can be represented over the field of complex scalars:
$$
 \vert\psi\rangle=  \begin{pmatrix} \psi_1  \\ \vdots\\ \psi_d \end{pmatrix}\in\CI^d,
$$
and linear operators are represented by $d\times d$ matrices. Then the physical observables are Hermitian $d\times d$ matrices with complex entries, and their algebra is denoted by $M_{d}(\CI)$,  the entries of $A\in M_{d}(\CI)$ being $A_{ij}=\langle i\vert A\vert j\rangle$.

\begin{theorem}\textbf{Spectral decomposition}

\label{sepcteral-decom}
Any normal operator $A$, $A^{\dagger}A=AA^{\dagger}$, on $\CI^d$ can be diagonalized with respect to the orthonormal basis of its eigenvectors:
$$
A=\sum_i a_i \vert a_i\rangle\langle a_i\vert, \quad A\vert a_i\rangle=a_i\vert a_i\rangle.
$$
\end{theorem}

Therefore, $A\in M_d(\CI)$ is positive semidefinite if all its eigenvalues are non-negative. Notice that matrix algebra $M_d(\CI)$ can be treated as a Hilbert space equipped with the Hilbert-Schmidt scalar product of two operators $A$ and $B$ in $M_d(\CI)$ which is defined to be :
\begin{equation}
\label{h-s-product}
\langle A\vert B\rangle=\Tr(A^{\dagger}B),
\end{equation}
where
$$
\Tr( A^{\dagger} B)=\sum_i \langle i\vert A^{\dagger} B\vert i\rangle
$$
is the trace of $A^{\dagger}B$ which is defined with respect to any orthonormal basis set $\{\vert i\rangle\}$.
The corresponding norm, called Hilbert-Schmidt norm:
\begin{equation}
\label{H-S norm}
\parallel A\parallel_2=\sqrt{\Tr( A^{\dagger}A)}=\sqrt{\sum_{i=1}^d a_i^2},
\end{equation}
can be expressed in terms of the eigenvalues $a_i^2$ of the positive operator $|A|^2=A^{\dagger}A$.
Another equivalent norm is the so called trace norm:
\begin{equation}
\label{trace-norm}
\parallel A\parallel_1:=\Tr|A|=\sum_{i=1}^d|a_i|.
\end{equation}

\begin{example}\textbf{Trace map}~\cite{benatti1}

Let $\HI=M_d(\CI)$, where $\{F_i\}_{i=1}^{d^2}$ is a set of $d\times d$ matrices which are orthogonal with respect to the Hilbert-Schmidt scalar product, $\Tr(F_l^{\dagger}F_k)=\delta_{lk}$. They form an orthonormal basis set, so that:
\begin{equation}
X=\sum_{i=1}^{d^2}\Tr(XF_i^{\dagger})F_i, \quad \forall X\in M_d(\CI).
\end{equation}
Now consider the Trace Map, $\Tr_d:M_d(\CI)\longrightarrow M_d(\CI)$ defined to be:
\begin{equation}
\label{trace-map}
M_d(\CI)\ni X\longmapsto\Tr_d[X]:=\Tr(X)\id_d.
\end{equation}
Given an orthonormal basis $\{\vert\alpha\rangle\}_{\alpha=1}^{d}$, one constructs the so called unit matrices $E_{\alpha\beta}:=\vert\alpha\rangle\langle\beta\vert$ in $M_d(\CI)$, which form an orthonormal basis in $M_d(\CI)$ with respect to the Hilbert-Schmidt scalar product. Therefore, using this orthonormal set, the Trace Map reads:
\begin{eqnarray}
\nonumber
 \Tr_d[X]=\Tr(X)\1_d&=&\sum_{\alpha=1}^d\langle\alpha\vert X\vert\alpha\rangle\sum_{\beta=1}^d\vert\beta\rangle\langle\beta\vert\\
 \nonumber
 &=&\sum_{\alpha,\beta=1}^d\vert\beta\rangle\langle\alpha\vert X\vert\alpha\rangle\langle\beta\vert=\sum_{\alpha,\beta=1}^d E^{\dagger}_{\alpha\beta}X E_{\alpha\beta}
\end{eqnarray}
since the two orthonormal basis $\{F_i\}_{i=1}^{d^2}$ and $\{E_{\alpha\beta}\}_{(\alpha,\beta=1)}^d$ can be transformed one into the other using a unitary matrix $U\in M_{d^2}(\CI)$: $E_{\alpha\beta}=\sum_{j=1}^{d^2}U_{\alpha\beta,j}F_j$. The Trace Map can thus be re-written as:
\begin{eqnarray}
\nonumber
\Tr_d[X]&=&\sum_{\alpha,\beta=1}^{d}E_{\alpha\beta}X E_{\alpha\beta}\\
\nonumber
&=&\sum_{i,j=1}^{d^2}\sum_{\alpha,\beta=1 \atop \gamma,\delta=1}^{d}U_{j,\alpha\beta}^*U_{\gamma\delta,i}F_j^{\dagger}XF_i\\
&=&\sum_{j=1}^{d^2}F_j^{\dagger}XF_j.
\end{eqnarray}
\end{example}

\begin{example}
The simplest quantum systems are the 2-level systems, which are described by $2\times2$ matrices: $A\in M_2(\CI)$. Most distinguished among them are the self-adjoint Pauli matrices:
\begin{equation}
\sigma_1=
\left( \begin{array}{cc}
        0&1\\
        1&0\\
        \end{array}
        \right), \quad
\sigma_2=
\left( \begin{array}{cc}
        0&-i\\
        i&0\\
        \end{array}
        \right),\quad
\sigma_3=
\left( \begin{array}{cc}
        1&0\\
        0&-1\\
        \end{array}
        \right),
\end{equation}
with respect to the orthonormal basis $\vert0\rangle$, $\vert 1\rangle$ of eigenvectors of $\sigma_3:$ $\sigma_3\vert0\rangle=\vert0\rangle$, $\sigma_3\vert1\rangle=\vert1\rangle$, where $\vert 0\rangle=\left(1\atop 0\right)$ and $\vert 1\rangle=\left(0\atop 1\right)$. For the other matrices we have $\sigma_1\vert0\rangle=\vert1\rangle$, $\sigma_1\vert1\rangle=\vert0\rangle$, $\sigma_2\vert0\rangle=-i\vert1\rangle$ and $\sigma_2\vert1\rangle=i\vert0\rangle$.

The algebra of Pauli matrices is summerized by:
\begin{equation}
\label{alg-Pauli}
\sigma_j\sigma_k=\delta_{jk}\1_2+i\varepsilon_{jkl}\sigma_l
\end{equation}
where $\varepsilon_{jkl}$ is the antisymmetric 3-tensor, $\1_2$ is the $2\times 2$ identity matrix which we also denote by $\sigma_0$. These four matrices are orthogonal in the sense of Hilbert-Schmidt scalar product, that is $\Tr(\sigma_i\sigma_j)=\delta_{ij}$ if $i\neq j$, when normalized, $\tilde{\sigma}_i:=\frac{\sigma_i}{\sqrt{2}}$ they form an orthonormal basis set in $M_2(\CI)$. Therefore any matrix $X\in M_2(\CI)$ can be written as:
$$
X=\sum_{i=0}^3(\Tr(\tilde{\sigma}_iX))\tilde{\sigma}_i.
$$
\end{example}

\begin{example}\textbf{Trace map in $M_2(\CI)$}~\cite{benatti2}
\label{ex-trace2}

Using the Pauli matrices as an orthonormal basis set we can write the Trace Map defined in~\eqref{trace-map} as:
\begin{equation}
\label{trace2}
\Tr[X]=\sum_{j=0}^{3}\sigma_jX\sigma_j.
\end{equation}
In this case Pauli matrices permit us to simplify this map even more. Let us define the map $S_{\alpha}:M_2(\CI)\longrightarrow M_2(\CI)$ such that:
\begin{equation}
\label{map-s}
S_\alpha[.]=\sigma_\alpha[.]\sigma_\alpha.
\end{equation}
The Trace Map is re-written as:
\begin{equation}
\label{trace-2}
\Tr_2[X]=\sum_{\alpha=0}^3 S_\alpha[X].
\end{equation}
We shall refer to this map in the next chapters.
\end{example}

\newpage

\chapter{States and Operators}

\paragraph*{}
In this chapter, we present some mathematical aspects concerning quantum states of multipartite systems. The presence of many parties introduces the notion of quantum correlations and entanglement. For the rest of this work, we consider only finite complex Hilbert spaces $\HI=\CI^d$, where $d$ denotes the dimension of the Hilbert space. States both in Classical and Quantum Mechanics are generic linear, positive, normalized functionals on the algebras of observables that fix their mean values. In Classical Mechanics one deals with probability distributions on phase-space; in Quantum Mechanics distributions are replaced by density matrices which extend the notion of quantum state beyond the single vectors in Hilbert spaces.

\section{Density Matrices}

\paragraph*{}
Usually one is accustomed to states of quantum systems only as vectors in a suitable Hilbert space. However, more general ones need to be introduced. Consider an ensemble of projectors $P_i:=\vert\psi_i\rangle\langle\psi_i\vert\in B(\HI)$ corresponding to the vector states $\vert\psi_i\rangle\in\HI$, with weights $p_i$. If the only knowledge about the physical system is that it can be found in such states with weight $p_i$, one is effectively dealing with a statistical ensemble. Its mathematical description is by a density matrix, namely:
$$\rho=\sum_i p_i\vert\psi_i\rangle\langle\psi_i\vert \in M_d(\CI).$$

\begin{definition}\textbf{Density Matrix}

Any positive operator $\rho\in M_d(\CI)$ with $\Tr(\rho)=1$ describes a mixed state; let $\rho=\sum_{j=1}r_j\vert r_j\rangle\langle r_j\vert$ be its spectral representation with $\quad 0\leq r_j\leq 1$ and $\sum_j r_j=1$. Then $\rho$ defines a positive, linear and normalized functional on $M_d(\CI)$~\cite{benatti1}:
\begin{equation}
M_d(\CI)\ni X\longmapsto\omega_\rho(X):=\Tr(\rho X)=\sum_j r_j\langle\varphi_j\vert X\vert\varphi_j\rangle.
\end{equation}
\end{definition}

\begin{remark}
 When the density matrix is a projector onto the corresponding state vector, which is its eigenvector with eigenvalue one, the state is said to be pure, hence the spectral representation of the density matrix reduces in to a one-dimensional projection.

The set of all the density matrices of a quantum system $S$, is denoted by $\SI(S)$. Since any convex combination of two density matrices is a density matrix as well, therefore $\SI(S)$ is a convex set. Its extreme points, which can not be written as a convex combination of other density matrices, correspond to pure states.
\end{remark}

\begin{example}\textbf{Bloch Sphere}

There is a very useful geometrical representation for density matrices $\rho\in M_2(\CI)$ of qubits:  the Bloch sphere. Taking the two eigenvectors $\vert 0\rangle$ and $\vert 1\rangle$ of the Pauli matrix $\sigma_3$ as the axes of this sphere:
\begin{center}

\includegraphics[scale=0.5]{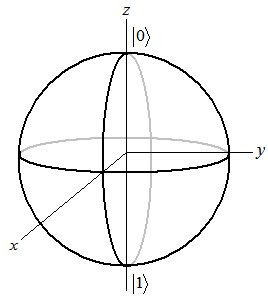}

\end{center}

Then any normalized vector $\vert\psi\rangle$ in a two-level system can be represented in terms of these two vectors:
$$
\vert\psi\rangle=\cos\theta\vert 0\rangle+e^{i\phi}\sin\theta\vert 1\rangle.
$$
It corresponds to a point on the surface of this sphere, which can be also represented using polar coordination by a new vector $\overrightarrow{\alpha}\in\RI^3$:

\begin{equation}
\overrightarrow{\alpha}=(\alpha_1,\alpha_2,\alpha_3)=(\sin2\theta\cos\phi,\sin2\theta\sin\phi,\cos2\theta).
\end{equation}

The vector $\overrightarrow{\alpha}$ is known as Bloch vector. In general a mixed density matrix $\rho\in M_2(\CI)$ can be represented in terms of Bloch vectors and Pauli matrices, called Bloch representation:

\begin{eqnarray}
\label{density-bloch}
&\rho&=\quad\frac{1}{2}(\1_2+\sum_{i=0}^3\alpha_i\sigma_i),\\
\nonumber
&where& \qquad \alpha_i=\Tr(\rho\sigma_i).
\end{eqnarray}

The positivity of $\rho$ is equivalent to asking that $\|\overrightarrow{\alpha}\|^2=\sum_{i=1}^3 \alpha_i^2\leq1.$
Note that pure qubit states correspond to Bloch vectors $\overrightarrow{\alpha}$ with norm one, i.e. to the points on the surface of the sphere. Mixed states are instead represented by points inside the sphere where $|\overrightarrow{\alpha}|<1$. The maximally mixed state $\frac{\1_2}{2}$ corresponds to the center of the sphere, while orthogonal states, in the sense of the Hilbert-Schmidt product, are connected by diameters.

\end{example}

Therefore a density matrix is a Hermitian, unit-trace operator with non-negative eigenvalues. For pure states $\Tr(\rho^2)=1$, while for mixed states $\Tr(\rho^2)<1$. Given two density matrices, there are different methods to judge about their mixedness.

The mixedness measure on a Hilbert space with dimension d, is defined to be ~\cite{pet-wei}:
\begin{equation}
M=\frac{d}{d-1}(1-\Tr(\rho^2)).
\end{equation}
This quantity is known as linear entropy and is 0 for pure states and maximal for maximally mixed states, i.e. $\omega=\frac{\1}{d}$.

However, the most reknown entropy is the von Neumann entropy~\cite{von-neu}.

\begin{definition}\textbf{von Neumann Entropy}
\label{von-neu entropy}

Given a density matrix $\rho\in\SI(S)$, the von Neumann entropy is defined:
\begin{equation}
S(\rho)=-\Tr(\rho\log\rho)=-\sum_i^d\lambda_i\log\lambda_i
\end{equation}
where $\lambda_i$ are the eigenvalues of $\rho$ and $d$ is the dimension of the system.
\end{definition}
Evidently this quantity is zero only for pure states, as the projectors have only eigenvalues equal to 0 or 1. For the maximally mixed states $\rho_{0}=\frac{\1_d}{d}$, the von Neumann entropy is maximum:
$$
S(\rho_{0})=-\sum_i\frac{1}{d}\log\frac{1}{d}=\sum_i\frac{1}{d}\log d=\log d.
$$
This shows that the von Neumann entropy can measure the mixedness of density matrices, and we have $0\leq S(\rho)\leq\log d$.

\section{Composite Systems}
In Quantum Information, one very frequently deals with systems consisting of several subsystems, called multi-partite systems: $S=S_1+S_2+...+S_n$. The Hilbert space of $S$ is the tensor product of the Hilbert spaces of its subsystems: $\HI^{\otimes n}=\bigotimes_{i=1}^n \HI_i$. On tensor product Hilbert spaces, operators act as tensor product of matrices.

\begin{definition}\textbf{Matrix Tensor Product}

Consider a bipartite system described by the Hilbert space $\CI^{d_1d_2}=\CI^{d_1}\otimes\CI^{d_2}$, together with the matrix algebra $M_{d_1}(\CI)$ for the first party and $M_{d_2}(\CI)$ for the second one. Thus the algebra of the whole system consists of the tensor product of the form~\cite{{ho-jo},{brus1}}:
$$
A\otimes B=
      \begin{pmatrix}
  a_{1,1}B & a_{1,2}B & \cdots & a_{1,n}B \\
  a_{2,1}B & a_{2,2}B & \cdots & a_{2,n}B \\
  \vdots  & \vdots  & \ddots & \vdots  \\
  a_{m,1}B & a_{m,2}B & \cdots & a_{m,n}B
 \end{pmatrix}, \quad A\in M_{d_1}(\CI), B\in M_{d_2}(\CI).
$$
\end{definition}

When $\HI^{\otimes n}=\bigotimes_{i=1}^n\CI^{d_i}$, the algebra of $\HI^{\otimes n}$ is $M^{\otimes n}(\CI)=\bigotimes_{i=1}^n M_{d_i}(\CI)$.

\paragraph*{}
The opposite of combining the subsystems to get a bigger one, is partial tracing. By tracing over a subsystem $S_j$ with respect to an orthonormal basis $\vert\psi_k^{(j)}\rangle\in\CI^{d_j}$, one reduces a density matrix $\rho\in M^{\otimes n}(\CI)$ to another density matrix:
\begin{equation}
\rho_{\widehat{j}}:=\Tr_j\rho=\sum_k\langle\psi_k^{(j)}\vert\rho\vert\psi_k^{(j)}\rangle \in M_j^{\otimes (n-1)}(\CI)=\bigotimes_{i=1\atop i\neq j}^n M_{d_i}(\CI),
\end{equation}
acting on $\CI_j^{\otimes (n-1)}=\bigotimes_{i=1\atop i\neq j}^n\CI^{d_i}$.

\paragraph*{}
In the following, we will consider only finite-dimensional bipartite systems, i.e. $S=S_1+S_2$, with $\HI^{\otimes 2}=\HI_1\otimes\HI_2$, $\HI_1=\CI^{d_1}$ and $\HI_2=\CI^{d_2}$. Therefore for the whole system we have $\HI=\CI^{d_1}\otimes\CI^{d_2}=\CI^{d_1d_2}.$ Density matrices are positive $\rho_{12}\in M_{d_1d_2}(\CI)$ of trace 1. The partial trace yields:
\begin{eqnarray}
M_{d_1}(\CI)\ni\rho_1=\Tr_2(\rho_{12}):=\sum_{j=1}^{d_2}\langle f_j^2\vert\rho_{12}\vert f_j^2\rangle\\
\nonumber
M_{d_2}(\CI)\ni\rho_2=\Tr_1(\rho_{12}):=\sum_{i=1}^{d_1}\langle e_i^1\vert\rho_{12}\vert e_i^1\rangle,
\end{eqnarray}
where $\rho_1$ (resp. $\rho_2$) is called reduced density matrix for the system $S_1$ (resp. for the system $S_2$), the sets $\{\vert e_i^1\rangle\}_{i=1}^{d_1}$ and $\{\vert f_j^2\rangle\}_{j=1}^{d_2}$ are orthonormal basis of $\CI^{d_1}$ and $\CI^{d_2}$ respectively.

The reduced density matrices describe the states that two parties locally possess. Their meaning is easy to understand in terms of  local observables, of the first party, say:
$A\otimes\id$. The mean value of such a local observable, of the first party alone, with respect to the two party state $\rho_{12}$ reads
$$
\Tr_{12}(\rho_{12} A\otimes\1)=\sum_{i=1}^{d_1} \langle e^1_i\vert A\,\Big(\,\sum_{j=1}^{d_2}\langle f^2_j\vert\rho_{12}\vert f^2_j\rangle\,\Big)\,\vert e^1_i\rangle=\Tr_1(A\,\rho_1)\ ,
$$
showing that it is completely specified by the reduced density matrix $\rho_1$ of the considered party.

For bipartite systems, we have the following lemma:

\begin{lemma}\textbf{Schmidt decomposition}

Every vector $\vert\psi_{12}\rangle\in\CI^{d_1}\otimes\CI^{d_2}$, can be represented as~\cite{mintert}:
\begin{equation}
\label{schmidt}
\vert\psi_{12}\rangle=\sum_{i=1}^{M}r_{i}^{(1)}\vert r_i^{(1)}\rangle\otimes\vert r_i^{(2)}\rangle
\end{equation}
where $M\leq \min\{d_1,d_2\}$, and $\{\vert r_i^{(1)}\rangle\}_{i=1}^{d_1}$ and $\{\vert r_i^{(2)}\rangle\}_{i=1}^{d_2}$ are orthonormal sets of vectors in $\CI^{d_1}$ and $\CI^{d_2}$ respectively. The non-negative scalars $\{r_i\}$ are such that $\sum_{i=1}^M r_i^2=1$.
\end{lemma}

\proof
Without loss of generality let us assume $d_1\leq d_2$. Given two ONB $\{\vert e_i\rangle\}_{i=1}^{d_1}$ and $\{\vert f_j\rangle\}_{j=1}^{d_2}$, any vector $\psi_{12}\in\CI^{d_1}\otimes\CI^{d_2}$ can be written as:
\begin{equation}
 \label{s-1}
\vert\psi_{12}\rangle=\sum_{i,j}c_{ij}\vert e_i^1\rangle\otimes\vert f_j^2\rangle, \quad c_{ij}\in\CI.
\end{equation}
Its corresponding density matrix is $\rho_{12}=\vert\psi_{12}\rangle\langle\psi_{12}\vert$. The reduced density matrix $\rho_1$ is obtained by partial tracing over the second subsystem, which in its spectral representation is:

\begin{equation}
 \label{s-2}
\rho_1=\Tr_2(\vert\psi_{12}\rangle\langle\psi_{12}\vert)=\sum_{i=1}^{d_1} r_i^{(1)} \vert r_i^{(1)}\vert\langle r_i^{(1)}\vert.
\end{equation}

Choosing $\{\vert r_i^{(1)}\rangle\}_{i=1}^{d_1}$ as an ONB for $\CI^{d_1}$, we can re-write~\eqref{s-1} as:

\begin{eqnarray}
 \label{s-3}
\vert\psi_{12}\rangle&=&\sum_{i=1}^{d_1}\sum_{j=1}^{d_2} c_{ij} \vert r_i^{(1)}\rangle\otimes\vert f_j\rangle\\
\nonumber
&=&\sum_{i=1}^{d_1} \vert r_i^{(1)}\rangle\otimes\vert\phi_i^{(2)}\rangle,
\end{eqnarray}
where $\vert\phi_i^{(2)}\rangle=\sum_{j=1}^{d_2}c_{ij}\vert f_j\rangle$. Note that the vectors $\vert\phi_i^{(2)}\rangle$ are not in general orthonormal. Using this new ONB, partial tracing over the second subsystem of $\rho_{12}$, will be read as:
\begin{equation}
\rho_1=\Tr_2(\vert\psi_{12}\rangle\langle\psi_{12}\vert)=\sum_{i=1}^{d_1} \vert r_i^{(1)}\rangle\langle r_i^{(1)}\vert \langle\phi_i^{(2)}\vert\phi_j^{(2)}\rangle .
\end{equation}
Comparing this with~\eqref{s-2}, we see that:
$$
\langle\phi_i^{(2)}\vert\phi_j^{(2)}\rangle=\delta_{ij} r_i^{(1)}.
$$
 After normalization we have:
$$
\vert\phi_i^{(2)}\rangle=\sqrt{r_i^{(1)}}\vert r_i^{(2)}\rangle \in\CI^2.
$$
Replacing this in to~\eqref{s-3}, we get:
\begin{equation}
 \vert\psi_{12}\rangle=\sum_{i=1}^{d_1}\sqrt{r_i^{(1)}}\vert r_i^{(1)}\rangle\otimes\vert r_i^{(2)}\rangle.
\end{equation}
Which ends the proof.
\qed
\medskip

The coefficients $r_i$ are called Schmidt coefficients, and their multiplicity is known as Schmidt rank. Note that using the Schmidt form of a given vector $\psi_{12}$, the reduced density matrices are:
$$
\rho_1=\Tr_2(\rho_{12})=\sum_{i=1}^{M}r_{i}^{(1)}\vert r_i^{(1)}\rangle\langle r_i^{(1)}\vert,
$$
$$
\rho_2=\Tr_1(\rho_{12})=\sum_{i=1}^{M}r_{i}^{(1)}\vert r_i^{(2)}\rangle\langle r_i^{(2)}\vert,
$$
one sees that they have the same eigenvalues $r_{i}^{(1)}$, with the same multiplicity.

\section{Separable and Entangled States}

\paragraph*{}
Consider a vector state $\vert\Psi\rangle\in \CI^{d_1}\otimes\CI^{d_2}$, which is a tensor product of two vector states: $\vert\Psi\rangle=\vert\psi_1\rangle\otimes\vert\psi_2\rangle$, where $\vert\psi_1\rangle\in\CI^{d_1}$ and $\vert\psi_2\rangle\in\CI^{d_2}$. The corresponding density matrix is of the form:
$$
\rho_{\Psi}=\vert\Psi\rangle\langle\Psi\vert=\vert\psi_1\rangle\langle\psi_1\vert\otimes
\vert\psi_2\rangle\langle\psi_2\vert=\rho_1\otimes\rho_2,
$$
where $\rho_{1,2}$ are pure states projections. As we see the density matrix corresponding to a product vector state is itself a tensor product. Using this fact we can give a definition for a separable state:

\begin{definition}~\cite{horo1}
A bipartite pure state $\rho$ is separable if and only if it is the tensor product of the density matrices of its subsystems:
\begin{equation}
\rho=\rho_1\otimes\rho_2.
\end{equation}
Otherwise it is an entangled pure state.
\end{definition}

For a pure separable state, tracing over one subsystem gives a projector over the other subsystem, i.e. pure states:
$$
\rho_1=\Tr_2(\rho_{12})=\Tr_2(\rho_1\otimes
\rho_2).
$$

Examples of entangled pure states are the Bell states which are maximally entangled.

\begin{example}\textbf{Bell States}
\label{ExBell}

The following four symmetric vectors in $\CI^2\otimes\CI^2$ are known as Bell basis vectors~\cite{kaye}:
\begin{eqnarray}
\label{bell-states}
\vert\Phi^{\pm}\rangle&=&\frac{1}{\sqrt{2}}(\vert 00\rangle\pm\vert 11\rangle),\\
\vert\Psi^{\pm}\rangle&=&\frac{1}{\sqrt{2}}(\vert 01\rangle\pm\vert 10\rangle).
\end{eqnarray}

For instance consider the density matrix corresponding to the $\rho_{\Phi^+}$:
$$
\rho_{\Phi^+}=\frac{1}{2}(\vert 0\rangle\langle 0\vert\otimes\vert 0\rangle\langle 0\vert+\vert 1\rangle\langle 1\vert\otimes\vert 1\rangle\langle 1\vert)+
$$
$$
\frac{1}{2}(\vert 0\rangle\langle 1\vert\otimes\vert 0\rangle\langle 1\vert+\vert 1\rangle\langle 0\vert\otimes\vert 1\rangle\langle 0\vert),
$$
reduced density matrices are:
$$
\rho_1=\Tr_2(\rho_{\Phi^+})=\rho_2=\Tr_1(\rho_{\Phi^+})=\frac{\1_2}{2},
$$
which are maximally mixed states.
\end{example}

\paragraph*{}

The following theorem provides a necessary and sufficient condition of separability for bipartite pure states, based on their Schmidt decomposition:

\begin{theorem}
The pure vector state $\vert\psi_{12}\rangle$ is separable if and only if only one Schmidt coefficient is different from zero.
\end{theorem}

Therefore the pure states, whose Schmidt decomposition contains more than one Schmidt coefficient, or in other words, their Schmidt rank is greater than one, are entangled.

\begin{example}
According to Schmidt decomposition the following state is separable:
$$
\vert\psi_{12}\rangle\langle\psi_{12}\vert=\vert\psi_1\rangle\langle\psi_1\vert\otimes\vert\psi_2\rangle\langle\psi_2\vert
$$
One can easily see that the von Neumann entropy is zero for the density matrix $\rho_{12}$, as well as the reduced density matrices $\rho_1$ and $\rho_2$:
$$
S(\rho_{12})=S(\rho_1)=S(\rho_2)=0.
$$
\end{example}

\begin{example}
 The Schmidt rank of Bell states in~\eqref{bell-states} is 2, and their Schmidt coefficients are $\frac{1}{\sqrt{2}}$.

The symmetric states in~\eqref{psi+} are another example of entangled states, which Schmidt rank d.
 \end{example}

\subsection{Shannon entropy and von Neumann entropy}

Example \ref{ExBell} is a clear indication that entanglement embodies correlations that have no classical counterpart. Indeed, the von Neumann entropy is for quantum states what the Shannon entropy~\cite{cover} is for classical states, that is for probability distributions of stochastic variables.
Let $X=\{x_j\}_{j=1}^d$ be a $d$-valued stochastic variable whose outcomes $x_i$ occur with probabilities $p_i$, then the Shannon entropy of $X$ is defined by
\begin{equation}
\label{Shannon}
H(X)=-\sum_{i=1}^dp_i\log p_i\ .
\end{equation}
One thus sees that the von Neumann entropy of density matrix $\rho$ is nothing else than the Shannon entropy of its spectrum.
However, the Shannon entropy of a composite system, described say by two stochastic variables $X_1=\{x^{(1)}_i\}_{i=1}^{d_1}$ and $X_2=\{x^{(2)}_j\}_{j=1}^{d_2}$ is always larger than the Shannon entropy of each of its two parties, that is
\begin{equation}
\label{Shannon1}
H(X_1,X_2)\geq \max\{H(X_1),H(X_2)\}\ ,
\end{equation}
where $H(X_1,X_2)$ is the Shannon entropies of the joint probability distribution $\pi_{1,2}=\{p(x_i^{(1)},x_j^{(2)})\}$
and $H(X_{1,2})$ are the Shannon entropy of the reduced probability distributions $\pi_1=\{p(x_i^{(1)})\}_{i=1}^{d_1}$, $\pi_2=\{p(x_j^{(2)})\}_{j=1}^{d_2}$ with
\begin{equation}
\label{Shannon2}
p(x_i^{(1)})=\sum_{j=1}^{d_2}p(x_i^{(1)},x_j^{(2)})\ ,\quad p(x_j^{(2)})=\sum_{i=1}^{d_1}p(x_i^{(1)},x_j^{(2)})\ .
\end{equation}
This can be seen as follows: let us introduce the \textit{conditional probability}
\begin{equation}
\label{Shannon3}
p(x_i^{(1)}|x_j^{(2)})=\frac{p(x_i^{(1)},x_j^{(2)})}{p(x_j^{(2)})}
\end{equation}
that the outcome for $X_1$ be $x_i^{(1)}$ given that the outcome $x_j^{(2)}$ has been obtained and the \textit{conditional entropy} of $X_1$ given $X_2$, namely the following convex combination of Shannon entropies
\begin{eqnarray}
\label{Shannon3}
H(X_1|X_2)&=&-\sum_{j=1}^{d_2}p(x_j^{(2)})\sum_{i=1}^{d_1}p(x_i^{(1)}|x_j^{(2)})\log p(x_i^{(1)}|x_j^{(2)})\\
\label{Shannon4}
&=&H(X_1,X_2)- H(X_2)\ .
\end{eqnarray}
Then, the result follows because $H(X_1|X_2)$ and $H(X_2|X_1)$ are $\geq 0$ and
\begin{equation}
\label{Shannon4}
H(X_1,X_2)=H(X_2)+H(X_1|X_2)=H(X_1)+H(X_2|X_1)\ .
\end{equation}

\paragraph*{}
\begin{remark}
Unlike the Shannon entropy, the von Neumann entropy of subsystems can be larger than that of the total system.
This fact indeed signals entanglement but, as we shall see, is not always implied by entanglement.
\end{remark}

One could now guess that quantum states either carry entanglement or classical correlations. It is not so: there can be separable states which are not simply classical
correlated.

In order to understand this fact one has to introduce the so-called Quantum Discord of a given state. One starts by recalling that,
using the conditional entropy, one can define the mutual information:
\begin{equation}
\label{mutual-j}
\fJ(X_1:X_2)=H(X_1)-H(X_1|X_2)
\end{equation}
which measures the amount of information that is common to both classical stochastic variables. It can also be rewritten as:
\begin{equation}
\label{mutual-i}
\fI(X_1:X_2)=H(X_1)+H(X_2)-H(X_1,X_2).
\end{equation}

One can extend the notion of mutual information to quantum systems~\cite{{brod1},{brod2}}; To do so, starting from~\eqref{mutual-i}, one can simply replace the probability distributions of individual systems by reduced density matrices of subsystems:
\begin{equation}
\label{mutual-i-q}
 \fI_{A:B}(\rho_{AB})=S(\rho_A)+S(\rho_B)-S(\rho_{AB}).
 \end{equation}
 However this generalization, in the case of expression~\eqref{mutual-j} for classical mutual information, is not trivial; in fact, the definition of conditional entropy of quantum systems involves quantum measurement processes.
Expressing the state of system $A$, when the state of system $B$ is known, is equivalent to having done a set of measurements on system $B$. Therefore the post-measurement state of system $A$, after performing measurement $\Pi_i^B$ will be:
 \begin{equation}
 \label{cond-state}
 \rho_{A| \Pi_i^B}=\frac{1}{p_i}(\1_d\otimes\Pi_i^B\rho_{AB}\1_d\otimes\Pi_i^B),\quad p_i=\Tr(\rho_B\Pi_i^B).
 \end{equation}

 Given a complete set of orthogonal measurements $\Pi=\{\Pi_i,i=1,...,d\}$, $\sum_i \Pi_i=\1_d$, $\Pi_i\Pi_j=\delta_{ij}\Pi_i$, on system $B$, the conditional entropy will be defined as~\cite{oliv}:
 \begin{equation}
 \label{cond-ent}
 S(\rho_A|\Pi^B):=\sum_i p_i S(\rho_A|\Pi_i^B).
 \end{equation}

With  this definition of quantum conditional entropy, $\fJ$ in~\eqref{mutual-j} reads:
 \begin{equation}
 \label{mutual-j-q}
 \fJ_{A:B}(\rho_{AB}):=S(\rho_A)-S(\rho_A| \Pi^B).
 \end{equation}
\medskip

 Unlike in the classical case where the two expressions for mutual information, $\fI$ and $\fJ$, are equal, in quantum case, the two expressions are not in general the same~\cite{{oliv},{hend}}.
Their difference depends on the set of measurements performed on system $B$ which can be removed by minimizing it over all possible sets $\{\Pi^B\}$, and define the minimum as the Quantum Discord of the given state $\rho_{AB}$:
 \begin{eqnarray}
\nonumber
 D_{A:B}(\rho_{AB})&=&\min_{\{\Pi^B\}} [ \fI_{A:B}(\rho-{AB})-\fJ_{A:B}(\rho_{AB})]\\
  \label{discord}
 &=&\min_{\{\Pi^B\}} [S(\rho_B)+S(\rho_A| \Pi^B)-S(\rho_{AB})]\ .
 \end{eqnarray}
The Quantum Discord measures the non-classical correlations whhc are present in a quantum state: it is $0$ for separable states with only classical correlations, and maximal for entangled states which are extremely non-classical correlated. However,
there exist separable, that is non-entangled, states with non-zero Quantum Discord.

\subsection{Mixed entangled states}

For mixed states the notion of separability extends as follows:

\begin{definition}
\label{mix-ent}
A mixed state $\rho$ is separable if and only if it cannot be written as a convex combination of product states:
\begin{equation}
\rho=\sum_i p_i \rho_1^i\otimes \rho_2^i, \qquad \sum_i p_i =1,\quad p_i\geq0.
\end{equation}
Otherwise, it is an entangled mixed state.
\end{definition}

\begin{example}
Consider a bipartite system $S=S_1+S_2$, where the Hilbert spaces of the subsystems are of the same dimension. Consider the symmetric states~\cite{benatti2}:
\begin{equation}
\label{psi+}
\vert\Psi_{+}^d\rangle=\frac{1}{d}\sum_{j=1}^d\vert j\rangle\otimes\vert j\rangle,
\end{equation}
where $\vert j\rangle$, $j=1,2,...,d$ is any fixed orthonormal basis in $\CI^d$. The corresponding projectors are:
\begin{equation}
\label{P+}
P_{+}^{d}\equiv\vert\Psi_{+}^d\rangle\langle\Psi_{+}^d\vert=\sum_{i,j=1}^d\vert i\rangle\langle j\vert\otimes\vert i\rangle\langle j\vert.
\end{equation}
Also in this case reduced density matrices are maximally mixed:
$$
\rho_1=\rho_2=\frac{1}{2}\sum_{i=1}^d\vert i\rangle\langle i\vert=\frac{\1_d}{d}
$$
which means that $\vert\Psi_{+}^d\rangle$ is entangled. We shall refer to this set of symmetric states in the future.
\end{example}

\begin{example}\textbf{Werner States}
\label{ex-werner}

Werner states in $\CI^2\otimes\CI^2$ are constructed as follows~\cite{{pop},{werner}}:
\begin{equation}
\label{werner-state}
\rho_{\alpha}=\alpha\vert\Psi^-\rangle\langle\Psi^-\vert+\frac{1-\alpha}{4}\1_2\otimes\1_2, \quad -\frac{1}{3}\leq\alpha\leq 1.
\end{equation}
Where $\Psi^-$ is one of the Bell states, introduced in~\eqref{bell-states}. In the basis where $\sigma_3$ is diagonal, Werner states can be represented by the following matrix:
$$
\rho_{\alpha}=\frac{1}{4}\begin{pmatrix}
  1-\alpha & 0& 0 & 0 \\
 0 & 1+\alpha & -2\alpha & 0 \\
  0  & -2\alpha  & 1+\alpha & 0  \\
  0 & 0 & 0 & 1-\alpha
 \end{pmatrix}
$$
The above range for $\alpha$ is to guarantee the positivity of $\rho_{\alpha}$. Werner states are entangled for $\alpha\geq\frac{1}{3}$. The reduced density matrices are:
$$
(\rho_\alpha)_1=(\rho_\alpha)_2=\frac{1}{2}\begin{pmatrix}
  1 & 0 \\
  0 & 1
 \end{pmatrix}.
$$
As we see, also in this case the reduced density matrices are maximally mixed and therefore their von Neumann entropy is maximum. Indeed, by checking how the function $\Delta^{\alpha}= S(\rho_{\alpha})- \log 2$ behaves:
\begin{center}

\includegraphics[scale=0.75]{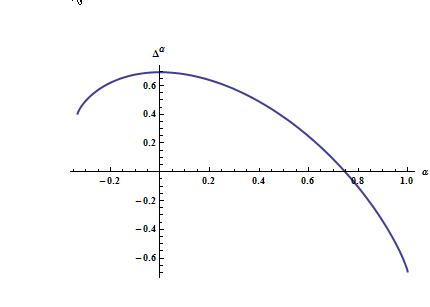}

\end{center}

one might then think that entangled Werner states are detected by checking whether their entropy is smaller than the entropy of one of the parties. We shall later see a counterexample to such a claim.
\end{example}

However the situation in case of pure states is completely clear thanks to Schmidt decomposition, deciding if a given mixed state is entangled or separable is still an open problem, even in case of bipartite systems. In the following chapters we will study how to partially answer this question.

\newpage

\chapter{Positive and Completely Positive Maps}

Physical transformations of Quantum systems are described by linear maps, either on the space of states or, dually, on the algebra of observables. If a linear map on the convex space of states describes a physical transformation (that, of course, transforms physical states into physical states), then it must transform density matrices into density matrices. It must thus be positivity preserving and trace preserving. However, though necessary, positivity is not sufficient to guarantee the physical consistency of a linear map. Via coupling the system affected by the transformation with a suitable unaffected ancilla, the existence of entanglement between the two subsystem requires the map to be also completely positive. In this section we will see how these constraints can take part in the challenge of detecting entangled states.

\section{Positivity and Complete Positivity}

\begin{definition}\textbf{Positive Map}
\label{def3-2}
 A linear map $\Lambda: M_m(\CI)\rightarrow M_n(\CI)$ is positive if and only if $M_n(\CI)\ni\Lambda[X]\geq 0$ for all $X\in M_m(\CI)$ ~\cite{{horo2},{niel-chu}}. The set of positive maps is denoted by $\cP$.
\end{definition}

\begin{example}\textbf{Trace Map}
\label{ex3-1}

The Trace Map, introduced in~\eqref{trace-map}, is a positive map, as it is the sum of the eigenvalues of an operator. Therefore if $ M_d(\CI)\ni X=\sum_i x_i \vert X_i\rangle\langle x_i\vert$, $x_i\geq 0$, then
$$
\Tr: X\longmapsto\Tr(X)\1_d=(\sum_i x_i)\1_d\geq0, \quad \forall X\in M_d(\CI).
$$
\end{example}

\begin{example}\textbf{Transposition Map}
\label{ex3-2}

Transposition map is another example of positive maps:
\begin{equation}
\label{trans-map}
\rT_d: M_d(\CI)\ni X\longmapsto X^{\rT}\in M_d(\CI)
\end{equation}
with respect to a fixed orthonormal basis. Given a matrix $X=[X_{ij}]$, its transposed is $X^{\rT}=[X_{ji}]$. Since transposing an operator does not affect its spectrum, it preserves positivity, which means that it is a positive map.
\end{example}

\begin{example}\textbf{Transposition Map on $M_2(\CI)$}
\label{ex-trans2}

Using the Pauli matrices as an orthonormal basis set in $M_2(\CI)$, similarly to the Trace Map in Example~\ref{ex-trace2}, we can also write the Transposition Map as~\cite{benatti2}:
\begin{equation}
M_2(\CI)\ni X\longmapsto \rT_2[X]:=\sum_{\alpha=0}^3\varepsilon_{\alpha}\sigma_{\alpha} X\sigma_{\alpha}
\end{equation}
where $\varepsilon_{\alpha}=1$ when $\alpha\neq2$, whereas $\varepsilon_2=-1$. In terms of the map $S_{\alpha}$ as in ~\eqref{map-s}, the Transposition Map is of the following form:
\begin{equation}
\label{trans-2}
\rT[X]=\sum_{\alpha=0}^3\varepsilon_\alpha S_\alpha[X].
\end{equation}

One can thus check that, in the standard representation, $T[\sigma_{\alpha}]=\sigma_{\alpha}$, $\alpha\neq2$ and $T[\sigma_{2}]=-\sigma_{2}$, which is exactly what transposition does in the chosen representation.
\end{example}

\begin{definition}\textbf{K-positivity}

A linear map $\Lambda: M_m(\CI)\rightarrow M_n(\CI)$ is k-positive if and only if $\id_k\otimes\Lambda: M_k(\CI)\otimes M_m(\CI)\rightarrow M_k(\CI)\otimes M_n(\CI)$ is a positive map, where $k\in N$.
$\cP_k$ denotes the set of k-positive maps~\cite{{woro1},{zycz1}}.
\end{definition}

\begin{definition}\textbf{Complete Positivity}
\label{cp-map}

 If a linear map $\Lambda$ is k-positive for every $k\in N$, it is called completely positive. The set of completely positive maps is denoted by $\cCP$.

 A linear map $\Lambda$ is a CPU map if it is both CP and unital, that is $\Lambda[\1]=\1$.
\end{definition}

The following theorem fully characterizes the completely positive maps:

\begin{theorem}\textbf{Kraus representation}
\label{theo-kraus}

A map $\Lambda:M_m(\CI)\rightarrow M_n(\CI)$ is completely positive if and only if there exists a set of operators $\{K_i\}$, such that~\cite{{kraus},{choi1},{benatti1}}:
\begin{equation}
\label{kraus-form}
\Lambda[X]=\sum_i K_i^{\dagger}X K_i, \quad \forall X\in M_m(\CI).
\end{equation}
The operators $M_{n\times m}(\CI)\ni K_i: \CI^n\rightarrow \CI^m$, are called Kraus operators. Furthermore, such a map is CPU if and only if $\sum_i K_i^{\dagger} K_i=\1$.
\end{theorem}

Theorem~\ref{theo-kraus} concerns the action of maps on operators (Heisenberg representation); however, one is often interested in how states transform (Schr\"{o}dinger representation). This fact can easily be derived by duality.

\begin{definition}\textbf{Dual map}

For any map $\Lambda:M_m(\CI)\rightarrow M_n(\CI)$, its dual map $\Lambda^T:M_n(\CI)\rightarrow M_m(\CI)$ is defined as~\cite{benatti2}:
\begin{equation}
\Tr(\Lambda[X]\rho)=\Tr(X\Lambda^T[\rho])
\end{equation}
for every $X\in M_m(\CI)$ and $\rho\in \SI_n$.
\end{definition}

The Kraus representation of the dual map $\Lambda^T:M_n(\CI)\rightarrow M_m(\CI)$ will be:
\begin{equation}
\label{kraus-dual}
\Lambda^T[\rho]=\sum_i K_i \rho K_i^{\dagger}.
\end{equation}
Therefore, one can see that CPU maps and trace-preserving maps are dual:
$$
\Lambda\quad\hbox{CPU}\quad \Longleftrightarrow \quad \Lambda^T\quad\hbox{Trace-preserving}.
$$

Consider a CP map $\Lambda:M_n(\CI)\rightarrow M_n(\CI)$, with its Kraus representation as in~\eqref{kraus-dual}. By choosing a Hilbert-Schmidt ONB set $\{F_j\}_{j=1}^{n^2}\in M_n(\CI)$, the Kraus operators $K_i$ can be written as:
$$
K_i=\sum_{j=1}^{n^2} c_{ij} F_j.
$$
Therefore the Kraus representation of $\Lambda$ will be:

\begin{eqnarray}
\nonumber
\Lambda[X]&=&\sum_i\sum_{j,k=1}^{n^2} c_{ij} \overline{c_{ik}} F_j X F_k^{\dagger}\\
\label{canon-kraus}
&=&\sum_{j,k=1}^{n^2} f_{jk} F_j X F_k^{\dagger}.
\end{eqnarray}

Where $f_{jk}=\sum_i c_{ij} \overline{c_{ik}}$ are the entries of a positive matrix $F=[f_{jk}]\in M_{n^2}(\CI)$. The representation~\eqref{canon-kraus} of a CP map $\Lambda$ is also known as Canonical Kraus Representation~\cite{zycz2}. Indeed the Kraus form~\eqref{kraus-dual}, is obtained by diagonalizing this general form.

One should also note that the set of Kraus operators $\{K_i\}$ is not unique, however, each two sets of Kraus operators corresponding to the same CP map, can be related by a unitary matrix. Indeed, given a canonical Kraus form, of a CP map $\Lambda$:
\begin{equation}
\label{k-1}
\Lambda[X]=\sum_{j,k=1}^{n^2} g_{jk} F_j X F_k^{\dagger}, \quad 0\leq G=[g_{jk}]\in M_{n^2}(\CI).
\end{equation}
Using the diagonal form of $G$ in terms of its eigenvectors:
$$
G=\sum_{l=1}^{n^2} g_l \vert g_l\rangle\langle g_l\vert, \quad g_l\geq 0,
$$
the matrix elements $g_{jk}$ are:
$$
g_{jk}=\langle j\vert G\vert k\rangle=\sum_{l=1}^{n^2} g_l g_{lj} \overline{g_{lk}}.
$$
Replacing this in~\eqref{k-1}, we get:
\begin{eqnarray*}
\Lambda[X]&=&\sum_{j,k,l=1}^{n^2} g_l g_{lj} \overline{g_{lk}} F_j X F_k^{\dagger}\\
&=&\sum_{l=1}^{n^2} (\sum_{j=1}^{n^2} \sqrt{g_l} g_{lj} F_j) X (\sum_{k=1}^{n^2} \sqrt{g_l} \overline{g_{lk}} F_k^{\dagger})\\
&=&\sum_{l=1}^{n^2} K_l X K_l^{\dagger}.
\end{eqnarray*}

\paragraph*{}
Any hermiticity and trace-preserving linear map $\Lambda: M_d(\CI)\rightarrow M_d(\CI)$ can be written as~\cite{koss2}:
\begin{equation}
\Lambda[\rho]=\sum_{i,j=1}^da_{ij} F_i^{\dagger}\rho F_j\ ,
\end{equation}
where $\sum_{i,j=1}^{d^2} a_{ij}F_i^{\dagger}F_j=\1_d$.
This can be shown as follows: consider the $d^2$ elementary linear maps $X\rightarrow \mathcal{F}_{ij}[X]=F_i\,X\,F_j^\dag$ and define on the linear space $\mathcal{L}$
of maps $\Lambda:M_d(\CI)\rightarrow M_d(\CI)$ the scalar product
\begin{equation}
\label{scalprod}
<<\Lambda_1|\Lambda_2>>=d^2\,\Tr\Big(\id_d\otimes\Lambda_1[P^{d}_+]^\dag\,\id_d\otimes\Lambda_2[P^{d}_+]\Big)\ ,
\end{equation}
where $P^d_+$ is the completely symmetric projection \eqref{P+}.
It turns out that $<<\mathcal{F}_{ij}|\mathcal{F}_{k\ell}>>=\delta_{ik}\delta_{j\ell}$; therefore, being orthogonal, the $\mathcal{F}_{ij}$ constitute an orthonormal basis in $\mathcal{L}$.
As we shall soon see, the scalar product defined above is strictly related to the notion of Choi matrix.

When the map $\Lambda$ is completely positive the matrix $A=[a_{ij}]$ is positive, while in case of positive maps, the matrix $A=[a_{ij}]$ is no longer positive, but with separating the positive and negative eigenvalues, one can write a positive map $\Lambda$ as the difference of two completely positive maps:
\begin{equation}
\label{kraus-gen-p}
\Lambda[\rho]=\sum_{a_l\geq 0} a_l K_l^{\dagger}\rho K_l - \sum_{a_l<0} |a_l| K_l^{\dagger}\rho K_l.
\end{equation}

\paragraph*{}
Based on the complete characterization of the structure of CP maps, the following concept provides a simple test to determine if a given map is CP or not~\cite{pau}.

\begin{definition}\textbf{Choi-Jamio{\l}kowski isomorphism}\hfill

Given a linear map $\Lambda: M_n(\CI)\mapsto M_m(\CI)$, its Choi matrix is the matrix on $\CI^n\otimes \CI^m$ defined by~\cite{{choi2},{belav},{pilis}}
\begin{equation}
\label{choi-matrix}
M_{n\times m}(\CI)=M_n(\CI)\otimes M_m(\CI)\ni C_\Lambda=\id_n\times\Lambda[P^{(n)}_+]\ ,
\end{equation}
where $P^{(n)}_+$ projects onto the completely symmetric state
$$
\vert\Psi^{(n)}\rangle=\frac{1}{n}\sum_{i=1}^n\vert e_i\rangle\otimes\vert e_i\rangle\ ,\quad \{\vert e_i\rangle\}_{i=1}^n\quad\hbox{ONB in $\CI^n$}\ .
$$

It is useful to see how the Choi matrix of the map $\Lambda$ looks like:
$$
C_\Lambda=\begin{pmatrix}
  \Lambda[\vert 1\rangle\langle 1\vert] & \Lambda[\vert 1\rangle\langle 2\vert] & \cdots & \Lambda[\vert 1\rangle\langle m\vert] \\
 \Lambda[\vert 2\rangle\langle 1\vert] & \Lambda[\vert 2\rangle\langle 2\vert] & \cdots & \Lambda[\vert 2\rangle\langle m\vert] \\
  \vdots  & \vdots  & \ddots & \vdots  \\
  \Lambda[\vert m\rangle\langle 1\vert] & \Lambda[\vert m\rangle\langle 2\vert] & \cdots & \Lambda[\vert m\rangle\langle m\vert]
 \end{pmatrix}.
$$

The Choi matrix, $C_\Lambda$,  is uniquely defined by $\Lambda$ since its entries ${C_\Lambda}_{(ij,pq)}$ with respect to an orthonormal basis of the form $\vert e_i\rangle\otimes\vert f_j\rangle\in\CI^n\otimes\CI^m$ are
\begin{equation}
\label{Choi2}
{C_\Lambda}_{(ij,pq)}=\langle e_i\otimes f_j\vert\, C_\Lambda\,\vert e_p\otimes f_q\rangle=
\frac{1}{n} \langle f_j\vert\Lambda[\vert e_i\rangle\langle e_p]\vert f_q\rangle\ .
\end{equation}
It turns out that the map $J:\Lambda\mapsto J[\Lambda]=C_\Lambda$ from the vector space $\mathcal{L}_{n,m}$ of linear maps onto the matrix algebra $M_{n\times m}(\CI)$ is invertible map and defines the so called Jamio{\l}kowski isomorphism. In fact, any $\Lambda: M_n(\CI)\mapsto M_m(\CI)$ is determined by its action on matrix units $\vert e_i\rangle\langle e_j\vert$ which is in turn completely characterized by the matrix elements
in (\ref{Choi2}). Therefore, given $C\in M_{n\times m}(\CI)$ this uniquely determines a linear map
$\Lambda_C: M_n(\CI)\mapsto M_m(\CI)$ such that
$$
\langle f_j\vert\Lambda_C[\vert e_i\rangle\langle e_p\vert] f_q\rangle=C(ij,pq)\ .
$$
\end{definition}

Equipped with these tools one can readily prove the following

\begin{theorem}
\label{choi-theo}
Every linear map $\Lambda: M_n(\CI)\mapsto M_n(\CI)$ with positive semi-definite Choi matrix $C_\Lambda\geq 0$ is completely positive and vice versa~\cite{{choi2},{horo3}}.
\end{theorem}

\proof
By definition of complete positivity, $\id_n\otimes\Lambda$ is positive if $\Lambda$ is completely positive, and thus $C_\Lambda=\id_n\otimes\Lambda[P^{(n)}]\geq 0$.

Vice versa, if $M_{n^2}(\CI)\ni C_\Lambda\geq 0$, then it can be spectralized as
$$
C_\Lambda=\sum_{j=1}^{n^2} c_j\,\vert c_j\rangle\langle c_j\vert\ ,\quad c_j\geq 0\ .
$$
The eigenvectors $\vert c_j\rangle\in\CI^n\otimes\CI^n$ can be conveniently rewritten using the completely symmetric
vector as follows:
$$
\vert c_j\rangle=\sum_{p,q=1}^n\,c_j^{pq}\,\vert e_p\rangle\otimes\vert e_q\rangle=
\frac{1}{n}\sum_{p=1}^n\,\vert e_p\rangle\otimes\, C_j\,\vert e_p\rangle=\1_n\otimes C_j\vert\Psi^{(n)}_+\rangle\ ,
$$
with the matrix $C_j\in M_n(\CI)$ defined by
$$
\langle e_q\vert\,C_j\,\vert e_p\rangle= n\, c_j^{pq}\ .
$$
Therefore, one can rewrite
$$
C_\Lambda=\sum_{j=1}^{n^2} \1_n\otimes K_j\,P^{(n)}_+\,\1_n\otimes K_j^\dag\ ,\quad K_j=\sqrt{L_j}\,\quad
C_j\in M_n(\CI)\ ,
$$
Then, by the Jamio{\l}kowski isomorphism, $\Lambda[X]=\sum_{j=1}^{n^2}K_j\,X\,K_j^\dag$ can be written in the Kraus-Stinespring form and is thus completely positive.\qed

\begin{remark}
In general, positive maps $\Lambda: B(\HI)\mapsto B(\HI)$ on bounded operators are completely positive if $\id_d\otimes\Lambda$ for all $d\geq2$. However, in the finite dimensional cases, when $B(\HI)=M_n(\CI)$, the above result shows that one needs check the positivity of $\id_n\otimes\Lambda$, only.
\end{remark}

\begin{example}
Consider the Trace Map $\Phi: M_d(\CI)\longmapsto M_d(\CI)$:
\begin{equation}
\label{trace-prese}
\Phi(X)=\frac{1}{d}\Tr(X)\1_d.
\end{equation}
It is easy to see that this map is CPU; indeed:
 $$
 \Phi[\1_d]=\1_d.
 $$
It's Choi matrix is:
\begin{eqnarray}
\nonumber
C_{\Phi}&=&\frac{1}{d}\sum_{ij} \vert i\rangle\langle j\vert \otimes \Tr(\vert i\rangle\langle j\vert) \vert i\rangle\langle j\vert\\
\nonumber
&=&\frac{1}{d}\sum_{ij} \vert i\rangle\langle i\vert \otimes \vert i\rangle\langle i\vert=\frac{1}{d}\1_{d^2}\ge0.
\end{eqnarray}
The positivity of Choi matrix shows the Trace Map is completely positive.
\end{example}
\medskip

\begin{definition}\textbf{Block Positive Operator}

Given a matrix $A\in M_d^2(\CI)$, using any fixed orthonormal basis, we can write it as $A=\sum_{i,j}^d \vert i\rangle\langle j\vert\otimes A_{ij}$, where $A_{ij}\in M_d(\CI)$, for $\forall 1\leq i,j \leq d$. Then matrix A is said to be block-positive if and only if for all $\vert \phi\rangle$ and $\vert \psi\rangle\in \CI^d$, the following holds:
\begin{eqnarray}
\nonumber
\langle \phi\otimes \psi\vert A\vert \phi\otimes \psi\rangle&=&\langle \phi\vert( \sum_{i,j} \langle \psi\vert A_{ij}\vert \psi\rangle  \vert i\rangle\langle j\vert)\vert \phi\rangle\\
\nonumber
&=&\langle \phi\vert \begin{pmatrix}
  \langle \psi\vert A_{11}\vert \psi\rangle & \langle \psi\vert A_{12}\vert \psi\rangle & \cdots & \langle \psi\vert A_{1d}\vert \psi\rangle \\
 \langle \psi\vert A_{21}\vert \psi\rangle & \langle \psi\vert A_{22}\vert \psi\rangle & \cdots & \langle \psi\vert A_{2d}\vert \psi\rangle\\
  \vdots  & \vdots  & \ddots & \vdots  \\
  \langle \psi\vert A_{d1}\vert \psi\rangle & \langle \psi\vert A_{d2}\vert \psi\rangle & \cdots & \langle \psi\vert A_{dd}\vert \psi\rangle
 \end{pmatrix} \vert \phi\rangle\\
 \nonumber
 &\geq& 0.
\end{eqnarray}
In another word, block-positivity of $A$ is equivalent to positivity of $\langle \psi\vert A_{ij}\vert \psi\rangle$ for $\forall \vert \psi\rangle\in\CI^d$.
\end{definition}

\begin{theorem}
\label{choi-block}
A linear map $\Lambda: M_d(\CI)\rightarrow M_d(\CI)$ is positive if and only if its Choi matrix is block-positive~\cite{{jami1},{koss1}}:
$$
\langle\psi\otimes\varphi\vert\1_d\otimes\Lambda[P^+_d]\vert\psi\otimes\varphi\rangle\geq0
$$
for every $\vert\psi\rangle$ and $\vert\varphi\rangle$ in $\CI^d$.
\end{theorem}

\proof
Given a positive map $\Lambda:M_d(\CI)\rightarrow M_d(\CI)$, by definition, it preserves positivity of matrices in $M_d(\CI)$:
\begin{equation}
\label{pos1}
\langle\phi\Lambda[\vert\psi\rangle\langle\psi\vert]\vert\phi\rangle\geq0\quad \forall\vert\psi\rangle,\vert\phi\rangle\in\CI^d.
\end{equation}
Given an ONB $\{\vert i\rangle\}_{i=1}^{d}\in\CI^d$, then any vector $\vert\psi\rangle$ is $\vert\psi\rangle=\sum_{i=1}^{d}\psi_i \vert i\rangle$, where $\psi_i=\langle i\vert\psi\rangle$. Using this~\eqref{pos1} is
\begin{eqnarray*}
\langle\phi\Lambda[\vert\psi\rangle\langle\psi\vert]\vert\phi\rangle&=&\sum_{i,j=1}^d \psi_i\psi_j^* \langle\phi\vert\Lambda[\vert i\rangle\langle j\vert]\vert\phi\rangle\\
&=&\langle\psi^*\otimes\phi\vert\sum_{i,j=1}^d \vert i\rangle\langle j\vert \otimes \Lambda[\vert i\rangle\langle j\vert]\vert\psi^*\otimes\phi\rangle\\
&=&\langle\psi^*\otimes\phi\vert\1_d\otimes\Lambda[P_d^+]\vert\psi^*\otimes\phi\rangle\\
&=&\langle\psi^*\otimes\phi\vert C_{\Lambda}\vert\psi^*\otimes\phi\rangle\\
&\geq0.
\end{eqnarray*}
Which shows that the positivity of $\Lambda$ is equivalent to the block positivity of its Choi matrix and vice versa.
\qed
\medskip

\begin{example}
Consider the Transposition Map $\rT: M_d(\CI)\rightarrow M_d(\CI)$, in~\eqref{trans-map}. It's Choi matrix is:
$$
C_{\rT}=\frac{1}{d}\sum_{i,j} \vert i\rangle\langle j\vert \otimes \vert j\rangle\langle i\vert.
$$
Which coincides with the definition of the Flip operator:
\begin{equation}
\label{flip}
V=\sum_{i,j=1}^{d} \vert i\rangle\langle j\vert\otimes \vert j\rangle\langle i\vert=d(\1_d\otimes\rT)P_d^+,
\end{equation}
which is also that
$$
V\vert \psi_1\otimes\psi_2\rangle=\vert\psi_2\otimes\psi_1\rangle.
$$
Since $V^2=\id_d$, its eigenvalues are $\pm1$.

For any vector $\vert \phi\rangle$ and $\vert \psi\rangle\in \CI^d$, we have:
\begin{eqnarray}
\nonumber
\langle\psi\otimes\varphi\vert C_{\rT} \vert\psi\otimes\varphi\rangle&=&\langle\psi\otimes\varphi\vert(\sum_{i,j} \vert i\rangle\langle j\vert \otimes \vert j\rangle\langle i\vert)\vert\psi\otimes\varphi\rangle\\
\nonumber
&=&\sum_{i,j}\langle\phi\vert i\rangle\langle j\vert \phi\rangle \langle\psi\vert j\rangle \langle i\vert\psi\rangle\\
\nonumber
&=& \sum_i \langle\phi\vert i\rangle \langle i\vert\psi\rangle \sum_j \langle j\vert \phi\rangle \langle\psi\vert j\rangle\\
\nonumber
&=&\vert\langle \psi\vert \phi\rangle\vert^2\\
\nonumber
&\geq&0.
\end{eqnarray}
Since $C_T$ has negative eigenvalues, $C_T\ngeq0$, Transposition Map is not completely positive.
\end{example}
\medskip

\begin{example}\textbf{Reduction Map}

In the next chapter we will introduce a number of positive maps, which are known to be able to detect some entangled states, one of them is called Reduction Map~\cite{horo3}, which has the following form:
\begin{equation}
\label{red-map}
\Lambda_R=\Tr_d-\id_d
\end{equation}
The Reduction map is positive: as the trace is the sum of the eigenvalues, which are non-negative as $\rho$ is a density matrix, therefore $\Tr_d[\rho]\geq 0$. Let $\vert\psi_i\rangle$ be eigenvectors of $\rho$ with eigenvalues $\lambda_i$, then using the spectral decomposition of $\rho$, we have:
\begin{eqnarray}
\nonumber
\Lambda_R[\rho]&=&(\sum_i\lambda_i)\1_d-\sum_j\lambda_j\vert\psi_j\rangle\langle\psi_j\vert\\
\nonumber
&=&\sum_j[(\sum_i\lambda_i)-\lambda_j]\vert\psi_j\rangle\langle\psi_j\vert.
\end{eqnarray}
Since $\sum_i\lambda_i\geq\lambda_j$, $\Lambda_R[\rho]$ is a positive matrix.
On the other hand, the Choi matrix of this map is:
\begin{eqnarray}
\nonumber
C_{\Lambda_{R}}=\id_d\otimes\Lambda_{R}[P_d^+]=\1_{d^2}-d P_d^+\\
\nonumber
\hbox{where}\qquad P_d^+=\frac{1}{d}\sum_{i,j=1}^{d^2} \vert i\rangle\langle j\vert\otimes \vert i\rangle\langle j\vert.
\end{eqnarray}
Thus, $C_{\Lambda_R}$ has one negative eigenvalue $1-d$ and is not positive semi-definite. Therefore, the Reduction Map is not completely positive.
\end{example}
\medskip

In what follows, we shall explain the results given in~\cite{storm2}, which is extensively used in the last chapter of this thesis.

\begin{theorem}
\label{theo-stormer}
Let $\Tr$ be the Trace Map. Then, given any positive map $\phi:M_d(\CI)\rightarrow M_d(\CI)$ can be written as
\begin{equation}
\phi=\frac{1}{c}(\Tr-\phi_{cp}),
\end{equation}
where $c\geq0$ is a positive constant and $\phi_{cp}: M_d(\CI)\rightarrow M_d(\CI)$ a completely positive map.
\end{theorem}

\proof

Let $\phi$ be a self-adjoint map, that is $\phi[A^{\dagger}]=(\phi[A])^{\dagger}$, then its Choi matrix $C_\phi$ is self-adjoint as well. Therefore $C_\phi$ can be written as difference of two positive matrices, i.e. $C_\phi=C_\phi^+-C_\phi^-$ where $C_\phi^+C_\phi^-=0$. Let $\mu\geq0$ be the smallest positive number such that
$$
\mu\1\geq C_\phi=C_\phi^+-C_\phi^-.
$$
The above inequality is satisfied by choosing $c=\|C_{\phi}^+\|$, which is the greatest eigenvalue of $C_{\phi}^+$.

Now consider the following positive matrix:
\begin{equation}
\label{eq5.1}
C_{\phi}^{\mu}=\1-\frac{1}{\mu}C_\phi\geq0.
\end{equation}

The completely positive map $\phi_{cp}$ of which $C_{\phi}^{\mu}$ is the Choi matrix reads

\begin{equation}
\label{eq5.2}
\phi_{cp}=\Tr-\frac{1}{\mu}\phi,
\end{equation}

Now, re-arranging the~\eqref{eq5.2},we have:
\begin{equation}
\label{eq5.3}
\phi=\frac{1}{c}(\Tr-\phi_{cp}).
\end{equation}
Note that the only assumption was to consider the map $\phi$ to be self-adjoint. Therefore any self-adjoint map can be written as in~\eqref{eq5.3}. Taking into account that the set of positive maps is a subset of self-adjoint maps completes the proof.
\qed
\medskip

\begin{example}
\label{ex1}
We have seen in Example~\ref{ex-trace2} that the transposition on $M_2(\CI)$ can be represented as
\begin{eqnarray}
\label{transp2}
M_2(\CI)\ni X\mapsto T[X]=\frac{1}{2}\sum_{\alpha=0}^3\,\varepsilon_\alpha\,S_{\alpha}[X], \\
\nonumber
S_{\alpha}[X]=\sigma_\alpha\,X\,\sigma_\alpha\ ,\quad \varepsilon_\alpha=(1,1,-1,1)\ .
\end{eqnarray}
Therefore, on $M_4(\CI)$,
\begin{eqnarray}
\label{transp4}
M_4(\CI)\ni X\mapsto T[X]=\frac{1}{4}\sum_{\alpha,\beta=0}^3\,\varepsilon_\alpha\varepsilon_\beta\,S_{\alpha\beta}[X],\\
\nonumber
S_{\alpha\beta}[X]=\sigma_{\alpha\beta}\,X\,\sigma_{\alpha\beta}\ ,\quad \sigma_{\alpha\beta}=\sigma_{\alpha}\otimes\sigma_\beta\ .
\end{eqnarray}
Notice that the transposition is not written in the Kraus form~(\ref{kraus-dual}) as the products $\varepsilon_\alpha\varepsilon_\beta=-1$ whenever $\alpha\neq\beta=2$ or $\beta\neq\alpha=2$.

On the other hand, the completely positive trace map $\Tr$ on $M_2(\CI)$ has the following Kraus representation:
\begin{equation}
\label{Tr2}
{\Tr}[X]=\frac{1}{2}\sum_{\alpha=0}^3S_\alpha[X]\ ,\qquad X\in M_2(\CI)\ ,
\end{equation}
Then, on $M_4(\CI)$, the trace map has the Kraus form~(\ref{kraus-dual}):
\begin{equation}
\label{Tr4}
M_4(\CI)\ni X\mapsto{\Tr}[X]=\frac{1}{4}\sum_{\alpha,\beta=0}^3\,S_{\alpha\beta}[X]\ .
\end{equation}

We would like to write the transposition map as in the form suggested by Theorem~\ref{theo-stormer}:
\begin{equation}
\label{stormer}
 \Lambda=\mu(\Tr -\Lambda^{CP}).
\end{equation}

When $\Lambda$ is the transposition map on $M_4(\CI)$, the CP maps $\phi_{CP}$ in~(\ref{stormer}) are easily found: the linear map
$$
\Tr-\frac{1}{\mu}\rT=\frac{1}{4}\sum_{\alpha,\beta=0}^3 \left(1-\frac{\varepsilon_\alpha\varepsilon_\beta}{\mu}\right) S_{\alpha\beta}\ ,
$$
is of the form~(\ref{kraus-dual}), thus CP, if and only if $\mu\geq 1$. The smallest choice, $\mu=1$, yields the completely positive map
\begin{equation}
\label{CPTRansp}
\Lambda^{CP}=\frac{1}{2}\Big(\sum_{\alpha\neq 2}\,S_{\alpha2}\,+\,\sum_{\beta\neq 2}\,S_{2\beta}\Big)\ .
\end{equation}
\end{example}
\medskip

\section{Entanglement and Positive Maps}

As we have seen the structure of completely positive maps is wholly understood via the Kraus representation. Moreover Choi's theorem provides a technique to sort them out. Instead, as soon as one is dealing with positive maps, the situation becomes much less clear due to the lack of a complete structural representation of them. As we will see, only in low dimensions, when $d_1\times d_2\leq 6$, the structure of positive maps is under control~\cite{{woro1},{zycz1}}. While in higher dimensions these maps are only partially understood~\cite{{koss3},{koss4},{hou},{take2},{tomi},{maje}}.

\paragraph*{}

From a physical point of view, it turns out that a linear map which describes a physical transformation needs not to be only positive but completely positive~\cite{lind}: since the eigenvalues of a system in statistical interpretation represent the probabilities, when a linear map describes a physical transformation, it has to preserve the positivity of the spectrum of the physical system, otherwise a negative eigenvalue representing a probability brings up a contradiction. As we have seen, by Definition~\ref{cp-map} any extension of completely positive maps does this job. However this might not be true in general if the transformation is described by a positive, not CP, map. While the extension of positive maps preserves the positivity of separable states, it can fail when the subsystems are entangled. Consider a bipartite separable state which is a convex combination of tensor product of states its subsystems:
\begin{equation}
\label{eq1}
 \rho=\sum_{i=1}p_i\rho_i^1\otimes\rho_i^2 \quad p_i\geq0,\quad \sum_i p_i=1.
\end{equation}
The action of $\id_k\otimes\Lambda$ when $\Lambda$ is a positive map gives:
$$
\rho'=(\id_k\otimes\Lambda)[\rho]=\sum_{i=1}p_i\id_k[\rho_i^1]\otimes\Lambda[\rho_i^2].
$$
The positivity of $\Lambda$ guarantees that $\Lambda(\rho_i^2)$ and therefore $\rho'$, remains a positive state. But if the state $\rho$ is not separable it cannot be written as~\eqref{eq1}, and consequently, for a generic $\rho$ and $k$, $(\id_k\otimes\Lambda)[\rho]$ is positive only if $\Lambda$ is completely positive.

\paragraph*{}
However, though positive maps cannot be used to describe physical transformations, they provide a new powerful theoretical approach to detect the entanglement. Indeed, if a given state $\rho$ does not remain positive under the action of an extended positive maps, it must be entangled:
$$
(\id_k\otimes\Lambda)[\rho]\ngeq0\quad\Rightarrow\rho\quad\hbox{ entangled.}
$$

\paragraph*{}

The prototype of positive, but not completely positive map is transposition.  In the following example we show that the extension of this map fails to be k-positive already when k is equal to 2, which means that it is not a complete positive map. In other words, though a positive matrix remains positive after being transposed, it might not be positive after being partially transposed.

\begin{example}
Take one of the maximally entangled Bell states, introduced in~\eqref{bell-states}, $\vert\Phi^+\rangle=\frac{1}{\sqrt{2}}(\vert 00\rangle+\vert 11\rangle)$, its density matrix is:
$$
\rho_{\Phi^+}=\vert\Phi^+\rangle\langle\Phi^+\vert=\frac{1}{2}
 \left (
 \begin{array}{cc|cc}
  1 & 0 & 0 & 1 \\
  0 & 0 & 0 & 0 \\ \hline
  0 & 0 & 0 & 0  \\
  1 & 0 & 0 & 1
 \end{array}
 \right )
$$
The action of $\1_2\otimes\rT$ on this state is to transposing each block of the matrix, the result is the following matrix:
$$
(\1_2\otimes\rT)\rho_{\Phi^+}=\frac{1}{2}
 \left (
 \begin{array}{cc|cc}
  1 & 0 & 0 & 0 \\
  0 & 0 & 1 & 0 \\ \hline
  0 & 1 & 0 & 0  \\
  0 & 0 & 0 & 1
 \end{array}
 \right )
$$
which has a negative eigenvalue $-\frac{1}{2}$.
\end{example}

According to Partial Transposition, the states can be divided into two classes: those which remain positive under this map, called PPT states, and those which turn out to have at least one negative eigenvalue after being partially transposed, known as NPT states.

The following class of positive maps has been used to classify these maps in low dimensional Hilbert spaces.

\begin{definition}\textbf{Decomposable Map}

A positive map $\Lambda: M_d(\CI)\longrightarrow M_d(\CI)$ is decomposable if and only if it can be written as~\cite{storm1}:
\begin{equation}
\label{decompose-map}
\Lambda=\Gamma_1+\Gamma_2\circ\rT_d
\end{equation}
where $\Gamma_1$ and $\Gamma_2$ are complete positive maps.
\end{definition}

\begin{theorem}
\label{decom-6}
All the positive maps $\Lambda: M_{d_1}(\CI)\longrightarrow M_{d_2}(\CI)$ where $d_1\times d_2\leq6$ are decomposable~\cite{{woro1},{storm1}}.
\end{theorem}

\begin{example}
The reduction map $\Lambda_R: M_d(\CI)\rightarrow M_d(\CI)$ in~\eqref{red-map} is decomposable~\cite{horo3}.

The Choi matrix of this map reads:
\begin{eqnarray}
\nonumber
C_{\Lambda_{R}}=\1\otimes\Lambda_{R}[P_d^+]=\1_{d^2}-d P_d^+\\
\nonumber
\hbox{where}\qquad P_d^+=\frac{1}{d}\sum_{i,j=1}^{d^2} \vert i\rangle\langle j\vert\otimes \vert i\rangle\langle j\vert.
\end{eqnarray}

Its partial transposition with respect to the second subsystem yields:
\begin{eqnarray}
\nonumber
C_{\Lambda_R}^{\rT_2}&=&(\1_d\otimes\rT_d)\circ(\1_d\otimes\Lambda_R)[P_d^+]\\
&=&(\1_d\otimes\rT\circ\Lambda_R)[P_d^+]=\frac{1}{d}(\1_{d^2}-V).
\end{eqnarray}
Since the flip operator $V$, introduced in~\eqref{flip}, has eigenvalues $\pm1$, it implies that $C_{\Lambda_R}^{\rT_2}$ has no negative eigenvalue and therefore is a positive matrix, whence the corresponding map $\Lambda'=\rT\Lambda_R$ is completely positive, according to Theorem~\ref{choi-theo}. Furthermore, since $\Lambda_R=\rT\Lambda'$ where $\Lambda'$ is CP, results decomposable.
\end{example}

In the next chapters we will see how and under which conditions positive maps can detect the entanglement.

\newpage

\chapter{Detecting Entanglement}

In the previous chapter we have seen formal necessary and sufficient conditions for separability, practically, establishing whether a given state is separable or not is in general a very hard task. Indeed apart from PPT criterion in low dimensions, the available theorems do not provide constructive tools to detect entangled states in general. In the present chapter we will review some of the entanglement detecting methods. Though none of them is able to detect all entangled states, they nevertheless work for particular classes of states.

\section{PPT Criterion}

In the last chapter, we have introduced Transposition Map. Since this map is not CP, it can be used to detect entangled states. In fact, partial transposition provides one of the most easily applicable tools to detect entanglement. Unfortunately in higher dimensions, being PPT is no longer a sufficient condition for separability.

\begin{theorem}\textbf{Horodecki Criterion}
\label{horo-cri}

A bipartite state $\rho\in\SI_{d\times d}$ is separable if and only if for any positive map $\Lambda:M_{d}(\CI)\rightarrow M_{d}(\CI)$, $(\id_{d}\otimes\Lambda)\rho$ is positive~\cite{horo2}.
\end{theorem}

As this must hold for any positive map $\Lambda$, so does it for Transposition Map:

\begin{theorem}\textbf{Peres Criterion}

If a bipartite state $\rho\in\SI_{d\times d}$ is separable, then $(\id_{d}\otimes\rT)\rho\geq0$~\cite{peres1}.
\end{theorem}

Obviously when $(\id_{d}\otimes\Lambda)\rho\ngeq0$, the state is NPT, and according to the above theorem, is entangled.  As we have already mentioned, Transposition Map is an exhaustive entanglement witness in low dimensions:

\begin{corollary}\textbf{PPT Criterion}
\label{peres}
A state $\rho$ acting on $\CI^2\otimes\CI^2$, $\CI^2\otimes\CI^3$ or $\CI^3\otimes\CI^2$ is separable if and only if its partial transposition is a positive matrix~\cite{horo2}
\end{corollary}

\proof

 If $\rho$ is separable, Theorem~\ref{horo-cri} tells us that its partial transposition $\rho^{\rT_2}$ is also positive.

 The sufficiency is proved as follows: In Theorem~\ref{decom-6} we have seen that all positive maps $\Lambda:M_{d_1}(\CI)\rightarrow M_{d_2}(\CI)$ when $d_1\times d_2\leq 6$ are decomposable and can be written as:
$$
\Lambda=\Gamma_1 +\Gamma_2\circ\rT,
$$
where $\Gamma_{1,2}$ are CP maps.
The extension of this map acting on $\rho$ gives:
$$
(\id_{d}\otimes\Lambda)[\rho]=(\id_{d}\otimes\Gamma_1)[\rho]+(\id_{d}\otimes\Gamma_2)[\rho^{\rT_2}].
$$
The first term is positive because $\Gamma_1$ is a CP map and by definition $(\id_{d}\otimes\Gamma_1)[\rho]\geq0$. If $\rho$ is PPT, then $\rho^{\rT_2}\geq0$, then the second term remains positive as well, due to complete positivity of $\Gamma_2$, which completes the proof.
\qed
\medskip

\begin{example}
 All Bell states introduced in~\eqref{bell-states}, have negative partial transposition and are NPT entangled.
\end{example}

\begin{example}
In~\eqref{werner-state}, we have introduced the Werner States in $\CI^2\otimes\CI^2$:
$$
(\1_2\otimes\rT_2)\rho_{\alpha}=\rho^{\rT_2}_{\alpha}=\alpha\vert\Psi^-\rangle\langle\Psi^-\vert+\frac{1-\alpha}{4}\1_2\otimes\1_2, \quad -\frac{1}{3}\leq\alpha\leq 1.
$$
Their partially transposed density matrix is:
$$
\rho_{\alpha}^{\rT_2}=\frac{1}{4}\begin{pmatrix}
  1-\alpha & 0& 0 & -2\alpha \\
 0 & 1+\alpha & 0 & 0 \\
  0  & 0  & 1+\alpha & 0  \\
  -2\alpha & 0 & 0 & 1-\alpha
 \end{pmatrix}.
$$
Its eigenvalues are:
$$
\lambda_1=\lambda_2=\lambda_3=\frac{1+\alpha}{4}, \lambda_4=\frac{1-3\alpha}{4}.
$$
As we see the first three eigenvalues are positive, while the fourth one when $\frac{1}{3}\leq\alpha\leq1$ is negative. Therefore:
\begin{eqnarray}
 \nonumber
\frac{-1}{3}\leq\alpha<\frac{1}{3}\quad\Rightarrow\rho_{\alpha}\quad \hbox{is separable,}\\
\nonumber
\\
\nonumber
\frac{1}{3}\leq\alpha\leq1\quad\Rightarrow\rho_{\alpha}\quad \hbox{is entangled.}
\end{eqnarray}
In Example~\ref{ex-werner}, we have seen the behavior of $\Delta^{\alpha}=S(\rho_{\alpha})-\log 2$, where we have mentioned, though it might look like that von Neumann entropy might be able to detect the entangled Werner states, however, it only detects entangled states for $0.75<\alpha\leq 1$, but it fails to detect the entangled Werner states for $\frac{1}{3}\leq\alpha\leq 0.75$.
\end{example}

\section{Entanglement Witness}

The question of whether a given density matrix $\rho$ of a higher system is separable or not is very difficult to answer due to the difficulties of reconstructing the whole density matrix in terms of tensor product of its subsystems, i.e. to write $\rho=\sum_i p_i\vert\psi_i^1\rangle\langle\psi_i^1\vert\otimes\vert\psi_i^2\rangle\langle\psi_i^2\vert$. However the fact that a set of such separable states is convex and closed, makes it possible to apply the Hahn-Banach separation theorem. This opens a new approach to detect entanglement, called entanglement witnessing.

\begin{theorem}\textbf{Hahn-Banach Theorem}
\label{hahn-banach}

Let $X$ be a finite-dimensional Banach space and S a convex compact set in it. Let $x_0$ be a point in $X$ such that $x_0\notin S$.
Then there exists a hyperplane which separates $x_0$ from S~\cite{{brus1},{zycz2}}.
\end{theorem}

\begin{center}
\label{fig2}
\includegraphics[scale=0.5]{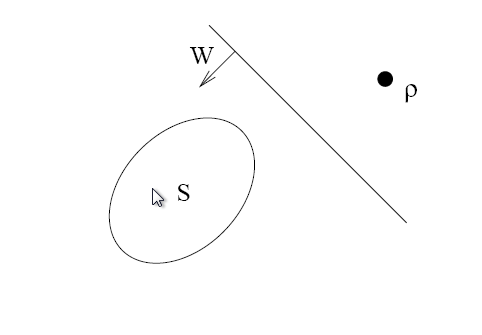}

\end{center}

Let the set $\SI_{sep}$ correspond to the set of separable states which by definition is convex, compact and closed. Therefore the Hahn Banach theorem~\ref{hahn-banach} says that there must exists a hyperplane separating this set from an entangled state which lies outside. This separator is called Entanglement Witness, which is a Hermitian operator~\cite{{horo2},{terhal}}.

\begin{definition}\textbf{Entanglement Witness}
\label{EW}

A Hermitian operator W acting on $\HI=\CI^{d_1}\otimes\CI^{d_2}$ is called an entanglement witness if
\begin{enumerate}
\item
$\exists \rho\in\SI_{d_1\times d_2} \quad\hbox{such that}\quad\Tr(\rho W)<0.$
\item
$\langle\varphi\otimes\psi\vert W\vert\varphi\otimes\psi\rangle\geq0\qquad \forall\vert\varphi\rangle\in\CI^{d_1},\vert\psi\rangle\in\CI^{d_2}.$
\item
$\Tr(W)=1$
\end{enumerate}
\end{definition}

The second property shows the block-positivity of the non-positive operator W, while the last one asks only for normalization. There are immediate consequences:

\begin{theorem}
 \label{ent-iff}
 $\rho\in\SI_{d\times d}$ is entangled if and only if there exists an entanglement witness such that $\Tr(\rho W)<0$~\cite{horo2}.
\end{theorem}

\begin{theorem} $\sigma\in\SI_{d\times d}$ is separable if and only if $\Tr(\sigma W)\geq 0$ for all entanglement witnesses W~\cite{horo2}.
\end{theorem}

\begin{definition}\textbf{Decomposable Entanglement Witness}
\label{DEW}

Let P and Q be positive operators acting on $\CI^d\otimes\CI^d$, then W is a decomposable entanglement witness if and only if~\cite{woro2}
$$\exists a,b\geq 0 \quad \hbox{such that}\quad W=a P+b Q^{\rT_1}$$
where $\rT_1$ is Partial Transposition.
\end{definition}

\begin{theorem}
\label{dew-2}
An entanglement witness, EW, is non-decomposable if and only if it detects PPT entangled states~\cite{lewe1}.
\end{theorem}

\begin{theorem}
$\rho\in\SI_{d\times d}$ is PPT entangled if and only if there exists a non-decomposable entanglement witness such that $\Tr(\rho W)<0$~\cite{{horo2},{woro2}}.
\end{theorem}

Given two entanglement witnesses $W_1$ and $W_2$, the latter one is said to be finer~\cite{lewe1} if all the entangled states detected by $W_1$ are also detected by $W_2$. An EW, $W_{opt}$ is optimal~\cite{{lewe1},{horo2}} if there exists a separable state $\tilde{\rho}$ such that $\Tr(\tilde{\rho}W_{opt})=0$. It means that the hyperplane corresponding to $W_{opt}$ is tangent plane to $\SI_{sep}$ (for illustrations see Fig~\eqref{fig3}).
Since optimal entanglement witnesses detect more entangled states than those non-optimal ones, they are highly interesting. Indeed the set of separable states can be completely characterized by the set of optimal EW~\cite{bert1}.

\begin{center}
\label{fig3}
\includegraphics[scale=0.5]{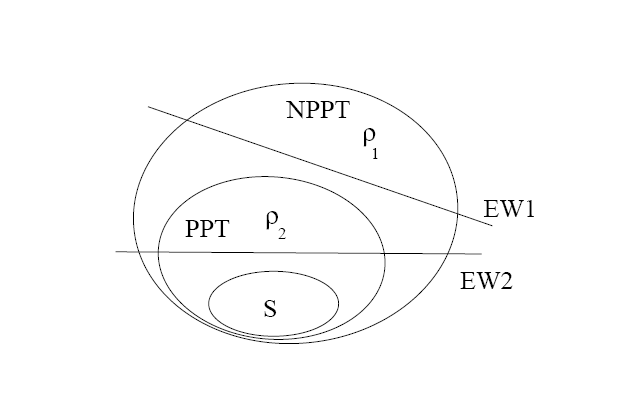}

\end{center}

It has been shown~\cite{lewe1} that a non-optimal EW can be optimized, but one should note that none of these entanglement witnesses can detect all the entangled states.

Both positive maps and entanglement witnesses provide necessary and sufficient conditions for separability. Indeed to some extent their behavior is quite similar toward entangled states; there exists a positive map and an EW for each entangled state. This brings up the conjecture that these connections can be explained through an isomorphism.

\begin{definition}\textbf{Jamio{\l}kowski Isomorphism}

Jamio{\l}kowski isomorphism says that there is a one-to-one relation between an EW as an operator and a positive map $\Lambda:M_{d}(\CI)\rightarrow M_{d}(\CI)$~\cite{jami1}:
$$
\Lambda_W\longleftrightarrow W_\Lambda\in M_{d^2}(\CI)
$$
Given an operator $W_\Lambda$ acting on $M_{d^2}(\CI)$ and choosing an orthonormal product basis set $\{ \vert e_i\otimes f_i\rangle\}$ in $\HI=\CI^{d}\otimes\CI^d$, the corresponding linear map $\Lambda_W: M_d(\CI)\longrightarrow M_d(\CI)$ is defined~\cite{brus1}:
$$
\Lambda_W[\rho]=\Tr_2[W\rho^{\rT_2}].
$$
Conversely from a given map $\Lambda$, the operator $W_\Lambda$ is obtained:
$$
W_\Lambda=(\1_d\otimes\Lambda)\vert\psi_d^+\rangle\langle\psi_d^+\vert
$$
where $\vert\psi_d^+\rangle$ is defined in~\eqref{psi+}.
\end{definition}

The following properties hold~\cite{{horo2},{jami1},{woro2}}
\begin{itemize}
\item
$W\geq0$ if and only if $\Lambda_W$ is completely positive.
\item
$W$ is EW, if and only if $\Lambda_W$ is positive.
\item
$W$ is decomposable if and only if $\Lambda_W$ is decomposable.
\item
$W$ is non-decomposable if and only if $\Lambda_W$ is non-decomposable.
\end{itemize}

\section{Reduction Map}

In the previous chapter, we have introduced a positive (not CP) map $\Lambda_R:M_d(\CI)\longrightarrow M_d(\CI)$, called Reduction map:
\begin{equation}
\Lambda_R[\rho]=\Tr_d[\rho]-\rho.
\end{equation}
Below we will see how this map can be used to detect the entanglement:

\begin{theorem}\textbf{Reduction Criterion}

A bipartite state $\rho_{12}$ acting on $\HI=\CI^{d_1}\otimes\CI^{d_2}$ with $d_1\times d_2\leq 6$, is separable if and only if~\cite{horo3}
$$
\id_{d}\otimes\rho_2-\rho_{12}\geq0.
$$
For states acting on higher dimensional Hilbert spaces, the criterion is only necessary for separability.
\end{theorem}

This map however can not detect PPT entangled states, as we have shown in the last chapter, that this map is decomposable.

\section{Range Criterion}

The range criterion is another approach to detect entanglement. This method is based on finding the product vectors in the range of the states. The range of $\rho$ is the set of all the vectors $\vert\psi\rangle$ for which there is another vector $\vert\phi\rangle$ such that:
$$
\vert\psi\rangle=\rho\vert\phi\rangle.
$$

\begin{definition}\textbf{Product Vector}
\label{PV}

Let $\HI=\bigotimes_{i=1}^m\CI^{d_i}$ be an m-partite Hilbert space. A vector $\vert\psi\rangle\in\HI$ is called product vector if and only if it has the form $\vert\psi\rangle=\bigotimes_{i=1}^m\vert\phi_i\rangle$ where $\vert\phi_i\rangle\in\CI^{d_i}$.
\end{definition}

Note that a product vector is not necessarily separable with respect to all the partitions of Hilbert space. For example consider the vector $\vert\psi\rangle=\frac{1}{\sqrt{2}}\vert 001\rangle+\vert 111\rangle$. This is a product vector when we look at it a bipartite state, i.e. in $\CI^4\otimes\CI^2$ it can be written as $\vert\psi\rangle=\frac{1}{\sqrt{2}}(\vert 00\rangle+\vert 11\rangle)\otimes\vert 1\rangle$. However in three-partite system in $\CI^2\otimes\CI^2\otimes\CI^2$ the vector $\vert\psi\rangle$ is no longer a product vector.

\begin{theorem}\textbf{Range Criterion}

If a bipartite density matrix $\rho_{12}$ acting on Hilbert space $\HI=\CI^{d_1}\otimes\CI^{d_2}$, is separable then there exits a set of product vectors $\{e_i\otimes f_i\}$ where $\vert e_i\rangle\in \CI^{d_1}$, $\vert f_i\rangle\in\CI^{d_2}$, where $\vert f_i^*\rangle$ is complex conjugate , such that $\{e_i\otimes f_i\}$ spans the range of $\rho_{12}$ and $\{e_i\otimes f_i^*\}$ spans the range of $\rho_{12}^{\rT_2}$\cite{horo5}.
\end{theorem}

\begin{example}
The following PPT state in $\CI^3\otimes\CI^3$ has been shown ~\cite{horo5} to be entangled:
$$
\varrho_a={1 \over 8a + 1}
\left[ \begin{array}{ccccccccc}
          a &0&0&0&a&0&0&0& a   \\
           0&a&0&0&0&0&0&0&0     \\
           0&0&a&0&0&0&0&0&0     \\
           0&0&0&a&0&0&0&0&0     \\
          a &0&0&0&a&0&0&0& a     \\
           0&0&0&0&0&a&0&0&0     \\
           0&0&0&0&0&0&{1+a \over 2}&0&{\sqrt{1-a^2} \over 2}\\
           0&0&0&0&0&0&0&a&0     \\
          a &0&0&0&a&0&{\sqrt{1-a^2} \over 2}&0&{1+a \over 2}\\
       \end{array}
      \right ]
$$
where $0<a<1$.
\end{example}

Note that the range criterion is neither stronger nor weaker than PPT criterion, as it can detect PPT entangled states, while there are NPT states which do not violate this criterion either.

There are some states which extremely violate this criterion, such states are called Edge states~\cite{{lewe2},{horo6}}.

\begin{definition}\textbf{Edge state}
\label{edge}

An edge state $\delta$ is a PPT entangled state such that, for all product vectors $\vert e\otimes f\rangle$ $(\vert e\rangle\in\CI^{d_1}$ and $\vert f\rangle\in\CI^{d_2})$ and $\epsilon >0$, $\delta-\epsilon\vert e\otimes f\rangle\langle e\otimes f\vert$ is not positive or does not have a positive partial transposition~\cite{{horo5},{horo6},{lewe1}}.
\end{definition}

Obviously the edge states lie between PPT entangled states and NPT states.
\begin{example}
Indeed the first PPT entangled state in $\CI^2\otimes\CI^4$ found in~\cite{horo5} is an example of an edge state:
$$
\sigma_b={1 \over 7b + 1}
\left[ \begin{array}{cccccccc}
           b&0&0&0&0&b&0&0   \\
           0&b&0&0&0&0&b&0     \\
           0&0&b&0&0&0&0&b     \\
           0&0&0&b&0&0&0&0     \\
           0&0&0&0&{1+b \over 2}&0&0&{\sqrt{1-b^2} \over 2} \\
           b&0&0&0&0&b&0&0     \\
           0&b&0&0&0&0&b&0     \\
           0&0&b&0&{\sqrt{1-b^2} \over 2}&0&0&{1+b \over 2}\\
       \end{array}
      \right ]
$$
where $0\leq b\leq 1$.
\end{example}

Using an edge state $\delta$, one can construct a non-decomposable entanglement witness as follows~\cite{lewe1}:

Let us denote the range and kernel of $\delta$ by $R(\delta)$ and $K(\delta)$ respectively.
Let $P$ and $Q$ denote the projector onto $K(\delta)$ and $K(\delta^{\rT_2})$ respectively and define
\begin{eqnarray}
\label{ndew-lew}
W_{\delta}&=&a(P+Q^{\rT_2})\\
\nonumber
 where\quad a&=&\frac{1}{\Tr(P+Q)}.
\end{eqnarray}
Define also
\begin{equation}
\label{ndew-lew1}
\epsilon\equiv\inf_{\vert e\otimes f\rangle}\langle e\otimes f\vert W_{\delta}\vert e\otimes f\rangle.
\end{equation}

\begin{theorem}~\cite{lewe1}
Given and edge PPT entangled state $\delta$, then $W\propto W_{\delta}-\epsilon\1$ is a non-decomposable entanglement witness, where $\epsilon$ and $W_{\delta}$ are defined in~\eqref{ndew-lew} and ~\eqref{ndew-lew1}.
\end{theorem}

\proof

 To show that W is an entanglement witness, we have to show that it satisfies the requirements in~\eqref{EW}. The block positivity of W requires that for any $\vert\psi\rangle\in\CI^{d_1}$ and $\vert\phi\rangle\in\CI^{d_2}$ we have:
\begin{eqnarray}
\nonumber
\langle \psi\otimes\phi\vert W\vert\psi\otimes\phi\rangle\geq&=&\langle \psi\otimes\phi\vert W_{\delta}\vert\psi\otimes\phi\rangle- \epsilon\langle \psi\otimes\phi\vert\1 \vert\psi\otimes\phi\rangle\\
\nonumber
&=&a(\langle \psi\otimes\phi\vert P\vert\psi\otimes\phi\rangle+\langle \psi\otimes\phi\vert Q^{\rT_2}\vert\psi\otimes\phi\rangle)-\epsilon.
\end{eqnarray}
The first two terms cannot vanish simultaneously, as $\delta$ is an edge state: If $\langle \psi\otimes\phi\vert P\vert\psi\otimes\phi\rangle=0$ then $\vert\psi\otimes\phi\rangle\in R(\delta)$, then by definition of an edge state, $\vert\psi\otimes\phi^*\rangle\notin R(\delta^{\rT_2})$, which implies that $\vert\psi\otimes\phi^*\rangle\in K(\delta^{\rT_2})$ and therefore $\langle \psi\otimes\phi^*\vert Q\vert\psi\otimes\phi^*\rangle=1$, and vice versa. Therefore $\langle \psi\otimes\phi\vert W_{\delta}\vert\psi\otimes\phi\rangle\geq 0$ for all $\vert\psi\rangle$ and $\vert\phi\rangle$.

On the other hand $W$ detects $\delta$:
\begin{eqnarray}
\nonumber
\Tr(W\delta)&=&a(\Tr(P\delta)+\Tr(Q^{\rT}_2)\delta))-\epsilon\Tr(\1\delta)\\
\nonumber
&=&-\epsilon.
\end{eqnarray}
The first two terms are zero due to fact that $P\delta=Q\delta^{\rT_2}=0$.
Since $W$ detects a PPT entangled states, by~\eqref{dew-2} it is a non-decomposable entanglement witness.
\qed
\medskip

\section{Unextendible Product Basis}

 Searching for the product states in the range of a density matrix was leading into a new method called Unextendible product basis, mentioned for the first time in~\cite{bennet2}.

\begin{definition}\textbf{Unextendible Product Basis}
\label{UPB}

Consider a multipartite quantum system $\HI=\bigotimes_{i=1}^m\CI^{d_i}$ with m parties.
An orthogonal product basis (PB) is a set S of pure orthogonal product states spanning a
subspace $\HI_s$ of $\HI$. An uncompletable product basis (UCPB) is a PB whose complementary
subspace $\HI^{\bot}$ i.e. the subspace in $\HI$ spanned by vectors that are orthogonal to all
the vectors in $\HI_s$, contains fewer mutually orthogonal product states than its dimension.
An unextendible product basis (UPB) is an uncompletable product basis for which $\HI_s$ contains no product state.
\end{definition}

As an example of product basis which does not span the whole space, one can take $\vert 0\rangle\otimes\vert 0\rangle$ and $\vert 0\rangle\otimes\vert 1\rangle$. As we see these two orthogonal vectors form a PB which spans a two dimensional subspace in $\CI^2\otimes\CI^2$.

\begin{example}
\label{tiles}
 Consider the following example of UPB given in~\cite{vinc1}, known as Tiles. It is a set of five product vectors in $\CI^3\otimes\CI^3$, spanning a 5-dimensional subspace:
\begin{eqnarray}
\label{tiles}
\nonumber \vert\psi_0\rangle&=&\vert 0\rangle\otimes\frac{1}{\sqrt{2}}(\vert 0\rangle-\vert 1\rangle)\\
\nonumber \vert\psi_1\rangle&=&\vert 2\rangle\otimes\frac{1}{\sqrt{2}}(\vert 1\rangle-\vert 2\rangle)\\
          \vert\psi_2\rangle&=&\frac{1}{\sqrt{2}}(\vert 0\rangle-\vert 1\rangle)\otimes\vert 2\rangle\\
\nonumber \vert\psi_3\rangle&=&\frac{1}{\sqrt{2}}(\vert 1\rangle-\vert 2\rangle)\otimes\vert 0\rangle \\
\nonumber \vert\psi_4\rangle&=&\frac{1}{3}(\vert 0\rangle+\vert 1\rangle+\vert 2\rangle)\otimes(\vert 0\rangle+\vert 1\rangle+\vert 2\rangle).
\end{eqnarray}

One sees that as all of these vectors are mutually orthogonal and each three of them span a full 3-dimensional Hilbert space, therefore it is not possible to find another product vector which would be orthogonal to all of them.
\end{example}

For the bipartite Hilbert space $\HI=\CI^{d}\otimes\CI^{d}$, when $d$ is even, one can construct UPB as follows~\cite
{vinc2}; Let's denote the set of $d$ orthonormal vectors by $\vert 0\rangle,...,\vert d-1\rangle$, then the set of vertical tiles are constructed as follows:
\begin{eqnarray}
\label{vert-tiles}
\vert\psi_{mn}\rangle&=&\vert n\rangle\otimes\vert\omega_{m,n+1}\rangle=\vert n\rangle\otimes\sum_{j=0}^{d/2-1}\omega^{jm}\vert j+n+1 \mod d\rangle,\\
\nonumber
m&=&1,...,\frac{d}{2}-1 \quad and\quad n=0,...,d-1,
\end{eqnarray}
where $\omega=e^{i4\pi/d}$. Similarly, the horizontal tiles are defined to be:
\begin{eqnarray}
\label{vert-tiles}
\vert\phi_{mn}\rangle&=&\vert\omega_{m,n}\rangle\otimes\vert n\rangle=\sum_{j=0}^{d/2-1}\omega^{jm}\vert j+n \mod d\rangle\otimes\vert n\rangle,\\
\nonumber
m&=&1,...,\frac{d}{2}-1 \quad and\quad n=0,...,d-1.
\end{eqnarray}

The following theorem illustrates the role of the UPB in the challenge of detecting entanglement.
\begin{theorem}
The state that corresponds to the uniform mixture on the space complementary of a UPB $\{\psi_i: i=1,...,n\}$ in a Hilbert space of total dimension d
\begin{equation}
\label{rho-upb}
\overline{\rho}=\frac{1}{d-n}(\1-\sum_{j=1}^n\vert\psi_j\rangle\langle\psi_j\vert)
\end{equation}
is a PPT entangled state~\cite{bennet2}.
\end{theorem}

\proof

Partial Transposition maps a set of UPB into another set of UPB, while leaving the identity unchanged, this shows that the state in~\eqref{rho-upb} is indeed a PPT state. On the other hand, $\overline{\rho}$ by construction contains no product state which means that it is entangled, all together it is a PPT entangled state.
\qed
\medskip

\section{Realignment}

We first introduce some new notions which will be used in the following.

\begin{definition}\textbf{Reshaping}

Consider an $m\times n$ matrix $A=[a_{ij}]$ where $a_{ij}$ are its entries. By putting the elements of each column one after another, one obtains a vector, which can be associated to this matrix. Conversely, a vector of length $mn$ can be turned into an $m\times n$ matrix~\cite{ch-wu}.
\begin{equation}
\label{reshape}
 A_{m,n} =
 \begin{pmatrix}
  a_{1,1} & a_{1,2} & \cdots & a_{1,n} \\
  a_{2,1} & a_{2,2} & \cdots & a_{2,n} \\
  \vdots  & \vdots  & \ddots & \vdots  \\
  a_{m,1} & a_{m,2} & \cdots & a_{m,n}
 \end{pmatrix} \longleftrightarrow \overrightarrow{a}=(a_{11},\cdots, a_{m1},\cdots, a_{mn})^{\rT}
\end{equation}
\end{definition}

Let $X$ be an $mn\times mn$ block matrix and $\overrightarrow{a_{kl}}=(a_{11},\cdots,a_{n1},\cdots,a_{nn})^{\rT}$ the vector corresponding to the block $kl$ of size n.
The realigned matrix $X^R$ of size $m^2\times n^2$, is defined by:
\[
 X^R=
 \left (\begin{array}{c}
 a_{11}^{\rT}\\
 \vdots\\
 a_{1m}^{\rT}\\
 \vdots\\
 a_{mm}^{\rT}
 \end{array}
 \right )
 \]

\begin{example} As an example let us take $m=n=2$:
\begin{equation}\label{matrix-x}
 X =
 \left (
 \begin{array}{cc|cc}
  x_{1,1} & x_{1,2} & x_{1,3} & x_{1,4} \\
  x_{2,1} & x_{2,2} & x_{2,3} & x_{2,4} \\ \hline
  x_{3,1} & x_{3,2} & x_{3,3} & x_{3,4}  \\
  x_{4,1} & x_{4,2} & x_{4,3} & x_{4,4}
 \end{array}
 \right )
\end{equation}

The vector corresponding to the first block is $\overrightarrow{a_{11}}=(x_{11},x_{21},x_{12},x_{22})^{\rT}$ and the realigned matrix is:

\[
 X^R =
 \begin{pmatrix}
  x_{1,1} & x_{2,1} & x_{1,2} & x_{2,2} \\
  x_{3,1} & x_{4,1} & x_{3,2} & x_{4,2} \\
  x_{1,3} & x_{2,3} & x_{1,4} & x_{2,4}  \\
  x_{3,3} & x_{4,3} & x_{3,4} & x_{4,4}
 \end{pmatrix}
\]
\end{example}

In~\cite{zycz2}, reshaping a matrix is performed by aligning its rows one by one in a vector, instead of its columns:
\[
 A_{m,n} =
 \begin{pmatrix}
  a_{1,1} & a_{1,2} & \cdots & a_{1,n} \\
  a_{2,1} & a_{2,2} & \cdots & a_{2,n} \\
  \vdots  & \vdots  & \ddots & \vdots  \\
  a_{m,1} & a_{m,2} & \cdots & a_{m,n}
 \end{pmatrix} \longleftrightarrow \overrightarrow{a'}=(a_{11},a_{12},\cdots, a_{1n},\cdots, a_{mn})^{\rT}
\]
Now any $mn\times mn$ square matrix can be reshuffled by reshaping each row into a square matrix and putting them block after block into a new matrix. The new reshuffled matrix, $X^{R'}$ obtained from matrix X in~\eqref{matrix-x} is:
\[
 X^{R'} =
 \left (
 \begin{array}{cc|cc}
  x_{1,1} & x_{12} & x_{2,1} & x_{2,2} \\
  x_{1,3} & x_{1,4} & x_{2,3} & x_{2,4} \\ \hline
  x_{3,1} & x_{3,2} & x_{4,1} & x_{4,2}  \\
  x_{3,3} & x_{3,4} & x_{4,3} & x_{4,4}
 \end{array}
 \right )
\]

In general, given a matrix $X\in M_d(\CI)\otimes M_d(\CI)=M_{d^2}(\CI)$, let us consider the orthonormal basis consisting of tensor products $\vert e_m\otimes f_\mu\rangle\in\CI^{d^2}$, where $\{\vert e_m\rangle\}_{m=1}^d$ and $\{\vert f_\mu\rangle\}_{\mu=1}^d$ are orthonormal bases in $\CI^d$. Further, let us denote as in \cite{zycz2} by
\begin{equation}
\label{shuffle}
X_{m\mu\atop n\nu}=\langle e_m\otimes f_\mu\vert\,X\,\vert e_n\otimes f_\nu\rangle\ .
\end{equation}
Then, the matrix elements of reshuffled matrix $X^{R'}$ are given by
\begin{equation}
\label{shuffle1}
X^{R'}_{m\mu\atop n\nu}=X_{\nu\mu\atop nm}\ .
\end{equation}

\begin{remark}
Note that the result of both methods are equivalent up to a permutation. In both cases after reshaping a matrix into a vector, the Hilbert-Schmidt product~\cite{zycz2} will be read as scalar product between two vectors:
$$
\langle A\vert\ B\rangle=\Tr A^{\dagger} B=\langle a'\vert b'\rangle .
$$
\end{remark}

Reshuffling the tensor product of two matrices turns to be very simple. Consider two matrices $X\in M_p(\CI)$ and $Y\in M_q(\CI)$. We denote the vectors assigned to each matrix by $vec(X)$ and $vec(Y)$. Now consider their tensor product:
\begin{eqnarray}
\nonumber
X\otimes Y&=&\begin{pmatrix}
  x_{11} & x_{12} & \cdots & x_{1p} \\
  x_{21} & x_{22} & \cdots & x_{2p} \\
  \vdots  & \vdots  & \ddots & \vdots  \\
  x_{p1} & x_{p2} & \cdots & x_{pp}
 \end{pmatrix}\otimes \begin{pmatrix}
  y_{11} & y_{12} & \cdots & y_{1q} \\
  y_{21} & y_{22} & \cdots & y_{2q} \\
  \vdots  & \vdots  & \ddots & \vdots  \\
  y_{q1} & y_{q2} & \cdots & y_{qq}
 \end{pmatrix}\\
&=&\left ( \begin{array}{c|ccc}
  \begin{array}{ccc}
  x_{11}y_{11}& \cdots & x_{11}y_{1q}\\
  x_{11}y_{21}& \cdots & x_{11}y_{2q}\\
  \vdots   & \ddots & \vdots  \\
  x_{11}y_{q1} & \cdots & x_{11}y_{qq}
  \end{array} & x_{12}Y & \cdots & x_{1p}Y \\
  \hline
  x_{21}Y & x_{22}Y & \cdots & x_{2p}Y \\
  \vdots  & \vdots  & \ddots & \vdots  \\
  x_{p1}Y & x_{p2}Y & \cdots & x_{pp}Y
 \end{array}\right ).
\end{eqnarray}
By the above method the vector corresponding to the first block is:
$$ \overrightarrow{a_1}=(x_{11}y_{11},\cdots,x_{11}y_{1q},\cdots,x_{11}y_{qq})^{\rT}=x_{11}\overrightarrow{Y},$$
where $\overrightarrow{Y}$ is the vector corresponding to the matrix $Y$ $($see~\eqref{reshape}$)$.
Putting all these vectors as rows of a new matrix, we have the reshuffled matrix of this tensor product:
$$
(X\otimes Y)^{R}=\begin{pmatrix}
                x_{11}(\overrightarrow{Y})^{\rT}\\
                \vdots\\
                x_{1p}(\overrightarrow{Y})^{\rT}\\
                \vdots\\
                x_{pp}(\overrightarrow{Y})^{\rT}\\
                \end{pmatrix} = \overrightarrow{X}(\overrightarrow{Y})^{\rT}
$$

We now show that reshuffling a density matrix can be useful to detect its entanglement.

\begin{theorem}\textbf{Reshuffling Criterion}
\label{ReshufTheo}
If a bipartite density matrix $\rho_{12}$ is separable, then the trace norm of its reshuffled matrix will not be increased~\cite{{ch-wu},{rodo1}}, i.e.:
$$
\| \rho_{12}^R\|_{\Tr}\leq\|\rho_{12}\|_{\Tr}=1.
$$
\end{theorem}

\proof

If $\rho_{12}$ is separable, then it can be written as $\rho_{12}=\sum_i p_i \rho_i^1\otimes\rho_i^2$ where $\sum_i p_i=1$. Consider the pure separable state: $\rho_{12}=\rho^1\otimes\rho^2$. As we have seen in the last remark, by reshuffling this state we have: $\rho_{12}^{R}=\overrightarrow{\rho^1}[\overrightarrow{\rho^2}]^{\rT}$.
But in this case, $\rho^1$ and $\rho^2$ are projectors and therefore their norm is equal to the norm of the corresponding vectors. We have:
$$
\|\rho_{12}^{R}\|\leq\|\overrightarrow{\rho^1}[\overrightarrow{\rho^2}]^{\rT}\|\leq\|\overrightarrow{\rho_1}\|\,\|\overrightarrow{\rho_2}\|=\|\rho_1\|\,
\|\rho_2\|=1\ .
$$
By properties of the norm, we can see that:
$$
\|\rho_{12}^{R}\|\leq\sum_i p_i \|(\rho_i^1\otimes\rho_i^2)^{R}\|\leq\sum_i p_i\leq 1.$$
\qed
\medskip

This theorem provides a necessary condition for separability, therefore if one finds that for a given density matrix $\rho$, after reshuffling $\|\rho^R\|>1$, this means that the state is entangled. Note that $\rho^R$ is not Hermitian. It was proved~\cite{{rodo2},{rodo3}} that realignment is independent of Partial Transposition, therefore none of them is neither stronger nor weaker than the other one.

\begin{example}
As an example one can consider the density matrix constructed as in theorem~\eqref{rho-upb}, using the UPB given in example~\eqref{tiles}, called tiles:
$$
\rho=\frac{1}{4}(\1-\sum_{i=0}^4\vert\psi_i\rangle\langle\psi_i\vert).
$$
However we know that this state is entangled but we can also see it by direct calculation of the norm of $\|\rho^R\|=1.32$. Since $\|\rho\|<\|\rho^R\|$ it reveals the entanglement of this state using the reshuffling method.
\end{example}

\newpage

\chapter{$\sigma$-diagonal States}

In the following chapter, we consider bipartite states that are diagonal in the basis generated by the action of tensor products of the form
$\1_{2^n}\otimes\sigma_{\vec{\mu}}$, $\sigma_{\vec{\mu}}=\otimes_{i=1}^n\sigma_{\mu_i}$, on the totally symmetric state $\vert\Psi^{2^n}_+\rangle\in  \CI^{2^n}\otimes\CI^{2^n}$.
We first characterize the structure of positive maps  detecting the entangled ones among them; then, the result will be illustrated by examining some entanglement witnesses for the case  $n=2$. Further, we will show how, for pairs of two qubits, being separable, entangled and bound entangled are state properties related to the geometric patterns of subsets of a $16$ point square lattice.

\section{$\sigma$-diagonal and lattice states}

In the following we shall freely call \textit{entanglement witness} both positive maps $\Lambda$ such that ${\rm id}_d\otimes\Lambda[\rho_{ent}]\ngeq0$ and their Choi matrices $M_\Lambda$.
\medskip

We shall consider a bipartite system consisting of two parties in turn comprising $n$ qubits; the corresponding matrix algebra $M_{2^{2n}}(\CI)$ is linearly spanned by $4^n$ tensor products of the form  $\sigma_{\vec{\mu}}:=\otimes_{i=1}^n\sigma_{\mu_i}=\sigma_{\mu_1}\otimes\sigma_{\mu_2}\otimes\cdots\sigma_{\mu_n}$.

Given the totally symmetric state, already introduced in~\eqref{psi+}:
$$
\vert\Psi_{+}^d\rangle=\frac{1}{d}\sum_{j=1}^d\vert j\rangle\otimes\vert j\rangle
$$
with $d=2^n$, the vectors
\begin{equation}
\label{ONV}
\vert\Psi_{\vec{\mu}}\rangle:=\1_{2^n}\otimes\sigma_{\vec{\mu}}\vert\Psi^{2^n}_+\rangle\in \CI^{2^n}\otimes\CI^{2^n} \ ,
\end{equation}
form orthogonal projectors
\begin{equation}
\label{ONB16}
P_{\vec{\mu}}:=\vert\Psi_{\vec{\mu}}\rangle\langle\Psi_{\vec{\mu}}\vert
=(\1_{2^n}\otimes\sigma_{\vec{\mu}})\vert\Psi^{2^n}_{+}\rangle\langle\Psi^{2^n}_{+}\vert(\1_{2^n}\otimes\sigma_{\vec{\mu}})\ .
\end{equation}
Orthonormality follows since
$$
\langle\Psi_{\vec{\nu}}\vert\Psi_{\vec{\mu}}\rangle=\langle\Psi^{2^n}_+\vert\1_{2^n}\otimes\sigma_{\vec{\mu}}\sigma_{\vec{\mu}}\vert\Psi^{2^n}_+\rangle=
\frac{1}{2^n}{\Tr}\left(\sigma_{\vec{\nu}}\sigma_{\vec{\mu}}\right)
=\frac{1}{2^n}\prod_{i=1}^n{\Tr}\left(\sigma_{\nu_i}\sigma_{\mu_i}\right)=\prod_{i=1}^n\delta_{\nu_i\mu_i}\ .
$$

\begin{definition}
\label{s-diagstates}
The class of bipartite states we shall study will consist of states of the form
\begin{equation}
\label{diagstates}
\rho=\sum_{\vec{\mu}}\  r_{\vec{\mu}}\ P_{\vec{\mu}}\ ,\qquad 0\leq r_{\vec{\mu}}\leq 1\ ,\quad\sum_{\vec{\mu}}\ r_{\vec{\mu}}=1\ ,
\end{equation}
that is diagonal with respect to the chosen orthonormal basis in $\CI^{2^n}\otimes\CI^{2^n}$: we shall call them $\sigma$-diagonal states.
\end{definition}

A particular sub-class of states of two pairs of two qubits, $n=2$ and $\sigma_{\vec{\mu}}=\sigma_\alpha\otimes\sigma_\beta=\sigma_{\alpha\beta}\in M_4(\CI)$, are considered in~\cite{benatti2,benatti3,benatti4} and called lattice states. In this case the orthonormal basis vectors and their corresponding projectors are the following:
\begin{eqnarray}
\nonumber
\vert\Psi_{\alpha\beta}\rangle&=&\1_4\otimes\sigma_{\alpha\beta}\vert\Psi^+\rangle\in\CI^{16},\\
\nonumber
P_{\alpha\beta}&\equiv&\vert\Psi_{\alpha\beta}\rangle\langle\Psi_{\alpha\beta}\vert=
(\1_4\otimes\sigma_{\alpha\beta})P_4^+(\1_4\otimes\sigma_{\alpha\beta}),
\end{eqnarray}
where $P_{+}^4$ is defined in~\eqref{P+}.

For them partial results concerning their entanglement properties have been obtained.
We shall tackle them again in the following: the class of lattice-states is defined as follows.

\begin{definition}\textbf{Lattice States}
\label{def:lattstate}

Taking a subset $I\subseteq L_{16}$ of cardinality $N_{I}$, then the corresponding lattice state $\rho_{I}$ is defined by:
\begin{equation}
\label{lattstate}
\rho_{I}=\frac{1}{N_{I}}\sum_{\alpha,\beta\in I }P_{\alpha\beta}\ .
\end{equation}
Let $L_{16}$ denote the set of pairs $(\alpha,\beta)$, where $\alpha$ and $\beta$ run from $0$ to $3$: it corresponds to a $4\times 4$ square lattice, whereas the subsets
\begin{eqnarray}
\nonumber
C_{\alpha}&:=&\{(\alpha,\beta)\in L_{16} : \beta=0,1,2,3\}\\
\label{ColRows}
 R_{\beta}&:=&\{(\alpha,\beta)\in L_{16} : \alpha=0,1,2,3\}
\end{eqnarray}
correspond to the columns and rows of the lattice, respectively.
\end{definition}

If compared with those in~\eqref{diagstates}, the lattice  states are uniformly distributed over, and thus completely characterized by chosen subsets of $L_{16}$.
\begin{example}
\label{ex6.1}
As an example consider the lattice states associated with the following subsets:
\begin{eqnarray}
\nonumber
\rho_8&=&\frac{1}{8}(P_{00}+P_{02}+P_{11}+P_{13}+P_{20}+P_{21}+P_{23}+P_{32})\\
&N_I=8&:\qquad
\begin{array}{c|c|c|c|c}
               3 & \quad\! & \times & \times &\quad\!   \\
               \hline
               2 & \times & \quad\! &  \quad\! &\times  \\
               \hline
               1 & \quad\! & \times& \times  &\quad\!   \\
               \hline
               0 & \times & \quad\! & \times  &   \\
               \hline
                 & 0 & 1 & 2 & 3
\end{array}\\
\nonumber\\
\nonumber
\rho_4&=&\frac{1}{4}(P_{00}+P_{11}+P_{22}+P_{33})\\
 &N_I=4&:\qquad
\begin{array}{c|c|c|c|c}
               3 & \quad\! & \quad\! & \quad\! &\times   \\
               \hline
               2 & \quad\! & \quad\! &\times &\quad\!  \\
               \hline
               1 & \quad & \times & \quad\!  &\quad\!   \\
               \hline
               0 & \times & \quad\! & \quad\!  &\quad\!   \\
               \hline
                 & 0 & 1 & 2 & 3
\end{array}\qquad .
\end{eqnarray}
\end{example}

In the following, we shall be interested in investigating the entanglement properties of such states. As an appetizer, we consider whether the von Neumann entropy of any $\rho_I$ can be smaller than that of its reuced density matrices. Indeed, we have seen that, if so, $\rho_I$ would be entangled.
By tracing over the first or the second party we get as reduced density matrices
$$
\rho_I^{(1),(2)}=\Tr_{2,1}\rho_I=\frac{\1_4}{4}
$$
with maximal von Neumann entropy $S(\rho_I^{(1),(2)})=2\log 2$. On the other, end each lattice state is already spectralized with equal non-zero eigenvalues $1/N_I$; therefore, $S(\rho_I)=\log N_I\geq 2\log 2$ whenever $N_I\geq 4$.

Furthermore, if we consider the Reshuffling Criterion in Theorem~\ref{ReshufTheo}, using the relations \eqref{shuffle} and~\eqref{shuffle1}, we obtain as reshuffled lattice states
$$
(\rho^{R'}_I)_{m\mu\atop n\nu}=(\rho^{R'}_I)_{\nu\mu\atop nm}=\frac{1}{4N_I}\sum_{(\alpha,\beta)\in I}\,
\langle \mu\vert\sigma_{\alpha\beta}\vert\nu\rangle\, \langle n\vert\sigma_{\alpha\beta}\vert m\rangle\ ,
$$
whence
$$
\rho_I^{R'}=\frac{1}{4N_I}\sum_{(\alpha,\beta)\in I}\,\sigma^\rT_{\alpha\beta}\otimes\sigma_{\alpha\beta}\ ,
$$
where we chose the ortonormal basis vectors in~\eqref{shuffle} to be the ones used in writing the Lattice States, namely
$\vert e_i\rangle=\vert f_i\rangle=\vert i\rangle$.
Thus, using the properties of the trace norm, one gets
$$
\|\rho_I^{R'}\|_{\Tr}\leq\frac{1}{4N_I}\sum_{(\alpha,\beta)\in I}\,\|\sigma_{\alpha\beta}\|_{\Tr}^2=\frac{4}{N_I}\ ,
$$
whence for $N_I\geq 4$ no entangled lattice state can be detected by the chosen criterion.

We now proceed to examine the lattice states from the point of view of partial transposition.
Positivity under partial transposition (PPT-ness) of lattice states $\rho_I$  is completely characterized by the geometry of $I$~\cite{benatti2}.

\begin{proposition}
\label{PPT}
A necessary and sufficient condition for a lattice state $\rho_I$ to be PPT is that for every
$(\alpha,\beta)\in L_{16}$ the number of points on $C_\alpha$ and $R_\beta$  belonging to
$I$ and different from $(\alpha, \beta)$ be not greater than $N_I/2$. In terms of the characteristic functions $\chi_I(\alpha,\beta) = 1$ if
$(\alpha, \beta)\in I$, $= 0$ otherwise, a lattice state $\rho_I$ is PPT if and only if for all $(\alpha, \beta)\in L_{16}$:
$$
\sum_{0=\delta\neq\beta}^3\chi_I(\alpha, \delta) + \sum_{0=\delta\neq\alpha}^3\chi_I({\delta,\beta})\leq\frac{N_I}{2}\ .
$$
\end{proposition}
\medskip

\begin{example}
\label{ex6.2}
Consider the following lattice states:
\begin{eqnarray}
 &N_I=5&:\qquad
\begin{array}{c|c|c|c|c}
               3 & \quad\! & \quad\! & \times &\quad\!   \\
               \hline
               2 & \times & \quad\! &\quad\! &\times  \\
               \hline
               1 & \quad & \quad\! & \times  &\quad\!   \\
               \hline
               0 & \quad\! & \quad\! & \quad\!  &\times   \\
               \hline
                 & 0 & 1 & 2 & 3
\end{array}\qquad\\
\nonumber\\
 &N_I=4&:\qquad
\begin{array}{c|c|c|c|c}
               3 & \quad\! & \quad\! & \times &\quad\!   \\
               \hline
               2 & \quad\! & \times &\quad\! &\quad\!  \\
               \hline
               1 & \quad & \quad\! & \times  &\quad\!   \\
               \hline
               0 & \times & \quad\! & \quad\!  &\quad\!   \\
               \hline
                 & 0 & 1 & 2 & 3
\end{array}\qquad
\end{eqnarray}
These states do not remain positive under partial transposition and are therefore entangled: in the first state $\rho_5$, the row and column passing through the point $(2,2)\notin I$ contains $4>5/2$ points in $I$, while in the second one $3>4/2=2$ points.
\end{example}

\begin{example}
\label{ex6.3}
By the same criterion, the following two states are instead PPT,

\begin{eqnarray}
N_I=6:\qquad
\begin{array}{c|c|c|c|c}
               3 & \quad\! & \quad\! & \times &\times   \\
               \hline
               2 & \times & \quad\! &  \quad\! &\times  \\
               \hline
               1 & \quad\! & \times & \quad\!  &\times   \\
               \hline
               0 & \quad\! & \quad\! & \quad\!  &   \\
               \hline
                 & 0 & 1 & 2 & 3
\end{array}
\qquad N_I=8:\qquad
\begin{array}{c|c|c|c|c}
               3 & \quad\! & \times & \times &\times   \\
               \hline
               2 & \times & \quad\! &\times &\times  \\
               \hline
               1 & \quad\! & \quad\! & \times  &\times   \\
               \hline
               0 & \quad\! & \quad\! & \quad\!  &\quad\!   \\
               \hline
                 & 0 & 1 & 2 & 3
\end{array}\qquad .
\end{eqnarray}
\end{example}
They need not be separable as in lower dimension; indeed, a sufficient criterion devised in~\cite{benatti1} show them to be entangled.

\begin{proposition}
\label{PPT2}
A sufficient condition for a PPT lattice state $\rho_I$ to be entangled is that there exists at
least a pair $(\alpha, \beta)\in L_{16}$ not belonging to $I$ such that only one point on $C_{\alpha}$ and $R_\beta$ belongs to $I$.
Equivalently, $\rho_I$ is entangled if there exists a pair $(\alpha, \beta)\in L_{16}$, $  (\alpha, \beta) \notin I$, such that
$$
\sum_{0=\delta\neq\beta}^3\chi_I(\alpha, \delta) + \sum_{0=\delta\neq\alpha}\chi_I(\delta, \beta) = 1 \quad .
$$
\end{proposition}
\medskip

In both patterns of the states in Example~\eqref{ex6.3}, it is the point $(0,0)\notin I$ which satisfies the sufficient criterion.
\begin{example}
\label{ex6.4}
Unfortunately, this criterion fails in the case of the PPT lattice state characterized by the following subset
\begin{eqnarray}
N_I=10:\qquad
\begin{array}{c|c|c|c|c}
               3 & \times & \quad & \quad\! &\times  \\
               \hline
               2 & \quad\! & \times &\times &\quad\!  \\
               \hline
               1 & \times & \times & \quad\!  &\times   \\
               \hline
               0 & \times & \times & \quad\!  &\times   \\
               \hline
                 & 0 & 1 & 2 & 3
\end{array}\qquad .
\end{eqnarray}
Indeed, the only candidate point to fulfill the criterion in Proposition~\eqref{PPT2} is $(2,2)$; however, it belongs to $I$.
\end{example}
Luckily, a refined criterion~\cite{benatti4} based on~\cite{breuer} shows it to be entangled.

\begin{proposition}
\label{PPT3}
A $PPT$ lattice state $\rho_I$
is entangled  if the quantity
$$
k_I^{\mu\nu}=
\sum_{\alpha\neq\nu\oplus2}\chi_I(\alpha, \nu \oplus 2) + \sum_{\beta\neq\mu\oplus2}\chi_I(\mu \oplus 2,\beta)\ ,
$$
where $\oplus$ denotes summation modulo $4$,
is such that $k_I^{\mu\nu}= 1$ for a column $C_{\mu\oplus 2}$ and a row $R_{\nu\oplus 2}$, independently of whether $(\mu\oplus 2, \nu\oplus 2)$ belongs to $I$ or not.
\end{proposition}
\medskip

\begin{example}
\label{ex6.5}
Thus, the state in~\eqref{ex6.4} fulfills the sufficient condition $k_I^{00}=1$.

Another example of a lattice state detected by the last proposition is the following:
$$
N_I=11:\qquad
\begin{array}{c|c|c|c|c}
               3 & \times & \times & \times &\quad\!  \\
               \hline
               2 & \times & \times &\times &\times  \\
               \hline
               1 & \times & \times & \times  &\quad\!   \\
               \hline
               0 & \quad\! & \quad\! & \quad\!  &\times   \\
               \hline
                 & 0 & 1 & 2 & 3
\end{array}
$$
\end{example}

As it is apparent from these examples the various criteria for entanglement are closely related to the structure of the subsets that define the lattice states;
in the following we will try to clarify this correspondence.

\section{Entanglement detection for $\sigma$-diagonal states}

In this section, we will show that positive maps that be entanglement witnesses for states of the form~\eqref{diagstates} can be sought in a particular subclass of them.

Any linear map on $\Lambda: M_d(\CI)\mapsto M_d(\CI)$ can be written as~\cite{benatti1}:
$$
M_d(\CI)\ni X\mapsto \Lambda[X]=\sum^{d^{2}-1}_{i,j=0}\lambda_{ij}\ G_i\,X\,G^{\dagger}_{j}\ ,
$$
where the matrices $G_{i}\in M_d(\CI)$ form an orthonormal basis with respect to Hilbert-Schmidt scalar product, namely ${\Tr}(G_i^\dag\,G_j)=\delta_{ij}$ and the coefficients $\lambda_{ij}$ are complex numbers.

In the present  case, the normalized  tensor products $\displaystyle\frac{\sigma_{\vec{\mu}}}{\sqrt{2^n}}$ constitute such a basis in $M_{2^n}(\CI)$, whence linear maps $\Lambda:M_{2^n}(\CI)\mapsto M_{2^n}(\CI)$ can be expressed as
\begin{equation}
\label{genexp}
M_{2^n}(\CI)\ni X\mapsto\Lambda[X]=\sum_{\vec{\mu},\vec{\nu}}\,\lambda_{\vec{\mu}\vec{\nu}}\ S_{\vec{\mu}\vec{\nu}}[X]\ ,\qquad
S_{\vec{\mu}\vec{\nu}}[X]=\sigma_{\vec{\mu}}\,X\,\sigma_{\vec{\nu}}\ .
\end{equation}

The next one is a simple observation based on~\eqref{ONV} and~\eqref{ONB16}.

\begin{lemma}
\label{lemma1}
A $\sigma$-diagonal state $\rho=\sum_{\vec{\mu}}\ r_{\vec{\mu}}\ P_{\vec{\mu}}$ is entangled if and only if there exists a positive map $\Lambda$ as in ~(\ref{genexp}) such that
$\sum_{\vec{\mu}}\ \lambda_{\vec{\mu}\vec{\mu}}\, r_{\vec{\mu}}\,<\,0$.
\end{lemma}

\noindent
\proof
According to Horodecki Criterion in~\eqref{horo-cri}, the state $\rho=\sum_{\vec{\mu}}\ r_{\vec{\mu}}\ P_{\vec{\mu}}$ is entangled if and only if
$$
{\Tr}\left(\id_{2^n}\otimes\Lambda[P^{2^n}_+]\,\rho\right)<\, 0\ ,
$$
for some positive map $\Lambda: M_{2^n}(\CI)\mapsto M_{2^n}(\CI)$.

Using the orthogonality of the vectors $\vert\Psi_{\vec{\mu}}\rangle$ in~\eqref{ONV}, $\rho$ is entangled if and only if there is a positive map $\Lambda$ such that:
$$
{\Tr}\left(\id_{2^n}\otimes\Lambda[P^{2^n}_+]\,\rho\right)=\sum_{\vec{\mu},\vec{\nu}}\ \lambda_{\vec{\mu}\vec{\nu}}\ \langle\Psi_{\vec{\nu}}\vert\rho\vert\Psi_{\vec{\mu}}\rangle=
\sum_{\vec{\mu}}\ \lambda_{\vec{\mu}\vec{\mu}}\, r_{\vec{\mu}}\,<\, 0\ .
$$
\qed
\medskip

The lemma indicates that the class of diagonal positive maps of the form
$\Lambda=\sum_{\vec{\mu}}\,\lambda_{\vec{\mu}}\ S_{\vec{\mu}\vec{\mu}}$
might suffice to witness the entanglement of states of the form $\rho=\sum_{\vec{\mu}}\ r_{\vec{\mu}}\ P_{\vec{\mu}}$.
What is needed to show that it is indeed so is the following result.

\begin{lemma}
\label{lemma2}
Given a positive map of the form $\Lambda=\sum_{\vec{\mu},\vec{\nu}}\,\lambda_{\vec{\mu}\vec{\nu}}\ S_{\vec{\mu}\vec{\nu}}$ , the diagonal map
$\Lambda_{diag}=\sum_{\vec{\mu}}\,\lambda_{\vec{\mu}\vec{\mu}}\ S_{\vec{\mu}\vec{\mu}}$ is also positive.
\end{lemma}

\proof
From Choi's theorem~\eqref{choi-block}, we know that a given map is positive if and only if its Choi matrix is block positive. Considering the map $\Lambda=\sum_{\vec{\mu},\vec{\nu}}\,\lambda_{\vec{\mu}\vec{\nu}}\ S_{\vec{\mu}\vec{\nu}}$, the positivity assumption is equivalent to
\begin{equation}
\label{posconst1}
\sum_{\vec{\mu},\vec{\nu}}\lambda_{\vec{\mu}\vec{\nu}}\,\langle\varphi\vert
\sigma_{\vec{\mu}}\vert\psi\rangle\langle\psi\vert\sigma_{\vec{\nu}}\vert\varphi\rangle\,\geq\, 0\qquad \forall\  \vert\psi\rangle,\vert\varphi\rangle\in \CI^{2^n}\quad .
\end{equation}
Given a pair $\vert\psi\rangle,\vert\varphi\rangle\in \CI^{2^n}$, consider another pair
$\vert\psi_{\delta_i}\rangle=\sigma_{\delta_i}\vert\psi\rangle$, $\vert\varphi_{\delta_i}\rangle=\sigma_{\delta_i}\vert\varphi\rangle$,
where $\sigma_{\delta_i}$ denotes the tensor product $\sigma_{\vec{\mu}}$ where $\mu_j=0$ for $j\neq i$ and $\mu_i=\delta_i\neq 0$.
Inserting the new pair into~\eqref{posconst1}, we get
\begin{eqnarray}
\nonumber
&&
\sum_{\vec{\mu},\vec{\nu}}\ \lambda_{\vec{\mu}\vec{\nu}}\ \langle\varphi\vert\bigotimes_{j=1}^{i-1}\sigma_{\mu_j}\otimes
\,\Big(\sigma_{\delta_i}\sigma_{\mu_i}\sigma_{\delta_i}\Big)\,\bigotimes_{j=i+1}^{n}\sigma_{\mu_j}\vert\psi\rangle\,\times\\
\label{posconst2}
&&\hskip 2cm\times\,
\langle\psi\vert
\bigotimes_{j=1}^{i-1}\sigma_{\nu_j}\otimes\,\Big(\sigma_{\delta_i}\sigma_{\nu_i}\sigma_{\delta_i}\Big)\,
\bigotimes_{j=i+1}^{n}\sigma_{\nu_j}\vert\varphi\rangle\,\geq\, 0\ .
\end{eqnarray}
Consider $\mu_i\neq\nu_i$; because of the Pauli algebraic relations, one can always choose $\sigma_{\delta_i}$ in such a way that
$$
\sigma_{\delta_i}\sigma_{\mu_i}\sigma_{\delta_i}=\sigma_{\mu_i}\qquad\hbox{and}\qquad
\sigma_{\delta_i}\sigma_{\nu_i}\sigma_{\delta_i}=-\sigma_{\nu_i}\ ,
$$
whence all the terms in~\eqref{posconst2} corresponding to the chosen pair of indices $(\mu_i,\nu_i)$ contribute with
$$
-\,\lambda_{\vec{\mu}\vec{\nu}}\,\langle\varphi\vert\bigotimes_{j=1}^{i-1}\sigma_{\mu_j}\otimes
\,\sigma_{\mu_i}\,\bigotimes_{j=i+1}^{n}\sigma_{\mu_j}\vert\psi\rangle\langle\psi\vert
\bigotimes_{j=1}^{i-1}\sigma_{\nu_j}\otimes\,\sigma_{\nu_i}\,
\bigotimes_{j=i+1}^{n}\sigma_{\nu_j}\vert\varphi\rangle\ .
$$
Then, adding inequalities~\eqref{posconst1} and~\eqref{posconst2} yields
$$
\sum_{\vec{\mu},\vec{\nu}; \mu_i=\nu_i}\ \lambda_{\vec{\mu}\vec{\nu}}\ \langle\varphi\vert
\sigma_{\vec{\mu}}\vert\psi\rangle\langle\psi\vert\sigma_{\vec{\nu}}\vert\varphi\rangle\,\geq\,0\quad .
$$
By applying the very same argument for pairs $(\mu_i,\nu_i)$ with varying index $i$, one cancels all contributions from $\mu_i\neq \nu_i$ and remains with
\begin{equation}
\label{posconst3}
\sum_{\vec{\mu}}\ \lambda_{\vec{\mu}\vec{\mu}}\ \langle\varphi\vert
\sigma_{\vec{\mu}}\vert\psi\rangle\langle\psi\vert\sigma_{\vec{\mu}}\vert\varphi\rangle\, \geq\, 0\qquad\forall\ \vert\psi\rangle,\vert\varphi\rangle\in \CI^{2^n}\quad .
\end{equation}
This, by Choi's theorem~\eqref{choi-block} amounts to the positivity of the diagonalized map $\Lambda_{diag}=\sum_{\vec{\mu}}\lambda_{\vec{\mu}\vec{\mu}}\,S_{\vec{\mu}\vec{\mu}}$.
\qed
\medskip

The previous result allows us to conclude with

\begin{proposition}
\label{diagprop}
Entangled $\rho=\sum_{\vec{\mu}}\ r_{\vec{\mu}}\, P_{\vec{\mu}}$ can be witnessed by diagonal positive maps
$\Lambda=\sum_{\vec{\mu}}\,\lambda_{\vec{\mu}\vec{\mu}}\ S_{\vec{\mu}\vec{\mu}}$.
\end{proposition}
\medskip

\proof
By the previous lemma, diagonalizing a positive map
$$
\Lambda=\sum_{\vec{\mu},\vec{\nu}}\,\lambda_{\vec{\mu}\vec{\nu}}\ S_{\vec{\mu}\vec{\nu}}
$$
always yields
a positive map $\Lambda_{diag}=\sum_{\vec{\mu}}\,\lambda_{\vec{\mu}\vec{\mu}}\ S_{\vec{\mu}\vec{\mu}}$. Then, from Lemma~\eqref{lemma1} it follows that either the entanglement of $\rho$ is witnessed by an already diagonal map or, if by a non-diagonal one, also  by
the map obtained by diagonalizing the latter.
\qed
\medskip

\begin{remark}
\label{rem3}
From the above Proposition we only get that entangled states of the form~\eqref{diagstates} can be witnessed by diagonal maps $\Lambda_{diag}=\sum_{\vec{\mu}}\,\lambda_{\vec{\mu}}\ S_{\vec{\mu}\vec{\mu}}$; the main problem is of course to characterize the coefficients $\lambda_{\vec{\mu}}$ in such a way that~\eqref{posconst3} be satisfied and thus $\Lambda_{diag}$ be positive.
\end{remark}

Expression~\eqref{Tr4} for Trace map in $M_4(\CI)$ can be extended to the trace operation on $M_{2^n}(\CI)$,
$$
{\Tr}=\frac{1}{2^n}\sum_{\vec{\mu}}\ S_{\vec{\mu}\vec{\mu}}\ .
$$
In the second part of chapter 4, we have seen that positive maps can be related to the complete positive maps. This relation was given in Theorem~\eqref{stormer}, which allows us to write a positive map as:
$$
\Lambda=\mu(\rT-\Lambda_{cp}).
$$
Therefore diagonal positive maps $\Lambda_{diag}$ are also related to completely positive diagonal maps
$$
\Lambda_{diag}^{CP}=\sum_{\vec{\mu}}\ \lambda_{\vec{\mu}}\,S_{\vec{\mu}\vec{\mu}}\ ,\qquad \lambda_{\vec{\mu}}\geq 0\ ,
$$
by setting
\begin{equation}
\label{Stormer3}
\Lambda_{diag}=\mu\sum_{\vec{\mu}}\Big(\frac{1}{2^n}\,-\,\lambda_{\vec{\mu}}\Big)\,S_{\vec{\mu}\vec{\mu}}\ ,\quad\mu>0\ ,
\end{equation}
and asking for block positivity of its Choi matrix, which is
$$
\langle\varphi\vert\Lambda_{diag}[\vert\psi\rangle\langle\psi\vert]\vert\varphi\rangle\geq0,\quad \forall\vert\varphi\rangle,\vert\psi\rangle\in\CI^{2^n},
$$
\begin{equation}
\label{Stormer4}
\langle\varphi\vert\Big({\Tr}-\Lambda^{CP}_{diag}\Big)[\vert\psi\rangle\langle\psi\vert]\vert\varphi\rangle=
1\,-\,\sum_{\vec{\mu}}\ \lambda_{\vec{\mu}}\,\left|\langle\varphi\vert\sigma_{\vec{\mu}}\vert\psi\rangle\right|^2\,\geq \,0\qquad\forall\ \vert\varphi\rangle,\vert\psi\rangle\in\CI^{2^n}\quad .
\end{equation}

Using these observations, we get necessary and sufficient conditions for the separability of $\sigma$-diagonal states.

\begin{proposition}
\label{propdiagst}
A  $\sigma$-diagonal  state $\rho=\sum_{\vec{\mu}}\ r_{\vec{\mu}}\,P_{\vec{\mu}}$ is separable if and only if for all sets of $4^n$ positive real numbers $\lambda_{\vec{\mu}}\geq 0$,
such that
\begin{equation}
\label{crit1}
\sum_{\vec{\mu}}\ \lambda_{\vec{\mu}}\,\left|\langle\varphi\vert\sigma_{\vec{\mu}}\vert\psi\rangle\right|^2\,\leq1\, \qquad\forall\  \vert\psi\rangle,\vert\varphi\rangle\in\CI^{2^n}\ ,
\end{equation}
it holds that
\begin{equation}
\sum_{\vec{\mu}}\ \lambda_{\vec{\mu}}\,r_{\vec{\mu}}\,\leq\,\frac{1}{2^n}\quad  .
\label{crit2}
\end{equation}
Otherwise, if a set of $4^n$ positive real numbers $\lambda_{\vec{\mu}}\geq 0$ can be found that satisfy~\eqref{crit1} and such that
\begin{equation}
\sum_{\vec{\mu}}\ \lambda_{\vec{\mu}}\,r_{\vec{\mu}}\,>\,\frac{1}{2^n}\quad  ,
\label{crit3}
\end{equation}
then the $\sigma$-diagonal state $\rho=\sum_{\vec{\mu}}\ r_{\vec{\mu}}\,P_{\vec{\mu}}$ is entangled.
\end{proposition}

Before tackling the case of lattice states, that is of $\sigma$-diagonal states with $n=2$, as a simple application, we consider the case of only two qubits, $n=1$, for which we know PPT-ness to coincide with separability.

\begin{example}
\label{ex6.6}
In the case $n=2$, $\sigma$-diagonal states have the form
$$
\rho=\sum_{\mu=0}^3r_\mu P_\mu\ ,\quad r_\mu\geq0\ ,\quad \sum_{\mu=0}^3r_\mu=1
$$
and
the $P_{\mu}$'s project onto the Bell states, already introduced in~\eqref{bell-states}:
\begin{eqnarray}
\nonumber
&&
\vert\Psi_0\rangle=\vert\Psi_+^2\rangle=\frac{\vert00\rangle+\vert11\rangle}{\sqrt{2}}\ ,\qquad
\vert\Psi_1\rangle=\frac{\vert01\rangle+\vert10\rangle}{\sqrt{2}}\\
\label{Bellstates}
&&
\vert\Psi_2\rangle=\frac{\vert00\rangle-\vert11\rangle}{\sqrt{2}}\ ,\qquad\qquad
\vert\Psi_3\rangle=\frac{\vert01\rangle-\vert10\rangle}{\sqrt{2}}\ .
\end{eqnarray}
Under transposition $T[\sigma_\mu]=\varepsilon_\mu\,\sigma_\mu$, $\varepsilon_\mu=(1,1,-1,1)$ and under partial transposition the projection
$P_0$ goes into the flip operator $V\vert\psi\otimes\varphi\rangle=\vert\varphi\otimes\psi\rangle$:
$$
\id\otimes T[P_0]=\frac{1}{2}\,V=\frac{1}{2}\sum_{\mu=0}^3\ v_\mu\, P_\mu\ ,\qquad v_\mu=(1,1,1,-1)\ .
$$
Therefore, the action of partial transposition on a $\sigma$-diagonal state yields
\begin{eqnarray}
\nonumber
\id\otimes T[\rho]&=&\frac{1}{2}\sum_{\mu=0}^3r_\mu\,\1\otimes\sigma_\mu\,V\,\1\otimes\sigma_\mu\\
\nonumber
&=&\frac{1}{2}\sum_{\mu,\nu=0}^3r_\mu\,v_\nu\,\1\otimes\sigma_\mu\sigma_\nu\,P_0\,\1\otimes\sigma_\nu\sigma_\mu\\
\nonumber
&=&\frac{1}{2}\sum_{\mu,\nu=0}^3 \,v_\nu\,r_\mu\, P_{[\mu,\nu]}\\
\nonumber
&=&
\frac{1}{2}\sum_{\alpha=0}^3 \Big(\sum_{\nu=0}^3\,v_\nu\,r_{[\alpha,\nu]}\Big)\, P_\alpha\\
\nonumber
&=&\frac{1}{2}\sum_{\alpha=0}^3 \Big(r_{[\alpha,0]}+r_{[\alpha,1]}+r_{[\alpha,2]}-r_{[\alpha,3]}\Big)\, P_\alpha\\
\label{PartTransp}
&=&\frac{1}{2}\sum_{\alpha=0}^3 \Big(1-2\,r_{[\alpha,3]}\Big)\, P_\alpha\ .
\end{eqnarray}

\begin{definition}
\label{Lmap}
In the above expression, the following construction has been employed: given $(\alpha,\mu)$, $\alpha,\mu=0,1,2,3$, $[\alpha,\mu]$ is the unique index from $0$ to $3$ such that $\sigma_\alpha\sigma_\mu=\eta^{[\alpha,\mu]}_{\alpha\mu}\,\sigma_{[\alpha,\mu]}$, where
$\eta^{[\alpha,\mu]}_{\alpha\mu}$ is a phase $\pm 1$ or $\pm i$, they can be considered as non-zero elements of the following matrices:
\begin{eqnarray}
\nonumber
\eta^1 =
 \begin{pmatrix}
  0 & 1 & 0 & 0 \\
  1 & 0 & 0 & 0 \\
  0  & 0   & 0 & i   \\
  0 & 0 & -i & 0
 \end{pmatrix},\qquad
 \eta^2 =
 \begin{pmatrix}
  0 & 0 & 1& 0 \\
  0 & 0 & 0 & -i \\
  1  & 0   & 0 & 0   \\
  0 & i & 0 & 0
 \end{pmatrix}\\
 \nonumber
 \eta^3 =
 \begin{pmatrix}
  0 & 0 & 0 & 1 \\
  0 & 0 & i & 0 \\
  0  & -i   & 0 &  \\
  1 & 0 & 0 & 0
 \end{pmatrix},\qquad
 \eta^0 =
 \begin{pmatrix}
  1 & 0 & 0 & 0 \\
  0 & 1 & 0 & 0 \\
  0  & 0 & 1 & 0   \\
  0 & 0 & 0 & 1
 \end{pmatrix}.
\end{eqnarray}
Because of the Pauli algebraic relations, the symbol $[\cdot,\cdot]$ enjoys the following properties that can be used to derive~\eqref{PartTransp}:
$$
[\alpha,\mu]=[\mu,\alpha]\ ,\qquad [\alpha,\mu]=\gamma\Rightarrow [\alpha,\gamma]=\mu\Rightarrow \ [\mu,\gamma]=\alpha\ .
$$
Thus, $[\alpha,\cdot]$ is a one to one map from the set $(0,1,2,3)$ onto itself.
\end{definition}

Since positivity under partial transposition identifies all separable states of two qubits, a $\sigma$-diagonal state is separable if and only if
$\displaystyle r_\mu\leq\frac{1}{2}$ for all $\mu=0,1,2,3$.

On the other hand we have seen that all positive maps acting on $\CI^2\otimes\CI^2$ are decomposable, therefore using~\eqref{stormer},~\eqref{transp2} and~\eqref{Tr2},
we can write any diagonal positive map $\Lambda_{diag}:M_2(\CI)\mapsto M_2(\CI)$ as
\begin{eqnarray*}
\Lambda_{diag}&=&\sum_{\alpha=0}^3\lambda^{(1)}_\alpha\,S_\alpha\,+\,\sum_{\alpha=0}^3\lambda^{(2)}_\alpha\,S_\alpha\circ T\\
&=&\sum_{\alpha=0}^3\lambda^{(1)}_\alpha\,S_\alpha\,+\,\frac{1}{2}\sum_{\alpha,\beta=0}^3\varepsilon_\beta\lambda^{(2)}_\alpha\,S_\alpha\circ S_\beta\\
&=&\sum_{\gamma=0}^3\Big(\lambda^{(1)}_\gamma\,+\,\frac{1}{2}\sum_{\beta=0}^3\varepsilon_\beta\lambda_{[\beta,\gamma]}\Big)\,S_\gamma
\\
&=&\mu\,\sum_{\gamma=0}^3\ \Big(\frac{1}{2}-\lambda_\gamma\Big)\,S_\gamma\ ,
\end{eqnarray*}
where $\lambda^{(1,2)}_\alpha$ are, according to Theorem~\eqref{kraus-form}, positive numbers. Then, the coefficients
$$
\lambda_\gamma=\frac{1}{2}-\frac{1}{\mu}\left(\lambda^{(1)}_\gamma+\frac{1}{2}\sum_{\beta=0}^3\varepsilon_\beta\lambda^{(2)}_{[\beta,\gamma]}\right)
$$
can always be made positive and thus $\Lambda=\sum_{\alpha=0}^3\lambda_\alpha\,S_\alpha$ completely positive, by choosing $\mu$ large enough. Then, they fulfill the condition~\eqref{crit1} that corresponds to
$\Lambda_{diag}$ being positive. We can thus consider when and whether inequality~\eqref{crit2} which reads
$$
\frac{1}{2}\,-\,\sum_{\gamma=0}^3\lambda_\gamma\,r_\gamma=\frac{1}{\mu}\sum_{\gamma=0}^3\,r_\gamma\,
\left(\lambda^{(1)}_\gamma+\frac{1}{2}\sum_{\beta=0}^3\varepsilon_\beta\lambda^{(2)}_{[\beta,\gamma]}\right)\,\geq\, 0\ ,
$$
is satisfied. By choosing $\lambda^{(1)}_\gamma=0$ for all $\gamma$ and $\lambda^{(2)}_\gamma=\delta_{\gamma\alpha}$, one gets
$$
\sum_{\beta=0}^3\,r_{[\alpha,\beta]}\,\varepsilon_\beta=r_{[\alpha,0]}+r_{[\alpha,1]}-r_{[\alpha,2]}+r_{[\alpha,3]}=1-2\,r_{[\alpha,2]}\,\geq\, 0
$$
and thus, by varying $\alpha$, $\displaystyle r_\mu\leq\frac{1}{2}$ for all $\mu=0,1,2,3$. Vice versa, if $\displaystyle r_\mu\leq\frac{1}{2}$ for all $\mu=0,1,2,3$, one obtains
\begin{eqnarray*}
\sum_{\gamma=0}^3\,r_\gamma\,\sum_{\beta=0}^3\varepsilon_\beta\lambda^{(2)}_{[\beta,\gamma]}&=&
\sum_{\alpha=0}^3\,\lambda^{(2)}_\alpha\sum_{\beta=0}^3\varepsilon_\beta\,r_{[\alpha,\beta]}\\
&=&\sum_{\alpha=0}^3\,\lambda^{(2)}_\alpha\,\Big(r_{[\alpha,0]}+r_{[\alpha,1]}-r_{[\alpha,2]}+r_{[\alpha,3]}\Big)\\
&=&\sum_{\alpha=0}^3\,\lambda^{(2)}_\alpha\,\Big(1-2\,r_{[\alpha,2]}\Big)\geq 0\ .
\end{eqnarray*}
\end{example}

\section{Lattice states}

In this section we shall restrict ourselves to the lattice states $\rho_I\in M_{16}(\CI)$ in~\eqref{lattstate}, namely to uniformly distributed $\sigma$-diagonal states with $n=2$ .
Proposition~\eqref{propdiagst} now reads

\begin{corollary}
\label{corLS}
A lattice state $\rho_{I}$ is separable if and only if
\begin{equation}
\label{crit2LS}
\sum_{(\alpha,\beta) \in I}\lambda_{\alpha\beta}\,\leq\,\frac{N_{I}}{4}
\end{equation}
for all choices of $16$ coefficients $\lambda_{\alpha\beta}\geq 0$ such that
\begin{equation}
\label{crit1LS}
\sum_{(\alpha,\beta)\in I}\lambda_{\alpha\beta}\vert\langle\varphi\vert\sigma_{\alpha\beta}\vert\psi\rangle\vert^2\,\leq\,1\qquad\forall\
\vert\varphi\rangle,\vert\psi\rangle\in\CI^4\ .
\end{equation}
Otherwise, if a choice of positive coefficients exists that satisfy~\eqref{crit1LS} and for which
\begin{equation}
\label{crit3LS}
\sum_{(\alpha,\beta) \in I}\lambda_{\alpha\beta}\,>\,\frac{N_{I}}{4}\ ,
\end{equation}
then a lattice-state $\rho_I$ is entangled.
\end{corollary}

Before drawing concrete conclusions from this result, we examine the entanglement criteria in Propositions~\eqref{PPT2} and~\eqref{PPT3} in the light of the diagonal structure of witnessing maps which is the main result of the previous section.

\begin{example}
\label{ex6.7}
The states considered in Example~\eqref{ex6.3} were found to be entangled by showing that they do not remain positive under the action of $\id\otimes\Gamma_t$ where
\begin{equation}
\Gamma^t=g_{00}(t)\,S_{00}+\sum_{i=1}^{3}\Big(g_{0i}(t)S_{0i}+g_{i0}(t)S_{i0}\Big),
\end{equation}
with
\begin{eqnarray*}
g_{00}(t)&=&\frac{1+3 e^{-4t}}{4}\frac{3+e^{-4t}}{4},\\
 g_{0i}(t)&=&\varepsilon_i\frac{1+3 e^{-4t}}{4}\frac{1-e^{-4t}}{4},\\
 g_{i0}(t)&=&\frac{1- e^{-4t}}{4}\frac{3+e^{-4t}}{4}\ ,
\end{eqnarray*}
was proved to be a positive map from $M_4(\CI)$ into itself.
This map, expressed by means of the notation of the last Example in chapter 4 ~\eqref{Tr4}, is already in diagonal form; the diagonal completely positive maps associated to it by St{\o}rmer Theorem~\eqref{stormer} have the form
\begin{eqnarray}
\nonumber
\Lambda^{cp}(t)&=&\Tr-\frac{\Gamma^t}{\mu}\\
\label{aid}
&=&\Big(\frac{1}{4}-\frac{g_{00}(t)}{\mu}\Big)\,S_{00}\\
\nonumber
&+&\sum_{i=1}^{3}\Big[\Big(\frac{1}{4}-\frac{g_{0i}(t)}{\mu}\Big)\,S_{0i}+
\Big(\frac{1}{4}-\frac{g_{i0}(t)}{\mu}\Big)\,S_{i0}\Big]+\frac{1}{4}\sum_{\alpha,\beta\neq 0}S_{\alpha\beta}\ ,
\end{eqnarray}
with $\mu$ which has to be adjusted taking into account that
$$
g_{0i}(t)\leq\frac{1}{16}\ ,\quad g_{i0}(t)\leq\frac{1}{16}\ ,\quad g_{00}(t)\leq 1\ .
$$
Then, complete positivity of the map $\Lambda^{cp}$ is guaranteed by $\mu \geq\frac{1}{4} $ which yields
\begin{eqnarray*}
\lambda_{00}(t)&=&\frac{1}{4}-\frac{g_{00}(t)}{\mu}\geq 0\\
 \lambda_{0i}(t)&=&\frac{1}{4}-\frac{g_{0i}(t)}{\mu}\geq 0,\\
 \lambda_{i0}(t)&=&\frac{1}{4}-\frac{g_{i0}(t)}{\mu}\geq 0\\
  \lambda_{ij}&=&\frac{1}{4}\ .
\end{eqnarray*}
These coefficients surely satisfy condition~\eqref{crit1LS} as the latter just reflects the positivity of the originating map $\Gamma^t$; they also satisfy condition~\eqref{crit3LS}. Indeed,
\begin{eqnarray}
\sum_{\alpha,\beta\in I}\lambda_{\alpha\beta}(t)&=&\frac{N_I}{4}\\
\nonumber
&-&\frac{1}{\mu}\sum_{\alpha,\beta\in I}\Big[g_{00}(t))\delta_{\alpha,0}\delta_{\beta,0}+g_{0\beta}(t))\delta_{\alpha,0}
+g_{\alpha 0}(t))\delta_{\beta,0})\Big]\\
\label{gamma2}
&\simeq\atop{t\to 0}&\frac{N_I}{4}-\frac{1}{\mu}\sum_{\alpha,\beta\in I}\Big[(1-4t)\delta_{\alpha,0}\delta_{\beta,0}+t(\delta_{\alpha,0}\varepsilon_{\beta}+\delta_{\beta,0})\Big].
\end{eqnarray}
For both subsets in \eqref{ex6.3}, the second term in ~\eqref{gamma2} is negative due to $\varepsilon_2=-1$. Thus, $\sum_{\alpha,\beta\in I}\lambda_{\alpha\beta}(t)>\frac{N_I}{4}$ for small times.
\end{example}

\begin{example}
\label{ex6.8}
Let us now consider the lattice state in Example~\eqref{ex6.4}:
 \begin{eqnarray}
 \label{n_10}
N_I=10:\qquad
\begin{array}{c|c|c|c|c}
               3 & \times & \quad & \quad\! &\times  \\
               \hline
               2 & \quad\! & \times &\times &\quad\!  \\
               \hline
               1 & \times & \times & \quad\!  &\times   \\
               \hline
               0 & \times & \times & \quad\!  &\times   \\
               \hline
                 & 0 & 1 & 2 & 3
\end{array}\qquad .
\end{eqnarray}
 In~\cite{benatti4}, it has been shown to be entangled by using the following positive map
\begin{equation}
\label{map3}
M_{4}(\CI)\ni X\mapsto\Phi_V[X]=\Tr[X] - \Big(\rT[X]+\cV[X]\Big) \ ,\quad \cV[X]=V^{\dag}\,X\,V\ ,
\end{equation}
consisting of the trace map to which one subtracts the transposition map and a completely positive map $\cV$ constructed with a $4\times 4$ matrix $V$ such that, in the standard representation,
\begin{equation}
V=\sum_{\alpha\neq2}v_{\alpha2}\sigma_{\alpha2}+\sum_{\beta\neq2}v_{2\beta}\sigma_{2\beta}=-V^T\ ,\quad \sum_{\alpha\neq 2}\Big(|v_{\alpha2}|^2+|v_{2\alpha}|^2\Big)=1\ .
\label{map3.1}
\end{equation}
In this way,
$$
\Phi_V[\vert\psi\rangle\langle\psi\vert]=1-\vert\psi^*\rangle\langle\psi^*\vert-V^\dag\,\vert\psi\rangle\langle\psi\vert\,V=1-P-Q\ ,
$$
where $P$ and $Q$ are orthogonal one-dimensional projections and thus ensure the positivity of the map.

Because of $\cV$, the map $\Phi_V$ is non-diagonal in the maps $S_{\alpha\beta}$: in order to illustrate the content of Proposition~\eqref{propdiagst} we diagonalize it. Using
\begin{equation}
M_4(\CI)\ni X\mapsto T[X]=\frac{1}{4}\sum_{\alpha,\beta=0}^3\,\varepsilon_\alpha\varepsilon_\beta\,S_{\alpha\beta}[X]\ ,\quad
S_{\alpha\beta}[X]=\sigma_{\alpha\beta}\,X\,\sigma_{\alpha\beta}\ .
\end{equation}
one gets
$$
\Phi^{diag}_V=\sum_{\alpha\neq 2}\Big(\Big(\frac{1}{2}-|v_{\alpha2}|^2\Big)S_{\alpha2}+\Big(\frac{1}{2}-|v_{2\beta}|^2\Big)S_{2\beta}\Big)\ .
$$
The mean value of the Choi matrix of $\Phi^{diag}_V$ with respect to the lattice state in~\eqref{n_10} reads
$$
\Tr\Big(\rho_I\,\id\otimes\Phi^{diag}_V[P^4_+]\Big)=\frac{1}{N_I}\Big(\frac{1}{2}-|v_{12}|^2\Big)
$$
and becomes negative choosing $|v_{12}|^2>1/2$ hence revealing the entanglement of $\rho_I$.

Proposition associates to $\Phi^{diag}_V$ completely positive maps of the form
\begin{eqnarray}
\nonumber
\Lambda^{CP}&=&\Tr-\frac{\Phi^{diag}_V}{\mu}=\sum_{\alpha\neq 2}\Big(\frac{1}{4}-\frac{1}{2\mu}\frac{|v_{\alpha2}|^2}{\mu}\Big)S_{\alpha2}\\
\nonumber
&+&\sum_{\beta\neq 2}\Big(\frac{1}{4}-\frac{1}{2\mu}+\frac{|v_{2\beta}|^2}{\mu}\Big)S_{2\beta}
+\frac{1}{4}\Big(S_{22}+\sum_{\alpha\neq 2,\beta\neq 2}S_{\alpha\beta}\Big)\ ,
\end{eqnarray}
whose coefficients can always be made positive by choosing $\mu\geq 2$.
The sum of the coefficients corresponding to the subset $I$ of the lattice state in~\eqref{n_10} yields
$$
\sum_{\alpha,\beta\in I}\lambda_{\alpha\beta}=\frac{N_I}{4}-\frac{1}{\mu}\Big(\frac{1}{2}-|v_{12}|^2\Big)\,>\,\frac{N_I}{4}
$$
for the choice of $v_{12}$ which exposes the entanglement of $\rho_I$, in agreement with~\eqref{crit3LS}.
\end{example}

\section{Separable Lattice States}

In absence of a positive map that witness the entanglement of a lattice state $\rho_I$, one can check its separability by trying to express it as a convex combination by other lattice states that are known to be separable.
For some $\rho_I$ this is feasible as shown in~\cite{benatti4}; for instance, consider the lattice state
$$
\rho_I=\frac{1}{8}\Big(P_{11}+P_{12}+P_{13}+P_{21}+P_{23}+P_{31}+P_{32}+P_{33}\Big)\ .
$$
According to Proposition~\eqref{PPT}, it is PPT and is also separable; indeed, the defining subset $I$ splits as follows
\begin{eqnarray*}
\underbrace{\begin{array}{c|c|c|c|c}
               3 & \quad\!  & \times & \times  & \times     \\
               \hline
               2 & \quad\!  &  \times   &  \quad\!  &  \times   \\
               \hline
               1 &  \quad\! & \times &  \times & \times     \\
               \hline
               0 &    &     &     &      \\
               \hline
                 & 0 & 1 & 2 & 3
 \end{array}}_{I}\ &=&\
 \underbrace{\begin{array}{c|c|c|c|c}
               3 &  \quad\!  &  \times  &  \times   &      \\
               \hline
               2 &  \quad\!  &  \quad\!  & \quad\!  &      \\
               \hline
               1 &  \quad\! & \times &  \times   &       \\
               \hline
               0 & \quad\!  &  \quad\!   &  \quad\! &      \\
               \hline
                 & 0 & 1 & 2 & 3
 \end{array}}_{I_1}\ +\
 \underbrace{\begin{array}{c|c|c|c|c}
               3 &  \quad\!  &  \quad\!  & \quad\!   &      \\
               \hline
               2 &  \quad\! & \times &  \quad\!   & \times     \\
               \hline
               1 &  \quad\! & \times &  \quad\!   & \times     \\
               \hline
               0 &  \quad\!  & \quad\!   &  \quad\!   &      \\
               \hline
                 & 0 & 1 & 2 & 3
 \end{array}}_{I_2}\ \\
 &+&\
 \underbrace{\begin{array}{c|c|c|c|c}
               3 &  \quad\!  &  \quad\!  &  \times &  \times    \\
               \hline
               2 & \quad\!  & \quad\! & \quad\!  &      \\
               \hline
               1 &  \quad\!  &  \quad\!  &  \times   &  \times     \\
               \hline
               0 & \quad\! & \quad\!  & \quad\! &      \\
               \hline
                 & 0 & 1 & 2 & 3
 \end{array}}_{I_3}
\ +\
\underbrace{\begin{array}{c|c|c|c|c}
               3 & \quad\!   & \times   &  \quad\! &  \times    \\
               \hline
               2 & \quad\!  & \times & \quad\!  &  \times    \\
               \hline
               1 &  \quad\!  & \quad\!   & \quad\! &       \\
               \hline
               0 & \quad\! & \quad\!  & \quad\! &      \\
               \hline
                 & 0 & 1 & 2 & 3
 \end{array}}_{I_4}\ .
 \end{eqnarray*}
The $4$-point subsets $I_i$ are not disjoint, but all points contribute exactly twice to $I$, thence one rewrites
$$
\rho_I=\frac{1}{4}\sum_{i=1}^4\rho_{I_i}\ ,
$$
in terms of rank-$4$ lattice states corresponding to the subsets $I_i$. The result follows since the criterion of Proposition~\eqref{PPT} ensures that they are all PPT~\cite{horo6}.

\subsection{Special Quadruples}

A more general sufficient condition for the separability of lattice states can be derived by introducing the notion of \textit{special quadruples}.

\begin{definition}\textbf{Special Quadruples}
\label{quadruples}

A special quadruple $\QI$ is any subset of the square lattice $L_{16}$ consisting of $4$ points $(\alpha,\beta)$ such that there exist $\vert\varphi\rangle,\vert\psi\rangle\in\CI^4$ for which
\begin{equation}
\label{quaterna0}
\frac{1}{4} \sum_{(\alpha,\beta)\in\QI} \vert\langle\varphi\vert\sigma_{\alpha\beta}\vert\psi\rangle\vert^2 =1\ .
\end{equation}
Given a lattice point $(\alpha,\beta)\in L_{16}$, we will denote by $Q_{\alpha\beta}\in\cQ$ any special quadruple containing $(\alpha,\beta)$, by $\cQ_{\alpha\beta}$ the set of such quadruples and by $n_{\alpha\beta}$ its cardinality.
\end{definition}

\begin{example}
\label{ex6.9}
Consider the lattice state:
 $$
 \rho_4=\frac{1}{4}(P_{02}+P_{03}+P_{12}+P_{13})
 $$
 $$
 N_I=4:\qquad
\begin{array}{c|c|c|c|c}
               3 & \times & \times & \quad\! &\quad\!   \\
               \hline
               2 & \times & \times &\quad\! &\quad\!  \\
               \hline
               1 & \quad & \quad\! & \quad\!  &\quad\!   \\
               \hline
               0 & \quad\! & \quad\! & \quad\!  &\quad\!   \\
               \hline
                 & 0 & 1 & 2 & 3
\end{array}.
 $$
By choosing $\vert\varphi\rangle=\vert+-\rangle$ and $\vert\psi\rangle=\vert++\rangle$, where $\sigma_2\vert\pm\rangle=\mp i\vert\mp\rangle$ and
$\sigma_3\vert\pm\rangle=\vert\mp\rangle$, we see that the set $\{\sigma_{02},\sigma_{03},\sigma_{12},\sigma_{13}\}$ satisfies~\eqref{quaterna0}.
\end{example}

\begin{example}
\label{ex6.10}
Some more examples of special quadruples are the following lattice states:
$$
\begin{array}{c|c|c|c|c}
               3 & \quad\! & \quad\! & \quad\! &\times   \\
               \hline
               2 & \quad\! & \quad\! &  \times &\quad\!  \\
               \hline
               1 & \quad\! & \times & \quad\!  &\quad\!  \\
               \hline
               0 & \times & \quad\! & \quad\!  & \quad\!  \\
               \hline
                 & 0 & 1 & 2 & 3
\end{array}
\quad,\quad
\begin{array}{c|c|c|c|c}
               3 & \times & \quad\! & \quad\! &\quad\!   \\
               \hline
               2 & \quad\! & \times &  \quad\! &\quad\!  \\
               \hline
               1 & \quad\! & \quad\! & \times  &\quad\!  \\
               \hline
               0 & \quad\! & \quad\! & \quad\!  & \times  \\
               \hline
                 & 0 & 1 & 2 & 3
\end{array}
$$

$$
\begin{array}{c|c|c|c|c}
               3 & \quad\! & \quad\! & \times &\quad\!   \\
               \hline
               2 & \quad\! & \quad\! &  \quad\! &\times  \\
               \hline
               1 & \quad\! & \times & \quad\!  &\quad\!  \\
               \hline
               0 & \times & \quad\! & \quad\!  & \quad\!  \\
               \hline
                 & 0 & 1 & 2 & 3
\end{array}
\quad,\quad
\begin{array}{c|c|c|c|c}
               3 & \quad\! & \times & \quad\! &\quad\!   \\
               \hline
               2 & \quad\! & \quad\! &  \times &\quad\!  \\
               \hline
               1 & \quad\! & \quad\! & \quad\!  &\times  \\
               \hline
               0 & \times & \quad\! & \quad\!  & \quad\!  \\
               \hline
                 & 0 & 1 & 2 & 3
\end{array}
$$
\end{example}

Since  $\vert\langle\varphi\vert\sigma_{\alpha\beta}\vert\psi\rangle\vert^2 =1$ if and only if  $\sigma_{\alpha\beta}\vert\psi\rangle=\eta\vert\phi\rangle$ with $\eta$ a pure phase, a set of $4$ points $\{(\alpha_j,\beta_j)\}_{j=0}^3\subset I$ is
a special quadruple if and only there exist $\vert\psi\rangle,\vert\phi\rangle\in\CI^4$ such that
\begin{equation}
\label{quaterna1}
\sigma_{\alpha_j\beta_j}\vert\psi\rangle={\rm e}^{i\chi_j}\,\vert\phi\rangle\qquad \forall j=0,1,2,3\ .
\end{equation}

Let us focus upon $\cQ_{00}$, the set of all special quadruples $\{(0,0),(\alpha_1,\beta_1),$ $(\alpha_2,\beta_2),(\alpha_3,\beta_3)\}$ containing  the point $(0,0)$.
Each of them  is obtained from the fact that, using~\eqref{quaterna1} with $(\alpha_0\beta_0)=(00)$,
$$
\vert\varphi\rangle=\vert\psi\rangle\Rightarrow\sigma_{\alpha_j\beta_j}\vert\psi\rangle={\rm e}^{i\chi_j}\,\vert\psi\rangle\ ,\quad j=1,2,3\ .
$$
Therefore, the $\sigma_{\alpha\beta}$ of a special quadruple must commute. Indeed,
\begin{eqnarray}
\nonumber
\Big[\sigma_{\alpha\beta}\,,\,\sigma_{\gamma\delta}\Big]
&=&\sigma_\alpha\sigma_\gamma\otimes\sigma_\beta\sigma_\delta\,-\,
\sigma_\gamma\sigma_\alpha\otimes\sigma_\delta\sigma_\beta\\
&=&\Big(1-\epsilon_{\alpha\gamma}\epsilon_{\beta\delta}\Big)\sigma_\alpha\sigma_\gamma\otimes\sigma_\beta\sigma_\delta\\
\label{comrel}
&=&\Big(1-\epsilon_{\alpha\gamma}\epsilon_{\beta\delta}\Big)
\eta^\mu_{\alpha\gamma}\eta^\nu_{\beta\delta}\sigma_{\mu\nu}\ ,
\end{eqnarray}
where $\eta^\mu_{\alpha\beta}$ are the coefficients $\pm1$ and $\pm i$, introduced in Definition~\eqref{Lmap}, such that $\sigma_\alpha\sigma_\beta=\eta^\mu_{\alpha\beta}\sigma_\mu$ and
the $16$ coefficients $\epsilon_{\alpha\gamma}=\pm1$ are given by the Pauli matrix commutation relations, which can be considered as the elements of the following matrix:
\begin{equation}
\label{matrsigma}
\epsilon=\begin{pmatrix}1&1&1&1\cr1&1&-1&-1\cr1&-1&1&-1\cr1&-1&-1&1\end{pmatrix}\ .
\end{equation}
It thus follows that
$\Big[\sigma_{\alpha\beta}\,,\,\sigma_{\gamma\delta}\Big]\vert\psi\rangle=0$ if and only if
$\epsilon_{\alpha\gamma}\epsilon_{\beta\delta}=1$, whence $\cQ_{00}$ consists of the following $15$ special quadruples (omitting the point $(0,0)$ common to all of them)
\begin{eqnarray}
\nonumber
&&\hskip-.5cm
\{(0,1);(1,0);(1,1)\}  \quad     \{(0,2);(2,0);(2,2)\} \quad   \{(0,3);(3,0);(3,3)\}\\
\nonumber
&&\hskip-.5cm
\{(0,1);(2,1);(2,0)\} \quad \{(0,2);(1,2);(1,0)\}  \quad \{(0,3);(1,3);(1,0)\}\\
\nonumber
&&\hskip-.5cm
\{(0,1);(3,1);(3,0)\}   \quad    \{(0,2);(3,2);(3,0)\}  \quad  \{(0,3);(2,3);(2,0)\}\\
\nonumber\\
&&\hskip-.5cm
\{(1,1);(2,2);(3,3)\} \quad \{(1,2);(2,3);(3,1)\}\\
\nonumber
&&\hskip-.5cm
\{(1,1);(2,3);(3,2)\} \quad \{(1,3);(2,2);(3,1)\}\\
\nonumber
&&\hskip-.5cm
\{(1,2);(2,1);(3,3)\}  \quad \{(1,3);(2,1);(3,2)\}.
\label{specQ00}
\end{eqnarray}

Knowledge of $\cQ_{00}$ is sufficient to derive the form of all $\cQ_{\alpha\beta}$; in order to prove this fact, let us introduce the following family of maps indexed by lattice points $(\alpha,\beta)\in L_{16}$:
\begin{equation}
\label{quadruplemap}
\tau_{\alpha\beta}\,:\,L_{16}\mapsto L_{16}\ ,\qquad \tau_{\alpha\beta}[(\gamma,\delta)]=([\alpha,\gamma],[\beta,\delta])\ ,
\end{equation}
where the map $(\alpha,\gamma)\mapsto[\alpha,\gamma]$ has been introduced in Definition~\eqref{Lmap}. It follows that the maps $\tau_{\alpha\beta}$ are invertible:
$\tau_{\alpha\beta}\circ\tau_{\alpha\beta}[(\gamma,\delta)]=(\gamma,\delta)$. Given a subset $I=\{(\alpha_i,\beta_i)\}\subseteq L_{16}$, $\tau_{\alpha\beta}[I]$ will denote the subset
$\{\tau_{\alpha\beta}[(\alpha_i,\beta_i)]\}$.

\begin{lemma}
\label{lemquadr1}
The map $\cQ_{\alpha\beta}\ni Q\mapsto\tau_{\alpha\beta}[Q]\in\cQ_{00}$ is one-to-one.
\end{lemma}

\proof
If $Q=\{(\alpha_j,\beta_j)\}_{j=0}^3$ is a special quadruple for $(\alpha,\beta)=(\alpha_0,\beta_0)$, then
$\displaystyle
\sigma_{\alpha_j\beta_j}\vert\psi\rangle={\rm e}^{i\chi_j}\vert\phi\rangle$
for some $\vert\psi\rangle,\vert\phi\rangle\in\CI^4$. Right multiplication by $\sigma_{\alpha\beta}$ yields:
\begin{eqnarray*}
\sigma_{\alpha\beta}\sigma_{\alpha_j\beta_j}\vert\psi\rangle&=&\sigma_{\alpha}\sigma_{\alpha_j}\otimes
\sigma_{\beta}\sigma_{\beta_j}\vert\psi\rangle\\
&=&\sigma_{[\alpha,\alpha_j],[\beta,\beta_j]}\vert\psi\rangle\\
&=&{\rm e}^{i\chi'_j}\sigma_{\alpha\beta}\vert\phi\rangle.
\end{eqnarray*}
Then, $\{([\alpha,\alpha_j],[\beta,\beta_j]\}_{j=0}^3=\tau_{\alpha\beta}[Q]$ is a special quadruple for $(0,0)$ exposed by the vectors
$\vert\psi\rangle$ and $\sigma_{\alpha\beta}\vert\phi\rangle$.
The one-to-one correspondence follows from the invertibility of the maps $\tau_{\alpha\beta}$.
\qed
\medskip

It is easy to check that all $Q\in\cQ_{00}$ in \eqref{specQ00} give rise to lattice states $\rho_Q$ that satisfy the criterion in Proposition~\eqref{PPT} for being PPT; this is also true for lattice states corresponding to $Q\in\cQ_{\alpha\beta}$: indeed,
they are obtained from the previous ones by the local action $\sigma_{\alpha\beta}\otimes\1\rho_Q\sigma_{\alpha\beta}\otimes\1$.

\begin{lemma}
\label{lem0}
\textbf{Properties of the special quadruples $Q_{00}$}

\begin{enumerate}
\item
Given two commuting $\sigma_{\alpha\beta}$ and $\sigma_{\gamma\delta}$ there is a unique $\sigma_{\mu\nu}$ commuting with both of them.
\item
Two quadruples cannot have more than one pair $(\alpha,\beta)$ in common; otherwise they would coincide.
\item
Each $\sigma_{\alpha\beta}$ commutes with $6$ $\sigma_{\gamma\delta}$ different from itself and from $\sigma_{00}$ \item
Each $\sigma_{\alpha\beta}$ belongs to three different quadruples.
\item
The pairs $(\gamma,\delta)\neq (\alpha,\beta)$ belonging to the $3$ quadruples with $(\alpha,\beta)$ in common correspond to anti-commuting $\sigma$'s.
\end{enumerate}
\end{lemma}

\noindent
\textbf{Proof:}\quad
The first and second property follow directly from~(\ref{comrel}, they implies the third and fourth ones, while the  last one is a consequence of the fourth property.
\qed
\medskip

The following Lemma shows that they are indeed separable as dictated by the general result in~\cite{horo6}.

\begin{lemma}
\label{lemmaquadr}
Let $I\subset L_{16}$; all positive coefficients $\lambda_{\alpha\beta}$ satisfying inequality~\eqref{crit2LS} are such that
$\sum_{(\alpha,\beta)\in Q}\lambda_{\alpha\beta}\leq 1$ for all special quadruples $Q\subseteq I$.
\end{lemma}

\proof
Recall the trace map~\eqref{Tr4}:
$$
M_4(\CI)\ni X\mapsto{\Tr}[X]=\frac{1}{4}\sum_{\alpha,\beta=0}^3\,S_{\alpha\beta}[X]\ .
$$
It follows that
$$
\label{eq6.2}
1=\langle\phi\vert\Tr[\vert\psi\rangle\langle\psi\vert]\vert\phi\rangle=\frac{1}{4}\sum_{(\alpha,\beta)\in L_{16}}\left|\langle\phi\vert\sigma_{\alpha\beta}\vert\psi\rangle\right|^2\ .
$$
Consider a given special quadruple $Q\in L_{16}$, then we can split~\eqref{eq6.2} into two terms:
\begin{eqnarray*}
1&=&\langle\phi\vert\Tr[\vert\psi\rangle\langle\psi\vert]\vert\phi\rangle\\
&=&\frac{1}{4}\sum_{(\alpha,\beta)\in L_{16}}\left|\langle\phi\vert\sigma_{\alpha\beta}\vert\psi\rangle\right|^2\\
&=&\frac{1}{4}\sum_{(\alpha,\beta)\in Q}\left|\langle\phi\vert\sigma_{\alpha\beta}\vert\psi\rangle\right|^2+\frac{1}{4}\sum_{(\alpha,\beta)\in L_{16}\backslash Q}\left|\langle\phi\vert\sigma_{\alpha\beta}\vert\psi\rangle\right|^2\ .
\end{eqnarray*}
Therefore, if $\vert \psi\rangle$ and $\vert\phi\rangle$ satisfy~\eqref{quaterna0} for a given special quadruple $Q$, then $\langle\phi\vert\sigma_{\alpha\beta}\vert\psi\rangle=0$ for all
$(\alpha,\beta)$ not belonging to $Q$, i.e. the last term in the above equality vanishes, on the other hand $|\langle\phi\vert\sigma_{\alpha\beta}\vert\psi\rangle|^2=1$ for all $(\alpha,\beta)\in Q$.

Consider now a set of positive coefficients
$\lambda_{\alpha\beta}$ satisfying
$$
\sum_{(\alpha,\beta)\in I} \lambda_{\alpha\beta}\,\left|\langle\phi\vert\sigma_{\alpha\beta}\vert\psi\rangle\right|^2\leq 1\qquad\forall \vert\psi\rangle,\vert\phi\rangle\in\CI^4\ .
$$
If $Q\subset I$ is a special quadruple and $\vert\psi\rangle$, $\vert\phi\rangle$ satisfy~\eqref{quaterna0}, then
$$
1\geq \sum_{(\alpha,\beta)\in I} \lambda_{\alpha\beta}\,\left|\langle\phi\vert\sigma_{\alpha\beta}\vert\psi\rangle\right|^2=
\sum_{(\alpha,\beta)\in Q} \lambda_{\alpha\beta}\ .
$$
\qed

\newpage

\begin{corollary}
\label{quadruplecor0}
Each rank $4$ lattice state $\rho_I$ with $I\in\cQ$ is separable.
\end{corollary}

\proof
Given any four positive coefficients $\{\lambda_{\alpha\beta}\}_{(\alpha,\beta)\in I}$ satisfying~\eqref{crit1LS}, from the assumption and the previous Lemma it follows that:

$$\displaystyle\sum_{(\alpha,\beta)\in I=Q_{\alpha_0\eta_0}}\lambda_{\alpha\beta}\,\leq\,1\ ,$$

since $I$ is a special quadruple for each of its points $(\alpha_0,\beta_0)$.
Summing over all $(\alpha_0,\beta_0)\in I$, each point is counted four times, thus
\begin{eqnarray*}
4\,\sum_{(\alpha,\beta)\in I}\lambda_{\alpha\beta}&=&\sum_{(\alpha_0,\beta_0)\in I}\sum_{(\alpha,\beta)\in \QI_{\alpha_{0}\beta_{0}}} \lambda_{\alpha\beta}\leq4\\
&\Longrightarrow&
\sum_{(\alpha,\beta)\in I}\lambda_{\alpha\beta}\leq 1=\frac{N_I}{4}\ ,
\end{eqnarray*}
whence the criterion~\eqref{crit1LS} for separability is satisfied.
\qed
\medskip

\subsection{Separability and Quadruples}

The main tool in the previous proof is the fact that all points in $I$ belongs to a same number, $1$, of special quadruples, $I$ itself; that is each point in $I$ belongs to a special quadruple contained in $I$.

Because of its importance in what follows, we now study this occurrence in more detail considering a higher rank lattice state corresponding to the subset
$$
I=\{(0,0),(1,1),(1,2),(1,3),(2,1),(2,2),(2,3),(3,1),(3,2),(3,3)\}
$$
\begin{equation}
\label{Ex10}
 \begin{array}{c|c|c|c|c}
               3 & \quad\!  & \times & \times  & \times     \\
               \hline
               2 & \quad\!  &  \times   &  \times  &  \times   \\
               \hline
               1 &  \quad\! & \times &  \times & \times     \\
               \hline
               0 & \times   &     &     &      \\
               \hline
                 & 0 & 1 & 2 & 3
 \end{array}  \qquad .
 \end{equation}
For $(0,0)$, of the set~\eqref{specQ00} the following quadruples are contained in $I$:
\begin{eqnarray}
\label{mincov0}
\nonumber
&\begin{array}{c|c|c|c|c}
               3 & \quad\! & \quad\! & \quad\! &\times   \\
               \hline
               2 & \quad\! & \quad\! &  \times &\quad\!  \\
               \hline
               1 & \quad\! & \times& \quad\!  &\quad\!  \\
               \hline
               0 & \times & \quad\! & \quad\!  &   \\
               \hline
                 & 0 & 1 & 2 & 3
\end{array}&\quad ,\quad
\begin{array}{c|c|c|c|c}
               3 & \quad\! & \quad\! & \times &\quad\!   \\
               \hline
               2 & \quad\! & \quad\! &\quad\! &\times  \\
               \hline
               1 & \quad & \times & \quad\!  &\quad\!   \\
               \hline
               0 & \times & \quad\! & \quad\!  &   \\
               \hline
                 & 0 & 1 & 2 & 3
\end{array}\\
&\begin{array}{c|c|c|c|c}
               3 & \quad\! & \times & \quad\!&\quad\!   \\
               \hline
               2 & \quad\! & \quad\! &  \quad\! & \times \\
               \hline
               1 & \quad & \quad\! & \times  &\quad\!   \\
               \hline
               0 & \times & \quad\! & \quad\!  &   \\
               \hline
                 & 0 & 1 & 2 & 3
\end{array}&\quad ,\quad
\begin{array}{c|c|c|c|c}
               3 & \quad\! & \quad\! & \times &\quad\!   \\
               \hline
               2 & \quad\! & \times &  \quad\! & \quad\! \\
               \hline
               1 & \quad\! & \quad\! & \quad\!  & \times  \\
               \hline
               0 & \times & \quad\! & \quad\!  &   \\
               \hline
                 & 0 & 1 & 2 & 3
 \end{array}\\
 \nonumber
 &\begin{array}{c|c|c|c|c}
               3 & \quad\! & \times & \quad\! &\quad\!   \\
               \hline
               2 & \quad\! & \quad\! &  \times & \quad\! \\
               \hline
               1 & \quad\! & \quad\! & \quad\!  & \times  \\
               \hline
               0 & \times & \quad\! & \quad\!  &   \\
               \hline
                 & 0 & 1 & 2 & 3
 \end{array}&\quad,\quad
 \begin{array}{c|c|c|c|c}
               3 & \quad\! & \quad\! & \quad\! &\times  \\
               \hline
               2 & \quad\! & \times &  \quad\! & \quad\! \\
               \hline
               1 & \quad\! & \quad\! & \times  & \quad\!  \\
               \hline
               0 & \times & \quad\! & \quad\!  &   \\
               \hline
                 & 0 & 1 & 2 & 3
 \end{array} .
\end{eqnarray}
Given another point $I\ni(\alpha,\beta)\neq (0,0)$, consider the lattice state
$\rho_{I_{\alpha\beta}}=\sigma_{\alpha\beta}\otimes\1\rho_I\sigma_{\alpha\beta}\otimes\1$, where $I_{\alpha\beta}\subseteq L_{16}$ is the set $I$ transformed into $\tau_{\alpha\beta}[I]$ by the map $\tau_{\alpha\beta}$ introduced in Lemma~\eqref{quadruplemap}. The point $(\alpha,\beta)\in I$ is mapped into $(0,0)\in I_{\alpha\beta}$ and, by Lemma~\eqref{lemquadr1}, it belongs to $6$ special quadruples contained in $I_{\alpha\beta}$ that are images under $\tau_{\alpha\beta}$
of the $6$ special quadruples relative to $(0,0)\in I$. Therefore, each point $(\alpha,\beta)\in I$ belongs to
$6$ special quadruples contained in $I$.

\begin{example}
\label{ex6.11}
As a concrete example, consider the following quadruples
\begin{equation}
\label{quadruplesneq00}
\underbrace{\begin{array}{c|c|c|c|c}
               3 &  \quad\!  &  \quad\!  &  \times &  \times    \\
               \hline
               2 & \quad\!  & \quad\! & \quad\!  &      \\
               \hline
               1 &  \quad\!  &  \quad\!  &  \times   &  \times     \\
               \hline
               0 & \quad\! & \quad\!  & \quad\! &      \\
               \hline
                 & 0 & 1 & 2 & 3
 \end{array}}_{Q_1}\quad,\quad
\underbrace{\begin{array}{c|c|c|c|c}
               3 & \quad\!   & \quad\!   &  \quad\! &  \quad\!    \\
               \hline
               2 & \quad\!  & \times & \quad\!  &  \times    \\
               \hline
               1 &  \quad\!  & \times   & \quad\! &  \times     \\
               \hline
               0 & \quad\! & \quad\!  & \quad\! &      \\
               \hline
                 & 0 & 1 & 2 & 3
 \end{array}}_{Q_2}\quad,\quad
\underbrace{\begin{array}{c|c|c|c|c}
               3 & \quad\!   & \times   &  \times &  \quad\!    \\
               \hline
               2 & \quad\!  & \times & \times  &  \quad\!    \\
               \hline
               1 &  \quad\!  & \quad\!   & \quad\! &  \quad\!     \\
               \hline
               0 & \quad\! & \quad\!  & \quad\! &      \\
               \hline
                 & 0 & 1 & 2 & 3
 \end{array}}_{Q_3}\quad .
\end{equation}
They are contained in $I$ but do not belong to $\cQ_{00}$; consider the point $(3,3)\in Q_1$, it is mapped into $(0,0)$ by $\tau_{33}$ which sends $I$ and $Q_1$
into
\begin{eqnarray*}
I_{33}=\tau_{33}[I]&=&\quad\begin{array}{c|c|c|c|c}
               3 &  \quad\!  &   &  &  \times    \\
               \hline
               2 & \times  & \times & \times  &      \\
               \hline
               1 &  \times  &  \times  &  \times   &      \\
               \hline
               0 & \times &\times   &\times  &      \\
               \hline
                 & 0 & 1 & 2 & 3
 \end{array}\\
\tau_{33}[Q_1]&=&\quad\begin{array}{c|c|c|c|c}
               3 &   &   &  &      \\
               \hline
               2 & \times  &  & \times  &      \\
               \hline
               1 &    &  &  &      \\
               \hline
               0 & \times &\quad\!   &\times  &\quad\!      \\
               \hline
                 & 0 & 1 & 2 & 3
 \end{array}\in\cQ_{00}\quad.
 \end{eqnarray*}
One thus checks that $\tau_{33}[Q_1]$ is a special quadruple contained in the transformed subset $I_{33}$.
Analogously, $\tau_{32}$, respectively $\tau_{12}$, map $(3,2)$, respectively $(1,2)$, into $(0,0)$ and $I$, $Q_2$, respectively $Q_3$, into
\begin{eqnarray*}
I_{32}=\tau_{32}[I]&=&\quad\begin{array}{c|c|c|c|c}
               3 &  \times  &\times   &\times  &\times     \\
               \hline
               2 &   &  &   &\times      \\
               \hline
               1 &    &  \times  &  \times   &      \\
               \hline
               0 & \times &\times   &\times  &      \\
               \hline
                 & 0 & 1 & 2 & 3
 \end{array}\\
\tau_{32}[Q_2]&=&\quad\begin{array}{c|c|c|c|c}
               3 &\times   &\quad\!   &\times  &\quad\!      \\
               \hline
               2 &   &  &   &      \\
               \hline
               1 &    &  &  &      \\
               \hline
               0 & \times &   &\times  &      \\
               \hline
                 & 0 & 1 & 2 & 3
 \end{array}\in\cQ_{00}\\
 I_{12}=\tau_{12}[I]&=&\quad\begin{array}{c|c|c|c|c}
               3 &\times   & \quad\!  &\times  &  \times    \\
               \hline
               2 &   & \times &  &      \\
               \hline
               1 &   &    &  \times   &\times      \\
               \hline
               0 & \times &\times   &\times  &\times      \\
               \hline
                 & 0 & 1 & 2 & 3
 \end{array}\\
\tau_{12}[Q_3]&=&\quad\begin{array}{c|c|c|c|c}
               3 &   &   &  &      \\
               \hline
               2 &   &  &   &      \\
               \hline
               1 &\times    & &  &\times      \\
               \hline
               0 & \times &\quad\!   &\quad\!  &\times      \\
               \hline
                 & 0 & 1 & 2 & 3
 \end{array}\in\cQ_{00}\quad.
 \end{eqnarray*}
\end{example}
\medskip

\begin{definition}
\label{defcovering}
Given a subset $I\subseteq L_{16}$ we shall term a covering of $I$ any collection $\cQ_I$ of special quadruples (not necessarily disjoint) contained in $I$ and denote by $N_{\cQ_I}$ its cardinality.
Further, we shall denote by $M^{\cQ_I}_{\alpha\beta}$ the number of special quadruples in $\cQ_I$ that contains the point $(\alpha,\beta)\in I$
and refer to them as to the multiplicities of $\cQ_I$. Finally, we shall call uniform any covering $\cQ_I$ of $I$ of constant multiplicity: $M^{\cQ_I}_{\alpha\beta}=M_{\cQ_I}$, for all $(\alpha,\beta)\in I$.
\end{definition}
\medskip

The usefulness of uniform coverings can be seen as follows: summations over $(\alpha,\beta)\in I$ can be split into sums of contributions from the special quadruples of  any covering by taking into account to how many
special quadruples $M_{\alpha\beta}^{\cQ_I}$ a point $(\alpha,\beta)$ does belong:
$$
\sum_{(\alpha,\beta)\in I} M_{\alpha\beta}^{\cQ_I}\,\lambda_{\alpha\beta}=\sum_{Q\in\cQ_I}\sum_{(\alpha,\beta)\in Q}\,\lambda_{\alpha\beta}\ ,
$$
whence, if the covering is uniform with multiplicity $M_{\cQ_I}$,
$$
\sum_{(\alpha,\beta)\in I}\,\lambda_{\alpha\beta}=\frac{1}{ M_{\cQ_I}}\sum_{Q\in\cQ_I}\sum_{(\alpha,\beta)\in Q}\,\lambda_{\alpha\beta}\ .
$$
\medskip

\begin{lemma}
\label{lem-mincov}
Let $I\subseteq L_{16}$ contain $N_I$ points and $\cQ_I$ be a uniform covering of $I$ of cardinality $N_{\cQ_I}$ and multiplicity $M_{\cQ_I}$; then,
\begin{equation}
\label{unif-mult}
N_{\cQ_I}=\frac{M_{\cQ_I}\, N_I}{4}\ .
\end{equation}
\end{lemma}

\proof
Each of the $N_I$ points in $I$ belongs to $n_I$ special quadruples each containing $4$ points of $I$.
\qed

In the case of ~\eqref{Ex10}, the special quadruples  in ~\eqref{mincov0} form a covering of $I$, but not a uniform one as its multiplicities are
\begin{eqnarray*}
&n^{\cQ_I}_{00}&=5\ ,\\
&n^{\cQ_I}_{11}&=n^{\cQ_I}_{22}=n^{\cQ_I}_{23}=n^{\cQ_I}_{32}=n^{\cQ_I}_{13}=n^{\cQ_I}_{31}=2\ ,\\ &n^{\cQ_I}_{33}&=n^{\cQ_I}_{21}=n^{\cQ_I}_{12}=1\ .
\end{eqnarray*}
A uniform covering $I$ is provided by $2$ special quadruples in ~\eqref{mincov0} and the $3$ in ~\eqref{quadruplesneq00}:
\begin{eqnarray}
\nonumber
\begin{array}{c|c|c|c|c}
3 & \quad\!  & \times & \times  & \times     \\
               \hline
               2 & \quad\!  &  \times   &  \times  &  \times   \\
               \hline
               1 &  \quad\! & \times &  \times & \times     \\
               \hline
               0 & \times   &     &     &      \\
               \hline
                 & 0 & 1 & 2 & 3
 \end{array} &=&
 \begin{array}{c|c|c|c|c}
               3 &  \quad\!  &  \quad\!  &  \quad\!   &  \times    \\
               \hline
               2 &  \quad\!  &  \quad\!  & \times  &      \\
               \hline
               1 &  \quad\! & \times &  \quad\!   &       \\
               \hline
               0 & \times  &  \quad\!   &  \quad\! &      \\
               \hline
                 & 0 & 1 & 2 & 3
 \end{array}\quad
  +\quad
 \begin{array}{c|c|c|c|c}
               3 &  \quad\!  &  \times  & \quad\!   &      \\
               \hline
               2 &  \quad\! & \quad\! &  \quad\!   & \times     \\
               \hline
               1 &  \quad\! & \quad\! &  \times   & \quad\!    \\
               \hline
               0 &  \times  & \quad\!   &  \quad\!   &      \\
               \hline
                 & 0 & 1 & 2 & 3
 \end{array}\\
 \nonumber
 &+&\begin{array}{c|c|c|c|c}
               3 &  \quad\!  &  \quad\!  &  \times &  \times    \\
               \hline
               2 & \quad\!  & \quad\! & \quad\!  &      \\
               \hline
               1 &  \quad\!  &  \quad\!  &  \times   &  \times     \\
               \hline
               0 & \quad\! & \quad\!  & \quad\! &      \\
               \hline
                 & 0 & 1 & 2 & 3
 \end{array}\quad +\quad
\begin{array}{c|c|c|c|c}
               3 & \quad\!   & \quad\!   &  \quad\! &  \quad\!    \\
               \hline
               2 & \quad\!  & \times & \quad\!  &  \times    \\
               \hline
               1 &  \quad\!  & \times   & \quad\! &  \times     \\
               \hline
               0 & \quad\! & \quad\!  & \quad\! &      \\
               \hline
                 & 0 & 1 & 2 & 3
 \end{array}\\
 \label{mincov}
 &+&\begin{array}{c|c|c|c|c}
               3 & \quad\!   & \times   &  \times &  \quad\!    \\
               \hline
               2 & \quad\!  & \times & \times  &  \quad\!    \\
               \hline
               1 &  \quad\!  & \quad\!   & \quad\! &  \quad\!     \\
               \hline
               0 & \quad\! & \quad\!  & \quad\! &      \\
               \hline
                 & 0 & 1 & 2 & 3
 \end{array}\quad .
\end{eqnarray}
Since all points of $I$ belongs to exactly $2$ special quadruples contained in $I$, this uniform covering  has multiplicity $n_I^{\cQ}=2$:
$$4\times N_{\cQ_I}=20=n_IN_I=2\times 10\ .$$
This is also a minimal covering among the uniform ones;
indeed,  by~\eqref{unif-mult}, as $N_{\cQ_I}$ is an integer, $4 N_{\cQ}=10\times n_I$ can only be satisfied by $n_I=2 m$, $\NI\ni m\geq 1$.

It thus follows that the lattice state corresponding to the subset~\eqref{Ex10} is separable as it can be decomposed into a convex combination of PPT separable rank-$4$ lattice states:
$$
\rho_I=\frac{1}{10}\Big(P_{00}+P_{11}+P_{12}+P_{13}+P_{21}+P_{22}+P_{23}+P_{31}+P_{32}+P_{33}\Big)=\frac{1}{5}\sum_{j=1}^5\rho_{Q_j}\ ,
$$
where $\{Q_j\}_{j=1}^5$ are the special quadruples of the minimal covering ~\eqref{mincov}.

The previous argument proves the separability of the lattice state~\eqref{Ex10}, an issue which left unsettled in \cite{benatti4}; it can be generalized as follows.

\begin{proposition}
 \label{theo3}
Suppose $I\subseteq L_{16}$ is a subset with minimal covering $\cQ_I$ of cardinality $N_{\cQ_I}$ and multiplicity $M_{\cQ_I}$, then $\rho_I$ can be convexly decomposed as
\begin{equation}
\label{cov-dec-ls}
\rho_I=\frac{1}{N_{\cQ_I}}\sum_{j=1}^{N_{Q_j}}\rho_{Q_j}\ ,
\end{equation}
where $Q_j\subseteq I$ are the special quadruples in $\cQ$,
and is thus separable.
\end{proposition}

\proof
Let $Q_j$, $1\leq i\leq N_{\cQ_I}$, be the elements of the minimal covering $\cQ_I$. From Lemma~\eqref{lemmaquadr} it follows that
\begin{eqnarray*}
N_{\cQ_I}&\geq&\sum_{j=1}^{N_{\cQ_I}}\sum_{(\alpha,\beta)\in Q_j} \lambda_{\alpha\beta}=M_{\cQ_I}\,\sum_{(\alpha,\beta)\in I}\lambda_{\alpha\beta}\\
&\Longrightarrow&\
\sum_{(\alpha,\beta)\in I}\lambda_{\alpha\beta}\,\leq\,\frac{N_{\cQ_I}}{M_{\cQ_I}}=\frac{N_I}{4}
\end{eqnarray*}
so that Corollary~\eqref{corLS} ensures separability. Furthermore, using~\eqref{unif-mult},
\begin{eqnarray*}
\rho_I&=&\frac{1}{N_I}\sum_{(\alpha,\beta)\in I}P_{\alpha\beta}\\
&=&\frac{4}{M_{\cQ_I}\,N_{\cQ_I}}\sum_{j=1}^{N_{\cQ_I}}\,\frac{1}{4}\sum_{(\alpha,\beta)\in Q_j}P_{\alpha\beta}\\
&=&\frac{1}{N_{\cQ_I}}\sum_{j=1}^{\cQ_I}\rho_{Q_j}\ .
\end{eqnarray*}
\qed
\medskip
\begin{example}
\label{ex6.12}
The following lattice states mentioned in~\cite{benatti4} were also unknown to be separable:
$$
\rho_8=\frac{1}{8}(P_{00}+P_{11}+P_{12}+P_{13}+P_{23}+P_{31}+P_{32}+P_{33}),
$$
\begin{eqnarray*}
  \begin{array}{c|c|c|c|c}
               3 & \quad\!  & \times & \times  & \times     \\
               \hline
               2 & \quad\!  &  \times   &  \quad\!  &  \times   \\
               \hline
               1 &  \quad\! & \times &  \quad\! & \times     \\
               \hline
               0 & \times   &     &     &      \\
               \hline
                 & 0 & 1 & 2 & 3
 \end{array}\quad&=&\quad
 \begin{array}{c|c|c|c|c}
               3 & \quad\!   & \quad\!   &  \times &  \quad\!    \\
               \hline
               2 & \quad\!  & \quad\! & \quad\!  &  \times    \\
               \hline
               1 &  \quad\!  & \times   & \quad\! &  \quad\!     \\
               \hline
               0 & \times & \quad\!  & \quad\! &      \\
               \hline
                 & 0 & 1 & 2 & 3
 \end{array}\quad +\quad
 \begin{array}{c|c|c|c|c}
               3 & \quad\!   & \quad\!   &  \times &  \quad\!    \\
               \hline
               2 & \quad\!  & \times & \quad\!  &  \quad\!    \\
               \hline
               1 &  \quad\!  & \quad\!   & \quad\! &  \times    \\
               \hline
               0 & \times & \quad\!  & \quad\! &      \\
               \hline
                 & 0 & 1 & 2 & 3
 \end{array}\\
 &+&\quad
 \begin{array}{c|c|c|c|c}
               3 & \quad\!   & \times   &  \quad\! &  \times    \\
               \hline
               2 & \quad\!  & \quad\! & \quad\!  &  \quad\!    \\
               \hline
               1 &  \quad\!  & \times   & \quad\! &  \times    \\
               \hline
               0 & \quad\! & \quad\!  & \quad\! &      \\
               \hline
                 & 0 & 1 & 2 & 3
 \end{array}\quad +\quad
 \begin{array}{c|c|c|c|c}
               3 & \quad\!   & \times   &  \quad\! &  \times    \\
               \hline
               2 & \quad\!  & \times & \quad\!  &  \times   \\
               \hline
               1 &  \quad\!  & \quad\!   & \quad\! &  \quad\!     \\
               \hline
               0 & \quad\! & \quad\!  & \quad\! &      \\
               \hline
                 & 0 & 1 & 2 & 3
 \end{array}\ .
\end{eqnarray*}

The next state is:
$$
\rho_9=\frac{1}{9}(P_{00}+P_{11}+P_{12}+P_{13}+P_{21}+P_{23}+P_{31}+P_{32}+P_{33}),
$$
\begin{eqnarray*}
\begin{array}{c|c|c|c|c}
               3 & \quad\!   & \times   &  \times &  \times    \\
               \hline
               2 & \quad\!  & \times & \quad\!  &  \times    \\
               \hline
               1 &  \quad\!  & \times   & \times &  \times     \\
               \hline
               0 & \times & \quad\!  & \quad\! &      \\
               \hline
                 & 0 & 1 & 2 & 3
 \end{array}\quad &=&\quad
 \begin{array}{c|c|c|c|c}
               3 & \quad\!   & \quad\!   &  \times &  \quad\!    \\
               \hline
               2 & \quad\!  & \times & \quad\!  &  \times   \\
               \hline
               1 &  \quad\!  & \times & \quad\! & \quad\!\\
               \hline
               0 & \times & \quad\!  & \quad\! &  \quad\!    \\
               \hline
                 & 0 & 1 & 2 & 3
 \end{array} \quad +\quad
 \begin{array}{c|c|c|c|c}
               3 & \quad\!   & \times   &  \quad\! &  \quad\!    \\
               \hline
               2 & \quad\!  & \quad\! & \quad\!  &  \times   \\
               \hline
               1 &  \quad\!  & \quad\!   & \times &  \quad\!     \\
               \hline
               0 & \times & \quad\!  & \quad\! &      \\
               \hline
                 & 0 & 1 & 2 & 3
 \end{array}\\
 &+&\quad
 \begin{array}{c|c|c|c|c}
               3 & \quad\!   & \quad\!   &  \quad\! &  \times    \\
               \hline
               2 & \quad\!  & \times & \quad\!  &  \quad\!   \\
               \hline
               1 &  \quad\!  & \quad\!   & \times &  \quad\!     \\
               \hline
               0 & \times & \quad\!  & \quad\! &      \\
               \hline
                 & 0 & 1 & 2 & 3
 \end{array}\quad +\quad
 \begin{array}{c|c|c|c|c}
               3 & \quad\!   & \quad\!   &  \times &  \quad\!    \\
               \hline
               2 & \quad\!  & \times & \quad\!  &  \quad\!   \\
               \hline
               1 &  \quad\!  & \quad\!   & \quad\! &  \times     \\
               \hline
               0 & \times & \quad\!  & \quad\! &      \\
               \hline
                 & 0 & 1 & 2 & 3
 \end{array}\\
 &+&\quad
 \begin{array}{c|c|c|c|c}
               3 & \quad\!   & \times   &  \times &  \quad\!   \\
               \hline
               2 & \quad\!  & \quad\! & \quad\!  &  \quad\!   \\
               \hline
               1 &  \quad\!  & \times   & \times &  \quad\!     \\
               \hline
               0 & \quad\! & \quad\!  & \quad\! &      \\
               \hline
                 & 0 & 1 & 2 & 3
 \end{array}\quad +\quad
 \begin{array}{c|c|c|c|c}
               3 & \quad\!   & \quad\!   &  \times &  \times    \\
               \hline
               2 & \quad\!  & \quad\! & \quad\!  &  \quad\!  \\
               \hline
               1 &  \quad\!  & \quad\!   & \times &  \times     \\
               \hline
               0 & \quad\! & \quad\!  & \quad\! &      \\
               \hline
                 & 0 & 1 & 2 & 3
 \end{array}\\
 &+&\quad
 \begin{array}{c|c|c|c|c}
               3 & \quad\!   & \times   &  \quad\! &  \times    \\
               \hline
               2 & \quad\!  & \quad\! & \quad\!  &  \quad\!   \\
               \hline
               1 &  \quad\!  & \times   & \quad\! &  \times     \\
               \hline
               0 & \quad\! & \quad\!  & \quad\! &      \\
               \hline
                 & 0 & 1 & 2 & 3
 \end{array}\quad +\quad
 \begin{array}{c|c|c|c|c}
               3 & \quad\!   & \quad\!   &  \quad\! &  \quad\!   \\
               \hline
               2 & \quad\!  & \times & \quad\!  &  \times   \\
               \hline
               1 &  \quad\!  & \times   & \quad\! &  \times    \\
               \hline
               0 & \quad\! & \quad\!  & \quad\! &      \\
               \hline
                 & 0 & 1 & 2 & 3
 \end{array}\\
 &+&
 \begin{array}{c|c|c|c|c}
               3 & \quad\!   & \times   &  \quad\! &  \times    \\
               \hline
               2 & \quad\!  & \times & \quad\!  &  \times   \\
               \hline
               1 &  \quad\!  & \quad\!   & \quad\! &  \quad\!     \\
               \hline
               0 & \quad\! & \quad\!  & \quad\! &      \\
               \hline
                 & 0 & 1 & 2 & 3
 \end{array}\ .
\end{eqnarray*}
\end{example}

From Proposition~\eqref{theo3} one draws the following conclusions:

\begin{corollary}
\label{cor2}
All the lattice states with $N_I\geq 14$ are separable.
\end{corollary}

\proof
\begin{enumerate}
 \item
$N_I$=16: is trivial.
\item
$N_I$=15: As we have seen in (section III of this chapter), every point $(\alpha,\beta)$ can take part in 15 quadruples. Therefore in $L_{16}$ we have:
$$
N_Q=\frac{16\times 15}{4}.
$$
By dropping out one point from the lattice, we decrease the number of quadruples by three for each point and therefore each point has 12 quadruples contained in I:
$$
N_Q=\frac{15\times 12}{4}.
$$
\item
$N_I=14$: Here there are two points missing from $L_{16}$, note that every two given points can share three quadruples. This is easy to verify it; by sending one of the two into the origin $(0,0)$, one sees that the other one appears in exactly three Q's (section III). Therefore, there are six pairs $(\alpha,\beta)\in I$, where the two points $(\alpha_1,\beta_1)$ and $(\alpha_2,\beta_2)$, remove only five quadruples, leaving 10 special quadruples contained in I. While for the other eight pairs $(\alpha,\beta)\in I$, they remove six quadruples. Therefore any $\rho_14$ would look like:
$$
\begin{array}{c|c|c|c|c}
               3 & \quad\!   & \quad\!   &  \times &  \times    \\
               \hline
               2 & \times & \times  & \times &   \times   \\
               \hline
               1 & \times & \times  & \times &   \times      \\
               \hline
               0 & \times & \times  & \times &   \times   \\
               \hline
                 & 0 & 1 & 2 & 3
 \end{array}
 $$
Then one can divide it into two subsets: $I_1:$ the six elements with 10 special quadruples contained in I, and $I_2:$ the eight elements with 9 special quadruples contained in I:
$$
 I_1=\quad\begin{array}{c|c|c|c|c}
               3 & \quad\!   & \quad\!   &  \quad\! &  \quad\!   \\
               \hline
               2 & \times & \times  & \quad\! &   \quad\!   \\
               \hline
               1 & \times & \times  & \quad\! &   \quad\!      \\
               \hline
               0 & \times & \times  & \quad\! &   \quad\!   \\
               \hline
                 & 0 & 1 & 2 & 3
 \end{array}\quad ,\quad
 I_2=\quad\begin{array}{c|c|c|c|c}
               3 & \quad\!   & \quad\!   &  \times &  \times    \\
               \hline
               2 & \quad\!  & \quad\!   & \times &   \times   \\
               \hline
               1 & \quad\!  & \quad\!   & \times &   \times      \\
               \hline
               0 & \quad\!  & \quad\!   & \times &   \times   \\
               \hline
                 & 0 & 1 & 2 & 3
 \end{array}
 $$

In order to show that all $\rho_14$ are separable, we need to show that they satisfy~\eqref{crit2LS}:
 $$
 \sum_{\alpha,\beta\in I}\lambda_{\alpha\beta}\leq\frac{14}{4}
 $$
for all choices of the coefficients $\lambda_{\alpha\beta}$ introduced in Corollary\ref{corLS}. But since we can divide $\rho_14$ into two subsets, we have:

\begin{eqnarray*}
\sum_{\alpha,\beta\in I}\lambda_{\alpha\beta}&=&\sum_{\alpha,\beta\in I_1}\lambda_{\alpha\beta}+\sum_{\alpha,\beta\in I_2}\lambda_{\alpha\beta}\\
&\leq&\frac{N_{I_1}}{4}+\frac{N_{I_2}}{4}\\
&\leq&\frac{N_{I_1}+N_{I_2}}{4}=\frac{14}{4}.
\end{eqnarray*}

Where we have used the separability of the two states $\rho_{I_1}$ and $\rho_{I_2}$. Indeed in case of $N_I=14$ the minimum number of special quadruples for every point to be included in I, is equal to $n^*=2$, which is much less than what happens.
\end{enumerate}
\qed
\medskip

\section{Entanglement and lattice geometry}

In this section we single out a family $\cI$ of geometric patterns such that, if $I\in\cI$, then $\rho_I$ is surely entangled.
Given a sub-set $I\subseteq L_{16}$, choose $\lambda_{\alpha\beta}$ such that for a certain $(\alpha_0,\beta_0)\in I$
\begin{equation}
\label{ent-suff-cond}
\lambda_{\alpha_{0}\beta_{0}}=\frac{1+\delta}{4}\ , \qquad \lambda_{\alpha\beta}=\frac{1}{4} \quad \forall (\alpha,\beta)\neq (\alpha_{0},\beta_{0})\ ,
\end{equation}
where $\delta> 0$ is a suitable parameter. Then,
$\displaystyle
\sum_{(\alpha\beta)\in I} \lambda_{\alpha\beta} = \frac{N_{I}+\delta}{4}>\frac{N_{I}}{4}
$
and inequality~\eqref{crit3LS} in Corollary~\ref{corLS} is satisfied.
According to the same corollary, given such a subset $I$, the corresponding lattice state $\rho_I$ is entangled, if also inequality~\eqref{crit1LS} is fulfilled, namely if
\begin{equation}
\label{pos-witn}
\exists\delta>0\,\hbox{such that}\,
\frac{\delta}{4}{\vert\langle\varphi\vert\sigma_{\alpha_{0}\beta_{0}}\vert\psi\rangle\vert}^2+
\frac{1}{4}\sum_{(\alpha,\beta)\in I}{\vert\langle\varphi\vert\sigma_{\alpha\beta}\vert\psi\rangle\vert}^2\leq1 ,\, \forall\vert\varphi\rangle,\vert\psi\rangle\in \CI^{4}\ .
\end{equation}
\medskip

\begin{remark}
\label{rem2}
Suppose that inequality~\eqref{pos-witn} can be satisfied by $\delta>0$ uniformly in the vectors $\vert\psi\rangle$ and $\vert\varphi\rangle$. Then, using~\eqref{Tr4} and~\eqref{stormer}, the choice of coefficients in~\eqref{ent-suff-cond} corresponds to a witness of the form
\begin{equation}
\label{ent-witn}
\Lambda=\Tr-\Lambda_{cp}=\frac{1}{4}\sum_{(\alpha,\beta)\notin I}S_{\alpha\beta}\,-\,\frac{\delta}{4}\,S_{\alpha_0\beta_0}\ .
\end{equation}
By comparison with the expression of the transposition in~\eqref{transp4}, one sees that, unlike the $6$ negative contribution of the latter, $\Lambda$ presents only one negative contribution. Yet, for the special subset $\cI$, it proves to be more flexible as an entanglement witness.
\end{remark}
\medskip

Clearly, if $(\alpha_0,\beta_0)$ belongs to a special quadruple contained in $I$, such a $\delta>0$ cannot exist; indeed, if $\vert\psi\rangle$ and $\vert\varphi\rangle$ satisfy~\eqref{quaterna0}, then
$\vert\langle\varphi\vert\sigma_{\alpha_{0}\beta_{0}}\vert\psi\rangle\vert>0$ and
$\displaystyle \frac{1}{4}\sum_{(\alpha,\beta)\in I}{\vert\langle\varphi\vert\sigma_{\alpha\beta}\vert\psi\rangle\vert}^2=1$, together with
inequality~(\ref{pos-witn}) yield $\delta=0$.

We now show that lattice state $\rho_I$ for which not all points of $I$ belong to special quadruples contained in $I$, such a $\delta>0$ can indeed be found and thus that they are entangled.
\medskip

\begin{theorem}
\label{theo2}
Given a lattice state $\rho_{I}$, if there exists a point $(\alpha_{0},\beta_{0})\in I$ such that
$\QI_{\alpha_{0}\beta_{0}}\nsubseteq I$ for all $\QI_{\alpha_{0}\beta_{0}}\in\cQ$, then $\rho_I$ is entangled.
We shall call any such $I$ a special sub-set and denote by $\cI$ their family.
\end{theorem}
\medskip

It is very constructive to present some examples before going through the proof.
\begin{example}
According to the last theorem the following state is entangled, as non of the special quadruples of $(0,0)$ are included in I:
\begin{equation}
 \label{ex6.13}
N_I=8 \quad
 \begin{array}{c|c|c|c|c}
               3 & \quad\! & \quad\! & \times & \times  \\
               \hline
               2 & \quad\! & \times &  \times & \quad\!  \\
               \hline
               1 & \quad\! & \quad\! & \quad\!  & \times  \\
               \hline
               0 & \times & \quad\! & \times  & \times  \\
               \hline
                 & 0 & 1 & 2 & 3
 \end{array}
\end{equation}
\medskip

Another example of such entangled lattice states is the following:
\begin{equation}
\label{ex6.14}
N_I=10 \quad
 \begin{array}{c|c|c|c|c}
               3 & \times & \times & \times  &  \times      \\
               \hline
               2 & \times & \quad\! &  \times &  \quad\!       \\
               \hline
               1 &  \quad\! & \quad\! &  \quad\! & \times       \\
               \hline
               0 & \times & \times  & \times  & \quad\!      \\
               \hline
                 & 0 & 1 & 2 & 3
 \end{array}
\end{equation}
As we observe the point $(3,3)$ has no special quadruple contained in I, it can be easily verified when this point is sent to $(0,0)$, the resulting state is:
$$
N_I=10 \quad
 \begin{array}{c|c|c|c|c}
               3 & \quad\! & \times & \times  &  \times      \\
               \hline
               2 & \times & \quad\! &  \quad\! &  \quad\!       \\
               \hline
               1 &  \quad\! & \times &  \quad\! & \times       \\
               \hline
               0 & \times & \times  & \times  & \times      \\
               \hline
                 & 0 & 1 & 2 & 3
 \end{array}
$$
\medskip

As the last example consider the following state:
\begin{equation}
\label{ex6.15}
N_I=11 \quad
 \begin{array}{c|c|c|c|c}
               3 & \quad\! & \times & \times  &  \times      \\
               \hline
               2 & \times & \quad\! &  \quad\! &  \quad\!       \\
               \hline
               1 &  \quad\! & \times &  \times & \times       \\
               \hline
               0 & \times & \times  & \times  & \times      \\
               \hline
                 & 0 & 1 & 2 & 3
 \end{array}
\end{equation}
Also in this case $(0,0)$ has no special quadruple contained in I.
\end{example}

\noindent
\textbf{Proof of Theorem~\ref{theo2}:}

According to Lemma~\ref{lemquadr1}, the point $(\alpha_0,\beta_0)$ can be transformed into $(0,0)$ and $I$ into a new set, that we shall denote again by $I$ for sake of simplicity, without altering the entanglement or separability of the transformed $\rho_I$ with respect to the initial one.
Then, the assumption of the theorem translates into the fact that no special quadruple in the list~\eqref{specQ00} is contained in $I$.
Having set  $(\alpha_0,\beta_0)=(00)$, inequality~\eqref{pos-witn} now reads
\begin{equation}
\label{pos-witn1}
\Delta^{\psi,\varphi}_{I,\delta}=\frac{\delta}{4}{\vert\langle\varphi\vert\psi\rangle\vert}^2+
\Delta^{\psi,\varphi}_I\ ,\quad\hbox{where}\quad
\Delta^{\psi,\varphi}_I=\frac{1}{4}\sum_{(\alpha\beta)\in I}{\vert\langle\varphi\vert\sigma_{\alpha\beta}\vert\psi\rangle\vert}^2\ .
\end{equation}
It proves convenient to introduce the following $\psi$-dependent $4\times 4$ matrices
\begin{equation}
\label{pos-witn2}
\hD^\psi_{I,\delta}=\frac{\delta}{4}\vert\psi\rangle\langle\psi\vert+\hD^\psi_I\ ,\quad \hD^\psi_I=\frac{1}{4}\sum_{(\alpha\beta)\in I}\,\sigma_{\alpha\beta}\vert\psi\rangle\langle\psi\vert\sigma_{\alpha\beta}\ ,
\end{equation}
so that one has to prove that
\begin{equation}
\label{pos-witn3}
\exists \delta>0\quad\hbox{such that} \quad \Delta^{\psi,\varphi}_{I,\delta}=\langle\varphi\vert\hD^\psi_{I,\delta}\vert\varphi\rangle\leq 1\quad
\forall\ \vert\psi\rangle,\,\vert\varphi\rangle\in\CI^4\ .
\end{equation}
Further, from~\eqref{Tr4}
\begin{equation}
\label{pos-witn4}
\hD^\psi_I+\hD^\psi_{I^c}=\1\Longrightarrow 0\leq \hD^\psi_I\leq \1\ .
\end{equation}
Clearly, the major obstruction to $\Delta^{\psi,\varphi}_{I,\delta}\leq1$ with $\delta>0$ arises when $\hD^\psi_I$ has eigenvalue $1$. Then, because of~\eqref{pos-witn4}, the corresponding eigenvectors, $\hD^{\psi}_{I}\vert\varphi\rangle=\vert\varphi\rangle$, satisfy
\begin{equation}
\label{D-comp}
\langle\varphi\vert\hD^{\psi}_{I^{c}}\vert\varphi\rangle=
\frac{1}{4}\sum_{(\alpha,\beta)\in I^c}|\langle\varphi\vert\sigma_{\alpha\beta}\vert\psi\rangle|^2=0
\Longleftrightarrow \vert\varphi\rangle\perp\sigma_{\alpha\beta}\vert\psi\rangle\quad\forall\ (\alpha,\beta)
\in I^c\ .
\end{equation}
Let the eigenvalues of $\hD^\psi_I$ be decreasingly ordered and consider the spectral decomposition
\begin{equation}
\label{pos-witn5}
\hD^\psi_I=P^\psi_I(1)+R^\psi_I\ ,\quad R^\psi_I=\sum_{j>1}d^\psi_I(j)\,P^\psi_I(j)\ ,
\end{equation}
where $P^\psi_I(1)$ projects onto the eigenspace relative to the eigenvalue $1$ and $P^\psi_I(j)$ are the other orthogonal spectral projections relative to the eigenvalues $0\leq d^\psi_I(j)<1$.
Then, if $P^\psi_I(1)\neq 0$, $\delta>0$ in~(\ref{pos-witn3}) is only possible with $P^\psi_I(1)\vert\psi\rangle=0$. These preliminary considerations allow us to prove~(\ref{pos-witn}) through a series of lemmas and corollaries.
\qed
\medskip

\begin{lemma}
\label{lem1}
With the notation of~(\ref{pos-witn5}), if $P^\psi_I(1)\vert\psi\rangle=0$ for all $\vert\psi\rangle\in\CI^4$, then~(\ref{pos-witn3}) is satisfied.
\end{lemma}
\medskip

\noindent
\proof
If $\displaystyle M_I=\sup_{\vert\psi\rangle\in\CI^4}\|R^\psi_I\|=1$, then, by compactness, there exists a converging sequence $\psi_n\to\psi^*$ of vectors in $\CI^4$ such that $R^{\psi_n}_I\perp P^{\psi_n}_I(1)$ and $\|R^{\psi_n}_I\|\to 1$. Then, $\hD^{\psi_n}_I$ converges in norm to $\hD^{\psi^*}_I$ with $\|R^{\psi^*}_I\|=1$ and $R^{\psi^*}_I\perp P^{\psi^*}_I(1)$ which is a contradiction. Therefore, $M_I<1$; hence, choosing $0<\delta\leq 4(1-M_I)$, from $P^\psi_I(1)\vert\psi\rangle=0$ and $P^\psi_I(1)R^\psi_I=0$, one gets
$$
\|\hD^\psi_{I,\delta}\|=
\max\left\{1,\left\|\frac{\delta}{4}\vert\psi\rangle\langle\psi\vert+R^\psi_I\right\|\right\}\leq
\max\left\{1,\frac{\delta}{4}+M_I\right\}\leq 1\ .
$$
\qed
\medskip

\begin{corollary}
\label{cor-ort}
Given a lattice state $\rho_I$ and $\vert\psi\rangle\in\CI^4$, let $V_{I^c}$ be the subspace spanned by
the vectors $\sigma_{\alpha\beta}\vert\psi\rangle$ with $(\alpha,\beta)\in I^c$, where $I^c$ is the complement of $I$. If $V^\psi_{I^c}=\CI^4$ for all $\vert\psi\rangle\in\CI^4$, then $\rho_I$ is entangled.
\end{corollary}
\medskip

\proof
From~(\ref{D-comp}) it follows that $P^\psi_I(1)=0$ for all $\vert\psi\rangle\in\CI^4$; thus Lemma~\ref{lem1} applies.
\qed
\medskip

Based on the previous two results, we now focus upon when $P^\psi_I(1)\neq 0$ and show that it projects onto a subspace orthogonal to $\vert\psi\rangle$.

\begin{lemma}
\label{lem2}
If $\sigma_{\mu\nu}\vert\psi\rangle=\pm\vert\psi\rangle$ for some $(\mu,\nu)\in L_{16}$, then, with the notation of~(\ref{pos-witn5}), $P^\psi_I(1)\vert\psi\rangle=0$.
\end{lemma}

\proof
If $(\mu,\nu)\in I^{c}$ and $\langle\varphi\vert\hD^{\psi}_{I}\vert\varphi\rangle=1$,
then, from~(\ref{D-comp}), $\vert\varphi\rangle\perp\sigma_{\alpha\beta}\vert\psi\rangle$ for all $(\alpha,\beta)\in I^{c}$, hence to $\pm\vert\psi\rangle=\sigma_{\mu\nu}\vert\psi\rangle$.
Suppose then that $(\mu,\nu)\in I$ and rewrite inequality~(\ref{pos-witn3}) as
$$
\frac{\delta}{4}{\vert\langle\varphi\vert\psi\rangle\vert}^{2}+\frac{1}{4}\sum_{(\alpha,\beta)\in I_{1}}{\vert\langle\varphi\vert\sigma_{\alpha\beta}\vert\psi\rangle\vert}^{2}+
\frac{1}{4}\sum_{(\alpha,\beta)\in I_{2}}{\vert\langle\varphi\vert\sigma_{\alpha\beta}\vert\psi\rangle\vert}^{2}
\leq 1\ ,
$$
where the index set $I$ has been split into
$$
I_{1}=\{(\alpha,\beta)\in I: [\sigma_{\alpha\beta},\sigma_{\mu\nu}]=0\}\quad\hbox{and}\quad
I_{2}=\{(\alpha,\beta)\in I: \{\sigma_{\alpha\beta},\sigma_{\mu\nu}\}=0\}\ .
$$
The vectors $\sigma_{\alpha\beta}\vert\psi\rangle$ from these two subsets are orthogonal; indeed,
$[\sigma_{\gamma\delta},\sigma_{\mu\nu}]=0$ and $\{\sigma_{\alpha\beta},\sigma_{\mu\nu}\}=0$ yield
$$
\langle\psi\vert\sigma_{\alpha\beta}\sigma_{\gamma\delta}\vert\psi\rangle=
\langle\psi\vert\sigma_{\mu\nu}\sigma_{\alpha\beta}\sigma_{\gamma\delta}
\sigma_{\mu\nu}\vert\psi\rangle
=-\langle\psi\vert\sigma_{\alpha\beta}\sigma_{\gamma\delta}\vert\psi\rangle=0\ .
$$
Therefore, the following two are orthogonal matrices:
\begin{eqnarray}
\label{ort1}
\hD^{\psi}_{I_{1}}&=&\frac{1}{2}\vert\psi\rangle\langle\psi\vert
+\frac{1}{4}\sum_{I_1\ni(\alpha,\beta)\neq(00),(\mu,\nu)}
\sigma_{\alpha\beta}\vert\psi\rangle\langle\psi\vert\sigma_{\alpha\beta}\\
\label{ort2}
\hD^{\psi}_{I_{2}}&=&\frac{1}{4}\sum_{(\alpha,\beta)\in I_2}
\sigma_{\alpha\beta}\vert\psi\rangle\langle\psi\vert\sigma_{\alpha\beta}\ .
\end{eqnarray}
Then, $\hD^\psi_I\vert\phi\rangle=\vert\phi\rangle$ can only be due to $\hD^{\psi}_{I_{2}}\vert\varphi\rangle=\vert\varphi\rangle$, in which case $\vert\varphi\rangle\perp\vert\psi\rangle$. Indeed, $\|\hD^{\psi}_{I_{1}}\|\leq 1$.

This can be seen as follows: at most three $\sigma_{\alpha\beta}$ may contribute to the sum in $\hD^{\psi}_{I_{1}}$ and they must anti-commute. In fact, if there were two commuting $\sigma_{\alpha\beta}$ contributing to the sum, then, as they would commute with $\sigma_{\mu\nu}$ and form a special quadruple $Q_{00}$ contained in $I$ which is excluded by hypothesis. Therefore, the $\sigma_{\alpha\beta}$ contributing to the sum must anti-commute and, from Lemma~\ref{lem0}, cannot be no more than three.
Suppose this is the case; denote by $S_\alpha$, $\alpha=1,2,3$, these three $\sigma_{\alpha\beta}$ such that $\{\sigma_{\alpha\beta},\sigma_{\mu\nu}\}=0$ and rewrite
$$
\hD^{\psi}_{I_{1}}=\frac{1}{4}\vert\psi\rangle\langle\psi\vert
+\frac{1}{4}\sum_{\alpha=0}^3S_\alpha\vert\psi\rangle\langle\psi\vert S_\alpha\ ,\quad S_0=\1_4\ .
$$
Without restriction, we choose $\vert\psi\rangle$ such that $\sigma_{\mu\nu}\vert\psi\rangle=\vert\psi\rangle$.
Each $S_\alpha\vert\psi\rangle$ is an eigenstate of $\sigma_{\mu\nu}$ belonging to the same twice degenerate eigenvalue $1$. Let $P$ project onto the corresponding eigenspace; then, $[S_\alpha,P]=0$ and the rank $2$ matrices $T_\alpha=P\,S_\alpha\,P=S_\alpha P=P\,S_\alpha$ satisfy the Pauli algebra~\eqref{alg-Pauli} over $\CI^4$. Thus,~\eqref{trace2} holds on $\CI^4$ with $T_\alpha$ replacing $\sigma_\alpha$, yielding
$$
\hD^{\psi}_{I_{1}}=\frac{1}{4}\vert\psi\rangle\langle\psi\vert
+\frac{1}{4}\sum_{\alpha=0}^3\,S_\alpha\vert\psi\rangle\langle\psi\vert S_\alpha=
\frac{1}{4}\vert\psi\rangle\langle\psi\vert+\frac{1}{2}P\ .
$$
If there are less than three anti-commuting contributions $S_\alpha$, $\alpha\neq 0$, then the second equality becomes a strict inequality. Therefore, $\|\hD^\psi_{I_1}\|\leq3/4<1$.
\qed
\medskip

Let us now consider the set of $\sigma_{\alpha\beta}$ indexed by the points $(\alpha,\beta)$ in the complement set $I^c$, that we list as $(\alpha_i,\beta_i)$, $1\leq i\leq N-N_I$.
The following facts are a consequence of the structure of $\cQ_{00}$ studied in Lemma~\ref{lem0} and concern
the points of $I^c$ and the special quadruples in $\cQ_{00}$ they belong to.
\begin{enumerate}
\item
Take the first point $(\alpha_1,\beta_1)\notin I$; then,  it belongs to three quadruples in~(\ref{specQ00}) which cannot be contained in $I$.
\item
Also the second point $(\alpha_2,\beta_2)\in I^c$ may eliminate three quadruples from those contained in $I$; it does so, only if it does not share any special quadruple with those of $(\alpha_1,\beta_1)$:
this can happen only if $\sigma_{\alpha_1\beta_1}$ and
$\sigma_{\alpha_2\beta_2}$ anti-commute. If they commute, then $(\alpha_1,\beta_1)$  and $(\alpha_2,\beta_2)$ share one special quadruple. In this case, $(\alpha_2,\beta_2)$ eliminates only two special quadruples from those contained in $I$.
It thus follows that the minimum cardinality of $I^c$ complying with the hypothesis of Theorem~\ref{theo2}, namely that no special quadruple of $(0,0)$ is contained in $I$, is five with the corresponding $\sigma_{\alpha\beta}$ forming an anti-commuting set.
\item
Consider the third point $(\alpha_3,\beta_3)\in I^c$: the number of special quadruples eliminated by this point from those contained in $I$ is equal to three minus the number of previous points whose $\sigma_{\alpha\beta}$ commute with
$\sigma_{\alpha_3\beta_3}$.
\item
Moving to the fourth point $(\alpha_4,\beta_4)$ and further on, the rule is the same; only, one has to take into account that $\sigma_{\alpha_4\beta_4}$  does not commute with more than three $\sigma_{\alpha\beta}$.
Therefore, if this happens, $(\alpha_4\beta_4)$ does not eliminate from those contained in $I$ any further special quadruple with respect to those already eliminated by the previous points.
\end{enumerate}

Based on these properties, we now distinguish the following cases.
\begin{description}
\item[Case 1.]
$\hbox{card}(I^c)=5$: the matrices $\{\sigma_{\alpha_i\beta_i}\}_{i=1}^5$, with $(\alpha_i,\beta_i)\in I^c$ anti-commute.
\item[Case 2.]
$\hbox{card}(I^c)>5$ and there is a set of anti-commuting $\{\sigma_{\alpha_i\beta_i}\}_{i=1}^5$, with $(\alpha_i,\beta_i)\in I^c$: let $(\alpha_6,\beta_6)\in I^c$ not belong to the index set of the five anti-commuting $\sigma$'s . The corresponding $\sigma_{\alpha_6\beta_6}$ must commute with at least one of them and may commute with at most three of the $\sigma_{\alpha_i\beta_i}$. Therefore, there must exist at least two of them that anti-commute with $\sigma_{\alpha_6\beta_6}$.
By putting together $\sigma_{\alpha_6\beta_6}$ with the one in the anti-commuting set which commutes with it and with the two which anti-commute with it, then one can construct a set $K$ of four
$\sigma_{\alpha\beta}$ where $3$ of them anti-commute among themselves and the fourth one commutes with only one of them.
\item[Case 3.]
$\hbox{card}(I^c)>5$, but there is no set of $5$ anti-commuting {$\sigma_{\alpha_i\beta_i}$} indexed by $(\alpha_i,\beta_i)\in I^c$.
If there are $4$ anti-commuting {$\sigma_{\alpha\beta}$} indexed by points in $I^c$, these latter correspond to twelve special quadruples in $\cQ_{00}$ not contained in $I$. Therefore, there are three special quadruples of $\cQ_{00}$ left that also can not be
contained in $I$. Thus, some of their points must be excluded from $I$.  These three quadruples share a common point that cannot be excluded from $I$, otherwise the corresponding $\sigma_{\alpha\beta}$, together with the four which
have already been assumed to anti-commute, would constitute a set of five anti-commuting $\sigma$'s contrary to the hypothesis of the present case.
Therefore, in order to eliminate the three remaining special quadruples from being contained in $I$, one point for each of the three special quadruples must be excluded from $I$.
The corresponding $\sigma_{\alpha\beta}$ will then anti-commute. Then, we have constructed two sets $U_{1,2}$ of $\sigma$'s, the first one containing four anti-commuting  $\sigma_{\alpha\beta}$ , the second one three anti-commuting
$\sigma_{\alpha\beta}$. We already know that each element from $U_2$ commutes at least with one from $U_1$; further, it cannot commute with more than two of them. Indeed, if this happened, one would have two anti-commuting
$\sigma_{\alpha\beta}$ in the same special quadruple which is impossible.

This makes it possible to contruct a set $K$ as in the previous case consisting of three anti-commuting $\sigma_{\alpha\beta}$ and a fourth one which commutes with only one of them. This set $K$ can be constructed as follows: consider the set $U_1$ with
$\sigma_{\alpha_i\beta_i}$, $i=1,2,3,4$, take the first three of them  and $\sigma_{\alpha'\beta'}$ from $U_2$: if  $\sigma_{\alpha'\beta'}$  commutes with only one $\sigma_{\alpha_i\beta_i}$, $i=1,2,3$, the set $K$ consists of these four $\sigma$'s.
Otherwise, if $\sigma_{\alpha'\beta'}$  commutes with two $\sigma_{\alpha_i\beta_i}$, then replace one of them with $\sigma_{\alpha_4\beta_4}$; as said before $\sigma_{\alpha'\beta'}$ does not commute with it because of the previous argument.
Therefore, these four elements form the set $K$ sought after.

The above procedure works also in the case of sets of anti-commuting $\sigma$'s with less than four elements.
\end{description}

The next two Lemmas concern Case $1$ above, where $I^c$ corresponds to a set of $5$ anti-commuting $\sigma_{\alpha\beta}$: the first lemma specifies the structure of the anti-commuting set, while the second one is about the sub-space generated by these matrices acting on a vector state $\vert\psi\rangle$.
\medskip

\begin{lemma}
\label{5anticom}
All sets of $5$ anti-commuting $\sigma_{\alpha_\ell\beta_\ell}$, $(\alpha_\ell,\beta_\ell)\in I^c$, must be of the form
$$
K_1=\{\sigma_{0i}\,,\,\sigma_{0j}\,,\,\sigma_{1k}\,,\,\sigma_{2k}\,,\,\sigma_{3k}\}\quad\hbox{or}\quad
K_2=\{\sigma_{i0}\,,\,\sigma_{j0}\,,\,\sigma_{k1}\,,\,\sigma_{k2}\,,\,\sigma_{k3}\}
$$
with $i,j,k$ such that the anti-symmetric tensor $\varepsilon_{ijk}\neq 0$.
\end{lemma}
\medskip

\proof
Since we have $5$ indices $\alpha_\ell$ and $\beta_\ell$ to choose among $0,1,2,3$, at least two $\alpha$ s
and two $\beta$ s must be equal: let us consider $\alpha_1=\alpha_2$.

If $\alpha_1=\alpha_2=0$, in order to have $\{\sigma_{0\beta_1}\,,\,\sigma_{0\beta_2}\}=0$, the corresponding $\beta_1$ and $\beta_2$ must be different and different from $0$; therefore, $\beta_1=i$, $\beta_2=j$, with $i\neq j$, $i\neq 0$, $j\neq 0$, so that the first two anti-commuting matrices are $\sigma_{0i}$ and $\sigma_{0j}$.
In the three $(\alpha_\ell\beta_\ell)$ left, there cannot appear $\alpha_\ell=0$, otherwise the corresponding $\beta_\ell$, which cannot be $0$ and must equal either $i$ or $j$, would lead to a $\sigma_{\alpha_\ell\beta_\ell}$ violating anti-commutativity. Then, with the three remaining $\alpha_\ell\neq 0$, the three corresponding $\beta_\ell$ must be different from $0$,  $i$ and $j$, otherwise $\sigma_{\alpha_\ell\beta_\ell}$ would commute with either $\sigma_{0i}$ or $\sigma_{0j}$. Thus, $\beta_3=\beta_4=\beta_5=\beta_k$ with $k\neq i$, $k\neq j$, while the corresponding $\alpha_\ell$ must be different, namely $\alpha_3=1$, $\alpha_4=2$ and $\alpha_5=3$. This gives the set $K_1$

If $\alpha_1=\alpha_2=k\neq 0$, the corresponding $\beta_{1,2}$ cannot be equal and must be both different from $0$. Thus, two $\sigma_{\alpha_\ell\beta_\ell}$ in the anti-commuting set are $\sigma_{ki}$ and
$\sigma_{kj}$, with $i\neq j=1,2,3$. Only one of the remaining $\alpha_\ell$ can be $0$, otherwise we would be back to the previous case: for $\sigma_{0\beta_\ell}$ to anti-commute with $\sigma_{ki}$ and $\sigma_{kj}$, $\beta_\ell$ must be equal to $s$ with $s\neq i$, $s\neq j$.
In order to anti-commute with $\sigma_{ki}$, $\sigma_{kj}$ and $\sigma_{0s}$, $\sigma_{\alpha_4\beta_4}$ and
$\sigma_{\alpha_5\beta_5}$, with $\alpha_{4,5}\neq 0$, must be of the form $\sigma_{pi}$ and $\sigma_{qj}$ with $p\neq q$ and both different from $0$ and $k$. But then, they would commute among themselves; thus, $\alpha_{3,4,5}\neq 0$. At least one of them, say $\alpha_3$ must equal $k$; then, anti-commutativity only holds if $\beta_3=r\neq 0$ with
$i\neq r$, $j\neq r$. Finally, since at least two $\beta$'s must be equal, in order to anti-commute among themselves and with $\sigma_{ki}$, $\sigma_{kj}$ and
$\sigma_{kr}$, $\sigma_{\alpha_4\beta_4}$ and $\sigma_{\alpha_5\beta_5}$ must have $\beta_4=\beta_5=0$ and $\alpha_4=m$, $\alpha_5=n$ such that the antisymmetric tensor $\epsilon_{kmn}\neq 0$.
This fixes the set $K_2$. The same result would obtain if one started
arguing with two equal $\beta$ s instead of $\alpha$ s.
\qed
\medskip

\begin{lemma}
\label{5lin-ind}
Let $K_{1,2}$ be the two sets of $5$ anti-commuting $\sigma_{\alpha\beta}$ as described in the previous lemma.
For any $\vert\psi\rangle\in\CI^4$, let
$$
K^\psi_{1,2}=\hbox{Linear Span}\{\sigma_{\alpha\beta}\vert\psi\rangle\,:\, \sigma_{\alpha\beta}\in K_{1,2}\}\ .
$$
Then, any vector orthogonal to $K_{1,2}^\psi$ is also orthogonal to $\vert\psi\rangle$.
\end{lemma}
\medskip

\proof
Consider the set $\{\sigma_{0i}\,,\,\sigma_{0j}\,,\,\sigma_{1k}\,,\,\sigma_{2k}\,,\,\sigma_{3k}\}$ and let $\vert0\rangle_2$ and $\vert1\rangle_2$ be the eigenvectors of $\sigma_k$. Furthermore, fix $i$ and $j$ so that
\begin{eqnarray*}
\sigma_k\vert0\rangle_2&=&\vert0\rangle_2\ ,\ \quad \sigma_k\vert1\rangle_2=\vert1\rangle_2\ ;\\ \sigma_i\vert0\rangle_2&=&\vert1\rangle_2\ ,\ \quad \sigma_i\vert1\rangle_2=\vert0\rangle_2\ ;\\
\sigma_j\vert0\rangle_2&=&i\vert1\rangle_2\ ,\ \,\,\sigma_j\vert1\rangle_2=-i\vert0\rangle_2\ .
\end{eqnarray*}
Given $\vert\psi\rangle\,,\,\vert\varphi\rangle\in\CI^4$, consider their expansions with respect to the orthonormal basis $\{\vert ij\rangle=\vert i\rangle_1\otimes\vert j\rangle_2\}_{i,j=0}^1$ where
$\vert i\rangle_1$, $i=0,1$, are eigenvectors of $\sigma_3$:  $\vert\psi\rangle=\sum_{i,j=0}^1\psi_{ij}\vert ij\rangle$ and
$\vert\varphi\rangle=\sum_{i,j=0}^1\varphi_{ij}\vert ij\rangle$.
Acting with the first four $\sigma_{\alpha\beta}$ on $\vert\psi\rangle$, we have:
\begin{eqnarray}
\nonumber
\sigma_{0i}\vert\psi\rangle&=&\psi_{00}\vert 01\rangle+\psi_{01}\vert 00\rangle+\psi_{10}\vert 11\rangle+\psi_{11}\vert 10\rangle\ ,\\
\nonumber
\sigma_{0j}\vert\psi\rangle&=&i(\psi_{00}\vert 01\rangle-\psi_{01}\vert 00\rangle+\psi_{10}\vert 11\rangle-\psi_{11}\vert 10\rangle)\\
\label{4vectors}
\sigma_{1k}\vert\psi\rangle&=&\psi_{00}\vert 10\rangle-\psi_{01}\vert 11\rangle+\psi_{10}\vert 00\rangle-\psi_{11}\vert 01\rangle\ ,\\
\nonumber
\sigma_{2k}\vert\psi\rangle&=&i(\psi_{00}\vert 10\rangle-\psi_{01}\vert 11\rangle-\psi_{10}\vert 00\rangle+\psi_{11}\vert 01\rangle)\ .
\end{eqnarray}
If $\vert\varphi\rangle$ is orthogonal to each of the previous vectors, it follows that
\begin{eqnarray*}
\overline{\varphi_{01}}\psi_{00}+\overline{\varphi_{11}}\psi_{10}=0\ ,&&
\overline{\varphi_{00}}\psi_{01}+\overline{\varphi_{10}}\psi_{11}=0\\
\overline{\varphi_{10}}\psi_{00}-\overline{\varphi_{11}}\psi_{01}=0\ ,&&
\overline{\varphi_{00}}\psi_{10}-\overline{\varphi_{01}}\psi_{11}=0\ .
\end{eqnarray*}
These relations recast in matrix equation read
$$
\begin{bmatrix}
  0 & \psi_{00} & 0 & \psi_{10}\\
  \psi_{01} & 0 & \psi_{11} & 0\\
  0 & 0 & \psi_{00} & -\psi_{01}\\
  \psi_{10} & -\psi_{11} & 0 & 0
 \end{bmatrix}
 \begin{bmatrix}
  \varphi_{00} \\
  \varphi_{01}  \\
  \varphi_{10} \\
   \varphi_{11}
 \end{bmatrix}
=0\ .
$$
The determinant of the matrix is $2\psi_{00}\psi_{01}\psi_{10}\psi_{11}$; thus, $\vert\varphi\rangle$ orthogonal to the linear span of the vectors in~(\ref{4vectors}) can only exist if all components of $\vert\psi\rangle$ are non-zero. Then,
\begin{eqnarray*}
\overline{\varphi_{01}}&=&-\frac{\overline{\varphi_{11}}\psi_{10}}{\psi_{00}}\ ,\ \overline{\varphi_{00}}=\frac{\overline{\varphi_{11}}\psi_{11}}{\psi_{00}}\ ,\ \overline{\varphi_{10}}=\frac{\overline{\varphi_{11}}\psi_{01}}{\psi_{00}}\quad\hbox{and}\\
\vert\varphi\rangle&=&\overline{\psi_{11}}\vert 00\rangle+\overline{\psi_{10}}\vert 01\rangle-\overline{\psi_{01}}\vert 10\rangle-\overline{\psi_{00}}\vert 1\rangle\ .
\end{eqnarray*}
Such a vector results orthogonal to both the fifth vector $\sigma_{3k}\vert\psi\rangle$ and $\vert\psi\rangle$ itself.
Similar considerations hold for the set $\{\sigma_{i0}\,,\,\sigma_{j0}\,,\,\sigma_{k1}\,,\,\sigma_{k2}\,,\,\sigma_{k3}\}$.
\qed
\medskip

The next Lemma concerns the dimensionality of the linear spans of vectors of the form $\sigma_{\alpha\beta}\vert\psi\rangle$ where the matrices belong to the sets discussed in Cases $2$ and $3$ in the proof of Theorem~\ref{theo2} where three anti-commute and the fourth one commutes with only one of them.
\medskip

\begin{lemma}
\label{lem3}
Consider a sub-set $K=\{\sigma_{\alpha_{i}\beta_{i}}:i=1,...,4\}$ consisting of three anti-commuting matrices $\sigma_{\alpha\beta}$ plus a fourth one which commutes with only one of these three, say the first:
\begin{equation}
\label{Lindalg}
\{\sigma_{\alpha_{i}\beta_{i}}\,,\,\sigma_{\alpha_{j}\beta_{j}}\}=0\ , \ i,j=1,3\quad
[\sigma_{\alpha_{4}\beta_{4}}\,,\,\sigma_{\alpha_1\beta_1}]=0\ .
\end{equation}
Then, unless $\vert\psi\rangle\in\CI^4$ is an eigenstate of some $\sigma_{\mu\nu}$, $V_K=\CI^4$, where $V_K$ is the linear span of $\{\sigma_{\alpha_i\beta_i}\vert\psi\rangle\}_{i=1}^4$.
\end{lemma}
\medskip

\proof
Suppose $\vert\psi\rangle$ is not an eigenstate of any $\sigma_{\mu\nu}$, then the vectors
$\sigma_{\alpha_{1}\beta_{1}}\vert\psi\rangle$, $\sigma_{\alpha_{2}\beta_{2}}\vert\psi\rangle$ cannot be proportional.
For the same reason, the vectors $\sigma_{\alpha_{1}\beta_{1}}\vert\psi\rangle$, $\sigma_{\alpha_{2}\beta_{2}}\vert\psi\rangle$ and $\sigma_{\alpha_{4}\beta_{4}}\vert\psi\rangle$ are linearly independent. Indeed, suppose
$$
\sigma_{\alpha_{4}\beta_{4}}\vert\psi\rangle=\alpha\sigma_{\alpha_{1}\beta_{1}}\vert\psi\rangle +  \beta\sigma_{\alpha_{1}\beta_{1}}\vert\psi\rangle\ .
$$
Then, acting on both sides with $\sigma_{\alpha_{4}\beta_{4}}$ we obtain
$\vert\psi\rangle=\alpha\,S_1\vert\psi\rangle +  \beta\,S_2\vert\psi\rangle$, where
$S_i=\sigma_{\alpha_4\beta_4}\sigma_{\alpha_i\beta_i}$, $i=1,2$.
When substituting this expression for $\vert\psi\rangle$ in the right hand side of the equality,
the relations~(\ref{Lindalg}) yield
$$
\vert\psi\rangle=(\alpha^{2}-\beta^{2})\vert\psi\rangle+ \alpha\beta\,[\sigma_{\alpha_1\beta_1}\,,\,\sigma_{\alpha_2\beta_2}]\vert\psi\rangle\ ,
$$
which, for non-trivial $\alpha\beta$, is only possible if $\vert\psi\rangle$ is an eigenstate of the
the $\sigma_{\mu\nu}$ proportional to the commutator in the previous equality.
The same conclusions can be drawn about the linear independence of the four vectors spanning $V_K$.
Indeed, if
$$
\sigma_{\alpha_{3}\beta_{3}}\vert\psi\rangle= \alpha\sigma_{\alpha_{1}\beta_{1}}\vert\psi\rangle + \beta \sigma_{\alpha_{2}\beta_{2}}\vert\psi\rangle+ \gamma \sigma_{\alpha_{4}\beta_{4}}\vert\psi\rangle\ ,
$$
the same argument of before obtains
$$
\vert\psi\rangle=-(\alpha^2+\beta^2+\gamma^2)\vert\psi\rangle\,+\,\alpha\gamma\,
\{\sigma_{\alpha_1\beta_1}\,,\,\sigma_{\alpha_4\beta_4}\}\vert\psi\rangle\ .
$$
This implies that, unless $\vert\psi\rangle$ is eigenstate of the $\sigma_{\mu\nu}$ arising from the anti-commutator in the above expression, the four vectors spanning $V_K$ are linearly independent.
\qed
\medskip

The previous results show that, as a consequence of the assumption that no special quadruple $Q\in\cQ_{00}$ is contained in $I$, given a vector $\vert\psi\rangle\in\CI^4$ the sub-space linearly generated by $\sigma_{\alpha\beta}\vert\psi\rangle$ with $(\alpha\beta)\in I^c$ is either $\CI^4$ or contains $\vert\psi\rangle$.
The proof of Theorem~\ref{theo2} is thus completed by using Lemma~\ref{lem1} and its corollary.
\qed
\medskip

\paragraph*{}
It might look like that, when all the points of the lattice have at least one special quadruple contained in I, the corresponding lattice state might be separable. But in the following examples, we will show that this is not in general true.

\begin{example}
\label{ex6.16}
Consider the following state:
$$
N_I=11 \quad
 \begin{array}{c|c|c|c|c}
               3 & \quad\! & \times & \times  &  \times      \\
               \hline
               2 & \quad\! & \quad\! &  \quad\! &  \times      \\
               \hline
               1 &  \quad\! & \times &  \times & \times       \\
               \hline
               0 & \times & \times  & \times  & \times      \\
               \hline
                 & 0 & 1 & 2 & 3
 \end{array}
$$
This looks very similar to the state in example~\ref{ex6.15}, where none of the special quadruples of $(0,0)$ were included in I. However in the above lattice state, $(0,0)$ has two special quadruples contained in I, which are:
$$
 \begin{array}{c|c|c|c|c}
               3 & \quad\! & \quad\! & \times  &  \quad\!      \\
               \hline
               2 & \quad\! & \quad\! &  \quad\! &  \times      \\
               \hline
               1 &  \quad\! & \times &  \quad\! & \quad\!      \\
               \hline
               0 & \times & \quad\!  & \quad\!  & \quad\!      \\
               \hline
                 & 0 & 1 & 2 & 3
 \end{array}\quad \hbox{and}\quad
  \begin{array}{c|c|c|c|c}
               3 & \quad\! & \times & \quad\!  &  \quad\!      \\
               \hline
               2 & \quad\! & \quad\! &  \quad\! &  \times      \\
               \hline
               1 &  \quad\! & \times &  \times & \quad\!      \\
               \hline
               0 & \times & \quad\!  & \quad\!  & \quad\!      \\
               \hline
                 & 0 & 1 & 2 & 3
 \end{array}
$$
But from Proposition~\ref{PPT}, we know that this state is NPT: the contribution to I from the row and the column for the point $(3,0)$ is $6>\frac{11}{2}$.
\end{example}

\newpage

\chapter*{Summary and Outlook}
\addcontentsline{toc}{chapter}{Summary and Outlook}

In this thesis work, we have studied the role of positive and completely positive maps in detecting entanglement. In the first chapters we have reviewed the necessary techniques that have been employed in the last chapter.

Indeed, in the final chapter of the thesis
we have considered a particular class of bipartite states $\rho_I$ on $\CI^4\otimes\CI^4$, introduced in~\cite{benatti2}, called Lattice States.
These are uniform mixtures of projections indexed by points $(\alpha,\beta)$  belonging to subsets $I$ of the finite square lattice of cardinality $16$. The projections are generated by the action of $\1_4\otimes\sigma_{\alpha\beta}$ on the completely symmetric vector in $\CI^4\otimes\CI^4$, where $\sigma_{\alpha\beta}$ is the tensor product of the Pauli matrices $\sigma_{\alpha}$, $\sigma_\beta$ .
We have generalized them to states on $\CI^{2^n}\otimes\CI^{2^n}$, referred to as $\sigma-$diagonal states.

One of the main issues was to decide whether a given $\sigma-$diagonal state, is entangled or separable. We have tackled this problem using the results of~\cite{storm2} and showed that, starting from a general non-diagonal positive map, possible entanglement of $\sigma-$diagonal states would be revealed using only a particularly simple sub-class of positive maps adapted to the diagonal structure of the states, that are combinations of elementary maps of the form $X\mapsto \sigma_{\alpha\beta}\,X\,\sigma_{\alpha\beta}$.

Back to the Lattice States in $\CI^4\otimes\CI^4$, we have shown the compatibility of the results given in~\cite{{benatti3},{benatti4}}, with ours.

As the next step, we have partially shown how the entanglement $($respec-\\tively, separability$)$ of the Lattice States can be explained through their geometrical structure. The results obtained are based on particular separable lattice states consisting of four points, called special quadruples, which play a crucial role in this game.

Using this notion, we have shown that if there exists a point in the set $I$, such that none of its special quadruples are contained in $I$, then the corresponding Lattice State $\rho_I$, is entangled. We have also provided a proper entanglement witness for such a state
based on the results obtained in the first part of the chapter.

Also, using the notion of uniform covering by special quadruples contained in the subset $I$, we could show that some of the Lattice States whose separability was left unknown in previous works, are indeed separable:
$$
 \begin{array}{c|c|c|c|c}
               3 & \quad\!  & \times  & \times   & \times   \\
               \hline
               2 & \quad\!  & \times  & \times   & \times \\
               \hline
               1 &  \quad\!  & \times  & \times   & \times   \\
               \hline
               0 & \times & \quad\! & \quad\!     \\
               \hline
                 & 0 & 1 & 2 & 3
 \end{array}\quad,\quad
  \begin{array}{c|c|c|c|c}
               3 & \quad\!  & \times  & \times   & \times   \\
               \hline
               2 & \quad\!  & \times  & \quad\!   & \times \\
               \hline
               1 &  \quad\!  & \times  & \times   & \times   \\
               \hline
               0 & \times & \quad\! & \quad\!     \\
               \hline
                 & 0 & 1 & 2 & 3
 \end{array}\quad,\quad
  \begin{array}{c|c|c|c|c}
               3 & \quad\!  & \times  & \times   & \times   \\
               \hline
               2 & \quad\!  & \times  & \quad\!  & \times \\
               \hline
               1 &  \quad\!  & \times  & \quad\!  & \times   \\
               \hline
               0 & \times & \quad\! & \quad\!     \\
               \hline
                 & 0 & 1 & 2 & 3
 \end{array}.
$$

\paragraph*{}
However, there are still some Lattice States whose separability could not be decided by the methods developed in this thesis; for instance,
$$
 \begin{array}{c|c|c|c|c}
               3 & \times  & \quad\!   &  \quad\! &  \times   \\
               \hline
               2 & \times  & \times & \quad\!  &  \times   \\
               \hline
               1 &  \times  & \times   & \times &  \quad\!    \\
               \hline
               0 & \times & \times & \times &      \\
               \hline
                 & 0 & 1 & 2 & 3
 \end{array}.
 $$
For this Lattice State, all points have at least three special quadruples contained in I, but no uniform covering could be found, nor could be excluded.

\paragraph*{}
Another interesting issue, still to be understood, is the relation between the special quadruples and partial transposition.The difficulties can be appreciated by looking at the following state:
$$
 \begin{array}{c|c|c|c|c}
               3 & \times  & \quad\!   &  \quad\! &  \times   \\
               \hline
               2 & \times  & \quad\! & \quad\!  &  \times   \\
               \hline
               1 &  \quad\!  & \quad\!   &  \times &  \times   \\
               \hline
               0 & \times & \times & \times &   \times   \\
               \hline
                 & 0 & 1 & 2 & 3
 \end{array},
 $$
which is NPT entangled, nevertheless all points belong to at least one special quadruples contained in $I$. Obviously, in this case there does not exist a uniform covering of special quadruples as the state is entangled.

\paragraph*{}
Further interesting issues to be tackled  are whether separable Lattice States carry quantum discord and whether the structure of PPT entangled Lattice States could be investigated in terms of
Edge States and Unextendible Product Bases.

\newpage

\appendix
\chapter{Convexity}

A good introduction of Convex Analysis is given in~\cite{{barv},{rock}}.

\begin{definition}\textbf{Convex Set}

Let $S$ be a finite set of points in $\RI^d$. $S$ is said to be convex if
\begin{equation}
\alpha x+(1-\alpha)y\in S\qquad \forall x,y\in S\quad 0\leq\alpha\leq1.
\end{equation}
Consequently a point $y\in\RI^d$ is said to be a convex combination of $\{x_i\}_{i=1}^n$ $\subseteq\RI^d$ if:
\begin{equation}
y=\sum_{i}p_i x_i, \quad \sum_i p_i=1,\quad p_i\geq0\quad\hbox{for }i=1,...,n.
\end{equation}
\end{definition}

\begin{example}
The set ofstates and the subset of separable states $\rho_s\in M_n(\CI)$ are convex sets.
\end{example}

\begin{definition}\textbf{Convex hull}

Convex hull of a convex set $S$, denoted by $conv(S)$ is the set of all the convex combinations of the elements of $S$:
$$
conv(S)=\{ \sum_{i}p_i x_i, \quad \sum_i p_i=1,\quad p_i\geq0\; x_i\in S\}.
$$
Convex hull of $S$ is the smallest convex set containing $S$.
\end{definition}

Let $S$ be a convex set. Then its elements split into two subsets:

\begin{definition}\textbf{Interior points}

The point $x\in S$ is an interior point if the line segment from any point in $S$ to $x$ can be extended within $S$.
\end{definition}

\begin{definition}\textbf{Extreme points}

A point $y\in S$ is said to be extreme if:
$$
y=\alpha x_1+(\1-\alpha) x_2,\quad x_1,x_2\in S\Longrightarrow (y=x_1 \wedge\alpha=1)\vee(y=x_2\wedge\alpha=0).
$$
\end{definition}
In other words, extreme points of $S$ do not belong to any proper segment defined by two distinct elements of $S$.

\begin{example}
  The pure states or projectors are extreme points for the set of density matrices, whereas the density states of mixed states are the interior points.
\end{example}

When an inner product $\langle.\vert.\rangle$ is defined on $\cC$, we have:

\begin{definition}\textbf{Hyperplane}

Let $a$ and $x_0$ be vectors in $\RI^d$. Then a hyperplane is a set of the form $\{x|\langle a\vert(x-x_0)\rangle=0\}$. The vector $a$ is called the normal vector for the hyperplane.
\end{definition}

The hyperplane divides $\RI^d$ into two halfspaces:
\begin{definition}\textbf{Halfspace}
 A halfspace is a set of the form:
$$
\{x|\langle a\vert(x-x_0)\rangle\leq0\}
$$
\end{definition}

The concept of hyperplanes and halfspaces are very useful as using them we can separate the convex sets which do not intersects:
\begin{theorem}

Let $C$ and $D$ be two convex sets such that $C\cap D=\varnothing$. Then there exists $a\neq0$ and $x_0$ such that $\langle a\vert(x-x_0)\rangle\leq b$ for all $x\in C$ and $\langle a\vert(x-x_0)\rangle\geq b$ for all $x\in D$. In another words, $\langle a\vert(x-x_0)\rangle$ is non-positive on $C$ and non-negative on $D$.
\end{theorem}

\begin{center}
\label{fig5.1}
\includegraphics[scale=0.5]{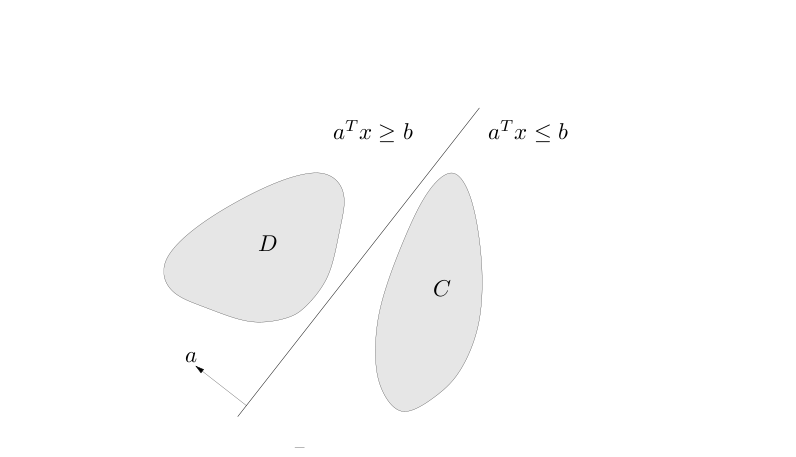}

\end{center}

If a hyperplane is tangent to the set, at some extreme point $x_0$, then it is called supporting hyperplane. Note also that the separating hyperplanes as well as supporting hyperplanes are not unique.

\begin{definition}\textbf{Cones}

A set $\cC\subseteq\RI^d$ is a cone if
$$
x\in\cC,\quad \alpha\geq0\Longrightarrow\alpha x\in\cC.
$$
$\cC$ is said to be a convex cone if:
$$
x,y\in\cC,\quad \alpha,\beta\geq0\Longrightarrow\alpha x+\beta y\in\cC.
$$
\end{definition}

To any convex cone one can associate a dual cone.
\begin{definition}\textbf{Dual Cone}

The dual cone  $\cC^{\circ}$ is defined to be:
$$
\cC^{\circ}=\{y:\langle y\vert x\rangle\geq0,\quad\forall x\in\cC\}.
$$
\end{definition}
The dual cone $\cC^{\circ}$ is always closed, even if $\cC$ is open.

\begin{example}

\end{example}

When the cone and dual cone coincide then the cone is said to be self-dual, i.e. :
$$
\langle y\vert x\rangle\geq0 \quad \forall x\in\cC \Longleftrightarrow y\geq 0.
$$

\begin{example}
 The cone of positive semidefinite $n\times n$ matrices is self-dual. This also holds for the cone of completely positive maps: $\cCP=\cCP^{\circ}$.
\end{example}

\begin{example}
All these concepts can be extended to the Hermitian operators and linear maps as well. Indeed one can easily verify that the set of positive operators is a convex set, and that positive (resp. complete positive) maps are unaffected by multiplication by positive scalars.

Let $\cC$ denote the cone of (positive, completely positive) linear maps, the dual cone $\cC^{\circ}$ in this case is defined via the Jamiolkowski Isomorphism. Let $\{E_{ij}\}$ be an orthonormal set of unit matrices, then the Choi matrix $C_{\phi}$, for the map $\phi:M_n(\CI)\longrightarrow M_n(\CI)$ is:
\begin{equation}
\label{choi-phi}
C_{\phi}=\sum E_{ij}\otimes\phi(E_{ij}).
\end{equation}
Using the Hilbert-Schmidt scalar product introduced in~\eqref{H-S norm}, the dual cone $\cC^{\circ}$ is defined to be:
\begin{equation}
\label{dual-cone}
\cC^{\circ}=\{\psi:M_n(\CI)\longrightarrow M_n(\CI): \Tr(C_{\phi}C_{\psi})\geq0,\quad\forall\phi\in\cC\}.
\end{equation}

Let $P(\CI^d)$ be the set of positive linear maps $\phi:M_n(\CI)\longrightarrow M_n(\CI)$, and $\cC$ be a closed cone in $P(\CI^d)$. Recall that the set of complete positive maps is denoted by $\cCP$.
\end{example}

\newpage

\bibliographystyle{ieeetr}

\end{document}